\documentclass[aps,longbibliography,showpacs,twocolumn,superscriptaddress,amsmath,amssymb,verbatim,floatfix]{revtex4-2}
\usepackage{graphicx}
\usepackage{lmodern}
\usepackage{epstopdf}
\usepackage{dcolumn}
\usepackage{bm}
\usepackage{subfigure}
\usepackage{amsmath} 
\usepackage[pdftex,colorlinks=true,citecolor=blue,linkcolor=blue,urlcolor=blue,bookmarks=true]{hyperref}
\usepackage{braket}
\usepackage{varwidth}
\usepackage{booktabs,siunitx}
\usepackage{tabularx}
\usepackage{blindtext} 
\usepackage{lipsum}
\usepackage{graphicx}
\usepackage{array}
\usepackage{tikz}
\usepackage{wasysym}
\usepackage{physics}
\usepackage{xcolor}
\usepackage{color}
\usepackage{bm}
\usepackage{setspace}
\usepackage{textcomp}
\usepackage{enumitem}
\usepackage[utf8]{inputenc}
\usepackage{booktabs}

\usepackage{natbib}

\usepackage{hyperref}
\hypersetup{
	colorlinks=true,
	urlcolor= blue,
	citecolor=blue,
	linkcolor= blue,
	bookmarks=true,
	bookmarksopen=false,
}

\allowdisplaybreaks[1]
\def\vec#1{{\bf #1}}
\usepackage{hyperref,mathtools}

\newcommand{\herb}{herbertsmithite\xspace} 

\usepackage{color}
\usepackage{textcomp}
\definecolor{red}{rgb}{1,0,0}

\definecolor{blue}{rgb}{0,0,1}

\definecolor{green}{rgb}{0,1,0}

\usepackage[normalem]{ulem} 

\definecolor{grAZ}{rgb}{0,0.5,0.5}

\begin{document}
	\preprint{APS}
\title{\textbf{Experimental signatures of quantum and topological states in frustrated magnetism}}
\author{J. Khatua}
\altaffiliation{Equally contributed}
\affiliation{Department of Physics, Indian Institute of Technology Madras, Chennai 600036, India}
\author{B. Sana}
\altaffiliation{Equally contributed}
\affiliation{Department of Physics, Indian Institute of Technology Madras, Chennai 600036, India}
	\author{A. Zorko}
	\email[Corresponding author. ]{andrej.zorko@ijs.si}
	\affiliation {Jo\v{z}ef Stefan Institute, Jamova c. 39, 1000 Ljubljana, Slovenia}
	\affiliation{Faculty of Mathematics and Physics, University of Ljubljana, Jadranska u. 19, 1000 Ljubljana, Slovenia}
\author{M. Gomil\v{s}ek}
 \email[Corresponding author. ]{matjaz.gomilsek@ijs.si}
	\affiliation {Jo\v{z}ef Stefan Institute, Jamova c. 39, 1000 Ljubljana, Slovenia}
		\affiliation{Faculty of Mathematics and Physics, University of Ljubljana, Jadranska u. 19, 1000 Ljubljana, Slovenia}
\author{K. Sethupathi}
\affiliation{Department of Physics, Indian Institute of Technology Madras, Chennai 600036, India}
\affiliation{Quantum Centre for Diamond and Emergent Materials, Indian Institute of Technology Madras,
	Chennai 600036, India.}
\author{M. S. Ramachandra Rao}
\affiliation{Quantum Centre for Diamond and Emergent Materials, Indian Institute of Technology Madras,
	Chennai 600036, India.}
\affiliation{Department of Physics, Nano Functional Materials Technology Centre and Materials Science Research Centre,
	Indian Institute of Technology Madras, Chennai-600036, India}
		\author{M. Baenitz}
	\affiliation{Max Planck Institute for Chemical Physics of Solids, 01187 Dresden, Germany}
	\author{B. Schmidt}
	\affiliation{Max Planck Institute for Chemical Physics of Solids, 01187 Dresden, Germany}
\author{P. Khuntia}
\email[Corresponding author. ]{pkhuntia@iitm.ac.in}
\affiliation{Department of Physics, Indian Institute of Technology Madras, Chennai 600036, India}
\affiliation{Quantum Centre for Diamond and Emergent Materials, Indian Institute of Technology Madras,
	Chennai 600036, India.}
\date{\today}

\begin{abstract}
Frustration in magnetic materials arising from competing exchange interactions can prevent the system from adopting long-range magnetic order and can instead lead to a diverse range of novel quantum and topological states with exotic quasiparticle excitations.
Here, we review prominent examples of such states, including magnetically-disordered and extensively degenerate spin ices,with emergent magnetic monopole
excitations, highly-entangled quantum spin liquids with fractional spinon excitations, topological order and emergent gauge fields, as well as complex particle-like topological spin textures known as skyrmions. 
We provide an overview of recent advances in the search for magnetically-disordered candidate materials on the three-dimensional pyrochlore lattice and two-dimensional triangular, kagome and honeycomb lattices, the latter with bond-dependent Kitaev interactions, and on lattices supporting topological magnetism.
We highlight experimental signatures of these often elusive phenomena and single out the most suitable experimental techniques that can be used to detect them.
Our review also aims at providing a comprehensive guide for designing and investigating novel frustrated magnetic materials, with the potential of addressing some important open questions in contemporary condensed matter physics. %
\end{abstract}

\maketitle

\tableofcontents{}

\section{\textbf{Introduction}}
\label{intro}
\textcolor{black}{Concerted efforts in science have been continuously devoted towards discovery and understanding of elementary particles and quasiparticles in various settings, ranging from particle physics to condensed matter~\cite{lee1956question, higgs1964broken, skyrme1962unified, dirac1928quantum, majorana1937teoria, pines1981elementary, wang2018experimental, nandkishore2019fractons}. 
In the latter case, emergent many-body states and excitations are found to be governed by complementary principles of symmetry, topology, and quantum mechanics that underlie some of the most profound fundamental discoveries and novel scientific paradigms in recent years. 
These include the appearance of high-$T_c$ superconductivity, massless fermions (electrons and holes) in
graphene, protected surface currents in topological insulators, exotic quasiparticles in spin systems, and the possibility of
quantum bits (qubits) in a range of condensed matter settings.
In addition to their importance in fundamental science, these novel phenomena hold the potential for enabling next-generation technologies that could address the world's ever increasing demands for energy  production, storage, and efficiency, high-density and low-power data storage, as well as
quantum computation, metrology, and sensing~\cite{Bednorz1986,Fert2013,divincenzo2000physical,Nayak_2008, KITAEV20062,giovannetti2006quantum, bollinger1996optimal}.
}

\textcolor{black}{Competition of interactions between constituents of matter, commonly referred to as frustration, is at the heart of many interesting phenomena, from protein folding to emergent electromagnetism, but reaches its purest manifestations in magnetic materials~\cite{lacroix2011introduction,Castelnovo2008,Bramwell2009,frauenfelder1991energy,Balents2010}. 
Even in some of the simplest frustrated two-dimensional (2D) spin lattices with triangular motifs, and three-dimensional (3D) lattices with tetrahedral motifs, competing interactions can conspire in a way to make it impossible for any spin state to simultaneously minimize the energy of all individual pair-wise interactions (Fig.~\ref{all_frustrated}).
As this destabilizes ‘‘simple’’ spin states (e.g., ordered states like collinear ferromagnetic or antiferromagnetic arrangements
of spins), frustrated spin lattices offer an ideal platform for realizing complex magnetic states with emergent quasiparticle excitations, which has far reaching implications in physics and materials science~\cite{lacroix2011introduction}. 
Prominent examples of novel unconventional magnetic phenomena arising from frustration include the appearance of spin ices~\cite{Bramwell1495,Castelnovo_2012} and quantum spin liquids (QSLs)~\cite{Balents2010,Savary_2016,RevModPhys.89.025003, Knolle_2019, wen2019choreographed,broholm2020quantum, hermanns2018physics, TREBST20221,Takagi2019}, their associated exotic excitations, such as magnetic monopoles and Dirac strings~\cite{Castelnovo2008}, fractionalized spinons \cite{Savary_2016,Knolle_2019}, anyons and Majorana fermions~\cite{KITAEV20062}, as well as topological properties~\cite{wen2019choreographed} and topologically non-trivial spin-textures, such as skyrmions~\cite{Fert2013, niitsu2022geometrically}. 
Exploiting and manipulating these exotic states is highly relevant for technological innovation in quantum computing and high-density data storage~\cite{KITAEV20062,Castelnovo_2012,Balents2010,hsieh2008topological,tokura2017emergent,Tokura_2014,witczak2014correlated}. 
Understanding the microscopic origin, and establishing paradigmatic theoretical models that capture, these frustration-induced phenomena, is, therefore, among the top priorities in condensed matter research~\cite{RevModPhys.89.040502,anderson1987resonating,wen2004quantum,RevModPhys.89.041004, powell2020emergent}. 
There is  currently a substantial effort focused on the discovery and investigation of novel materials that realize these phenomena,
and on the development of powerful experimental approaches that would be able to clearly and unambiguously characterize them. 
The aim of this review is to present both paradigmatic and novel examples of emergent phases and phenomena in frustrated quantum and topological magnetic materials. 
We focus on systems based on the most common frustrated lattices, the pyrochlore lattice in 3D, triangular, kagome and honeycomb lattices in 2D, and on lattices that stabilize topological spin textures.
Special emphasis is put on experimental signatures of frustration-stabilized phenomena via various complementary state-of-the-art experimental techniques.
}

\textcolor{black}{The first illustrative example we present are spin-ice states on the 3D pyrochlore lattice, where the orientations of magnetic moments can be mapped to a analogue water-ice structure~\cite{Ramirez1999}. 
Consequently, spin ices ground state are macroscopically degenerate, as an extensive number of spin configurations are found to have the same ground-state energy.
Furthermore, they are characterized by emergent magnetic monopole excitations that arise from the fractionalization of spin degrees of freedom~\cite{Castelnovo2008}, 
and which act as well-defined emergent quasiparticles with topological properties~\cite{Castelnovo_2012}.
In some cases, pronounced quantum-mechanical fluctuations in these systems can even lead to a quantum version of this state, called a quantum spin ice state, which is an example of a quantum spin liquid (QSL), which we tackle
next~\cite{PhysRevX.1.021002,gingras2014quantum}.
\textcolor{black}{Quantum spin liquids are especially prominent example of emergent states of frustrated magnets.
A QSL is a highly quantum-entangled state with non-local spin correlations, where strong quantum fluctuations competely suppress the onset of long-range magnetic ordering, even at zero temperature~\cite{Balents2010,Savary_2016}, leading to a highly dynamical (i.e., persistently
fluctuating) ground state.
In this state, an electron effectively splits into two collective excitations: a spinon (a spin-
1/2 quasiparticle with zero electric charge) and a holon (a spin-zero quasiparticle with unit electric charge), which
behave as independent quasiparticles.
As holons are bosons they could potentially condense, giving rise to exotic superconductivity~\cite{RevModPhys.89.041004}.
Spinons, on the other hand, are accompanied by emergent gauge fields that mediate interactions between them~\cite{powell2020emergent} and can possess either fermionic, bosonic, 
or even more exotic anyonic exchange statistics~\cite{RevModPhys.89.025003}. 
Emergent gauge fields can obey various local symmetries~\cite{RevModPhys.89.025003}, including $\mathrm{SU}(N)$, $\mathrm{U}(1)$, and $\mathbb{Z}_2$.
The low-energy effective theory for this state is thus in many cases a topological quantum field theory~\cite{RevModPhys.89.025003,kitaev2006anyons,wen2004quantum}.
With a broad array of possible effective emergent elementary quasiparticles (various flavour of spinons) and interactions between them (emergent gauge fields), QSLs can thus be understood as effectively simulating alternate, (usually) lower-dimensional universes,
with emergent particle physics that is described in the same theoretical language as the usual Standard model of
particle physics, but with a different (tunable) set of fundamental particle and interactions between them~\cite{wen2019choreographed,RevModPhys.86.1189,wen2004quantum}. 
As fundamentally different QSLs which arise from different models of geometrical frustration, number in the hundreds~\cite{RevModPhys.89.041004}, frustrated magnets provide an ideal test bed to study the complex behaviour of exotic interacting quasiparticles with fractional quantum numbers. 
Perhaps the  cleanest example of this is the exactly-solvable Kitaev model with bond-dependent and highly-anisotropic Ising interactions on a honeycomb lattice~\cite{KITAEV20062}.
Its resulting Kitaev QSL ground state is characterized by deconfined fractional Majorana fermions (fermionic particles that are their own antiparticles) and localized $\mathbb{Z}_2$ gauge flux excitations with mutual anyonic statistics~\cite{kitaev2006anyons}.
In real materials, such a state is the result of strong spin--orbit coupling and specific geometrical constraints~\cite{PhysRevLett.102.017205}.
}\\
\textcolor{black}{Next we turn to experimental signatures of these phenomena. Both quantum entanglement and fractional excitations are the defining characteristics of QSLs, and are therefore of particular interest.
However, being non-local in character,
they are generally highly elusive and therefore hard to unambiguously detect in experiment~\cite{Knolle_2019, Wen2019}. As QSL states
typically do not break any lattice symmetries, in contrast to magnetically ordered states, and remain dynamical at all
temperatures, one also cannot rely on usual order parameter measurements.
Perhaps the most informative experimental techniques that can be used to detect and characterise QSL candidates efficiently are those that use local magnetic probes, which includes muon spectroscopy ($\mu$SR), nuclear magnetic resonance (NMR), and electron spin resonance (ESR).
Complementary techniques like inelastic neutron scattering (INS) and thermal conductivity are crucial as well. 
The complementarity of all of these experimental techniques is of the utmost importance for unambiguous identification of novel quantum and topological states in frustrated magnetic systems.
}
\\ \textcolor{black}{Interacting systems usually undergo a phase transition at low $T$ that is governed by a local order parameter. 
These classical phase transitions are associated with the breaking of a certain symmetry, as described by the seminal theory of Landau.
Quantum phase transitions, on the other hand, may instead occur at zero $T$ where thermal fluctuations are absent.
These transitions are a manifestation of the competition between nearly-degenerate phases of the system near a quantum critical point (QCP).
They can be driven by external tuning parameters and provide an excellent route to realize exotic quantum phases that go beyond Landau's paradigm~\cite{sachdev1999quantum}. 
Here, quantum fluctuations play a decisive role and lead to unconventional scaling laws for thermodynamic and microscopic quantities in the vicinity of a QCP. 
The paradigmatic concepts of quantum phase transitions and quantum criticality thus offer a unique way of realizing novel and diverse quantum states, ranging from various QSLs to superconductivity, via a suitable choice of the host system and the values of its tuning parameters, like pressure or applied magnetic field~\cite{Sebastian2006,PhysRevLett.95.177001,baek2017evidence,banerjee2018excitations,vojta2003quantum,witczak2014correlated}.
}}\\
\textcolor{black}{Finally, the lack of classical phase transitions in QSLs often requires us to consider their topological aspects~\cite{wen2002quantum,wen2004quantum,RevModPhys.89.025003,wen2019choreographed}.
Topological phases of matter are those that host non-local order parameters or invariants and are thus robust against local perturbations such as impurities, defects, or applied fields.
They can be realized in many diverse settings such as in particle physics, metals,  superconductors, but also, very prominently, in frustrated magnetic systems~\cite{hasan2010colloquium,qi2011topological,armitage2018weyl,wen2019choreographed}.
The various unconventional phases of frustrated magnets that are characterized by topological order include spin ices~\cite{Castelnovo_2012}, QSLs~\cite{broholm2020quantum, Savary_2016}, and complex topological spin textures including skyrmions~\cite{niitsu2022geometrically}, which are described by topological Berry curvature and the related anomalous Hall effect~\cite{berry1984quantal,nakatsuji2015large,RevModPhys.82.1539}.
}\\
\textcolor{black}{
There are several extensive and high-quality reviews of frustrated magnets already available in the literature, including general reviews of quantum spin liquids~\cite{Balents2010,Savary_2016,RevModPhys.89.025003, Knolle_2019, wen2019choreographed,broholm2020quantum}, as well as more specialized reviews focused of Kitaev spin liquids~\cite{hermanns2018physics, TREBST20221, Takagi2019} and spin ices~\cite{Bramwell1495, Castelnovo_2012, gingras2014quantum, Bramwell_2020}. 
As the field is growing rapidly, the aim of the present review is to overview the variety of novel physical phenomena recently discovered in the field, focusing mainly on their quantum and topological aspects.
The selected cases portray some of the most extensively studied frustrated magnets in recent years, as well as highlight promising novel materials, with an emphasis on the experimental signatures of these phenomena.
Our review aims at providing deeper insights, which are vital in the quest for new design routes and further development of new materials with optimized functionalities.
The review is organized as follows. 
Section~\ref{Exchan} starts with an overview of magnetic interactions that are commonly present in correlated quantum materials and introduces the concept of geometrical frustration. 
Section~\ref{spinicc} describes the basic concepts of the spin-ice ground state and related magnetic monopole and Dirac-string excitations in 3D pyrochlores. 
Section~\ref{qsll} introduces the concept of quantum spin liquids and reviews the most 
extensively studied QSL spin lattices and their realizations in 2D.
Section~\ref{expsig} elucidates the main experimental signatures of QSL and gives a brief overview of complementary experimental techniques that are best suited for probing static and dynamical properties of frustrated quantum and topological magnets.
Selected examples demonstrating the specific strengths of each technique are presented. 
Section~\ref{top} is dedicated to important topological
aspects of frustrated magnets, with illustrative examples of topologically non-trivial spin textures with non-vanishing
Berry curvature.  
Section~\ref{out} concludes the review with an outlook on possible future directions of research in the field, including the most promising families of materials to study and the development of novel experimental techniques, theoretical approaches, and applications.
}

\section{\textbf{Magnetic interactions and external stimuli in frustrated magnets}} \label{Exchan}
\textcolor{black}{
Understanding the role of competing exchange interactions, anisotropy, and external stimuli on the magnetic ground state of materials is a fundamental problem addressed in the field of frustrated magnetism.} 
Before embarking on a discussions of unusual spin states, such as spin ices, spin liquids, and topological magnetic textures, we thus start with an introduction of the concept of geometric frustration and a brief discussion of its origin \textcolor{black}{and experimental signatures}. 
The idea of frustration originates from the seminal work of Linus Pauling on the quenched disorder of crystalline water ice~\cite{doi:10.1021/ja01315a102}, there it was shown that multiple energetically-equivalent (frustrated) arrangements of hydrogen atoms in ice crystals were possible, resulting in extensive ground-state degeneracy and thus predicted residual entropy at $T = 0$. 
Likewise, frustration in magnetic materials (defined as competition between individual magnetic interactions) can lead to a range of exotic effects, from extensive magnetic ground-state degeneracy in classical spin liquids, to novel quantum states with unusual excitations, which can be fundamentally different from those found in conventional magnets, (especially when
pronounced quantum spin fluctuations are present, e.g., in $S$ = 1/2 spin systems)~\cite{Balents2010}. 
\textcolor{black}{In most magnetic materials, the
	dominant interaction is the exchange interaction, which originates from the Pauli exclusion principle and the interplay
	of orbital overlap and Coulomb repulsion.}
If the spins of the $i^{\rm th}$ and $j^{\rm th}$ sites of a spin lattice are denoted by ${\bf S}_i$ and ${\bf S}_j$, respectively, the general interacting Hamiltonian up to second order in the spin operators, and in the possible presence of an applied magnetic field $H$ along $z$ direction, can we written as~\cite{blundell2001magnetism}
\begin{equation}
	\begin{aligned}
		\mathcal{H} = \sum_{\langle ij \rangle} 
		{\bf S}_i \cdot J_{ij} \cdot {\bf S}_j-g\mu_{\rm B}H\sum_iS_i^z,
	\end{aligned}
\label{Hei}
\end{equation}
where $J_{ij}$ in the first term is the exchange tensor and the second term represents the Zeeman energy with $g$ is the $g$ factor of the magnetic ion and $\mu_{\rm B}$ is the Bohr magneton. 
{If the $J_{ij}$ tensor is proportional to the identity tensor, we call the exchange isotropic or Heisenberg, if it only has two non-zero eigenvalues the exchange is of XY-type, and if there is only one non-zero eigenvalue (i.e., a single local axis along which the spins at $i$ and $j$ couple) we call the exchange Ising-type.}

\begin{figure}
\includegraphics[width=0.5\textwidth]{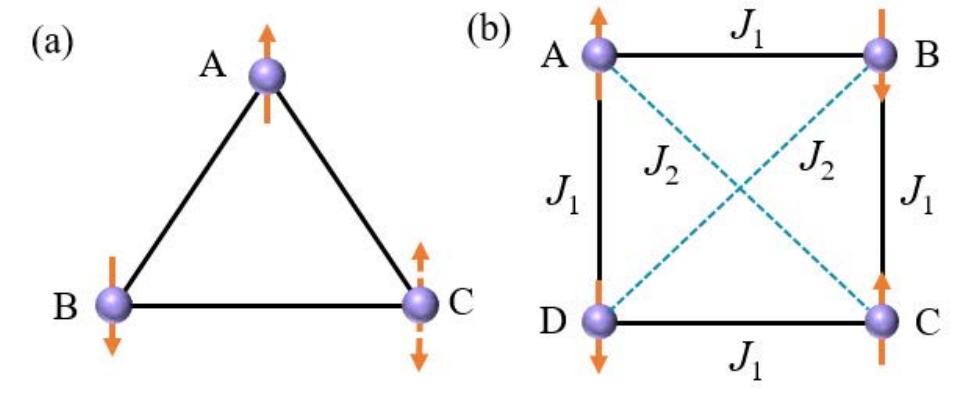} 
\caption{(a) The spins located at the corners A and B are arranged antiferromagnetically, however, the spin at corner C is geometrically frustrated because it can not simultaneously minimize its antiferromagnetic interaction with the spins in the A and B corners of the triangle. (b) On a square with antiferromagnetic exchange interactions $J_1$ and $J_2$ a similar situation (exchange frustration) is encountered due to competition between the two types of interactions.} {\label{all_frustrated}}
\end{figure}

\begin{figure*}
\includegraphics[width=\textwidth]{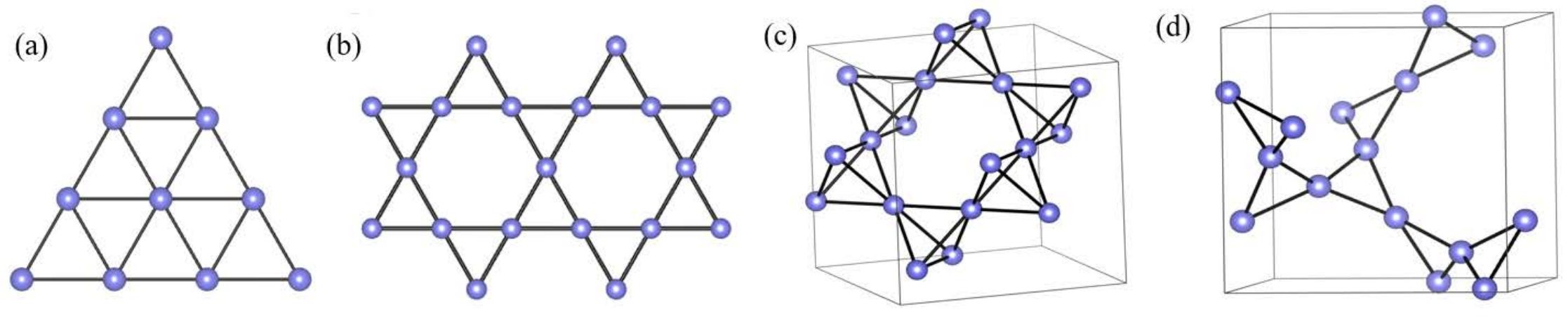} 
\caption{(a) 2D triangular lattice with edge-sharing spin triangles, where each spin has $z = 6$ nearest neighbors. 
(b) 2D kagome lattice with corner-sharing spin triangles and $z = 4$.
(c) 3D pyrochlore lattice with corner-sharing tetrahedra and $z = 6$. 
(d) 3D hyperkagome lattice with corner-sharing triangles and $z = 4$.} {\label{frustrated_lattices}}
\end{figure*}

To introduce the concept of geometrical frustration let us consider three spins residing at the corners of an equilateral triangle with antiferromagnetic exchange couplings between them (see Fig.~\ref{all_frustrated}(a)). 
If the spins at the corners A and B align antiferromagnetically (i.e., if they point in anti-parallel direction to each other) to satisfy their local exchange interaction, the third spin at corner C cannot choose a direction that is anti-parallel to both spins at A and at B at the
same time; i.e., it cannot satisfy its exchange interactions with both of its neighbors.
The geometry of the spin triangle thus prevents the simultaneous minimization of the pair-wise interaction energy of its three spins. 
We call such a system geometrically frustrated. 

A similar situation to frustration originating from geometry of the spin lattice can also occur due to competing interactions between spins (see Fig.~\ref{all_frustrated}(b)). 
For example, let us consider a square of spins where,  in addition to antiferromagnetic interactions $J_{1}$ between nearest-neighboring spins, we also have two antiferromagnetic next-nearest neighbor interactions $J_{2}$ across the two diagonals of the square. It is obvious that spins cannot align antiferromagnetically
on both nearest-neighbor and next-nearest-neighbor exchange bonds simultaneously, meaning that such a system is also
frustrated

A generic consequence of frustration is the suppression of long-range magnetic ordering. 
A general, quantitative
measure of the strength of magnetic frustration is thus given by the frustration parameter $f=|\theta_{\rm CW}|/T_N $, where $\theta_\mathrm{CW}$ is the high-$T$ Weiss temperature, proportional to the average magnetic interaction strength in the system, and $T_N \geq 0$ is the critical (N\'eel) temperature of a phase transition to a low-$T$ long-range ordered magnetic state. 
The frustration parameter measures the degree to which ordering transitions are suppressed due to the destabilizing effect of magnetic frustration, which promotes ground state degeneracy and strong spin fluctuations instead. 
If a particular material has $f>10$ it is usually considered to be highly frustrated~\cite{Balents2010}. 
In the case of the antiferromagnetic square lattice with additional diagonal bonds (see Fig.~\ref{all_frustrated}(b)) this occurs when the interaction strengths of the two competing exchange interactions $J_1$ and $J_2$ becomes comparable in magnitude. 
However, in the most frustrated systems long-range magnetic order can be completely absent even at $T = 0$, in which case $f \rightarrow \infty$. 
This most often occurs when the system has an extensive number of low-energy states, either as degenerate ground states as in classical spin liquids, or as a continuum of low-energy excited states in quantum spin liquids. As low-energy excited states in this case do not
arise as perturbations of long-range order, but as correlated fluctuations over a disordered spin background (which is,
nevertheless, strongly quantum entangled in the case of QSLs), they often have a highly exotic character.
Thus, besides merely suppressing long-range magnetic order, frustration can actually stabilize a range of unique and unusual magnetic
states with characteristic exotic excitations.

Effects of strong frustration are most often observed in frustrated 2D kagome and triangular lattices as well as three-dimensional (3D) pyrochlore and hyperkagome lattices (Fig.~\ref{frustrated_lattices}). 
While a spin in the highly frustrated kagome lattice has
$z$ = 4 nearest neighbors on the two neighboring corner-sharing spin triangles (Fig.~\ref{frustrated_lattices} (b)), a spin in the triangular lattice
has $z$ = 6 nearest neighbors on edge-sharing spin triangles (Fig.~\ref{frustrated_lattices} (a)). 
\textcolor{black}{As the number of local constraints on spins due to
	exchange couplings to nearest neighbors is thus larger for spins in the triangular lattice than for spins in the kagome lattice,
	the triangular lattice is actually more prone to magnetic ordering than the kagome lattice, and is thus less frustrated. 
In fact, in the case of the Heisenberg Hamiltonian, the ground-state of the triangular lattice exhibits long-range 120$^{\circ}$ order~\cite{PhysRevLett.60.2531, PhysRevLett.69.2590}, while the spin ground states for a Heisenberg Hamiltonian on the kagome lattice~\cite{PhysRevB.48.9539} is extensively-degenerate for classical spins~\cite{PhysRevB.48.9539} resulting in complete suppression of magnetic order in this case. 
Only via magnetic anisotropy, which in this case lowers the degree of frustration, can 120$^{\circ}$ long-range order be stabilized on the kagome lattice~\cite{Mondal_2021, zorko2019negative}.}
Among 3D frustrated spin lattices, a spin in the pyrochlore lattice with corner sharing tetrahedra has $z$ = 6 nearest
neighbors (Fig.~\ref{frustrated_lattices} (c)), whereas a spin in a 3D spin lattice with corner sharing triangles known as the hyperkagome lattice
has only $z$ = 4 nearest neighbors (Fig.~\ref{frustrated_lattices} (d)). 
\textcolor{black}{Due to their corner sharing spin building blocks, both of these 3D lattices
	support classically extensivelly degenerate ground states }\\
\begin{figure}[b]
\includegraphics[height=7cm, width=8.3cm]{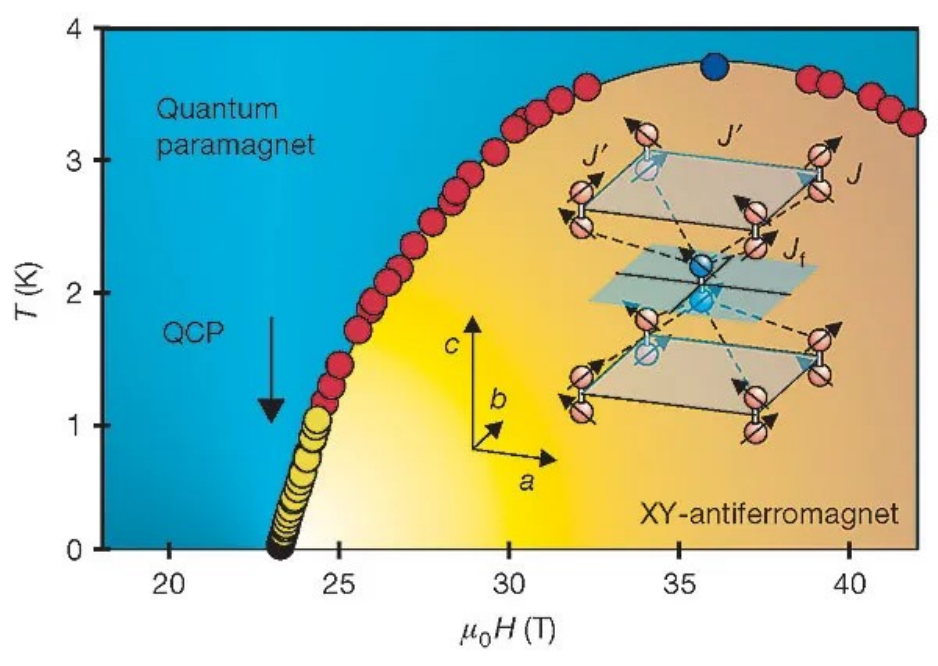}
\caption{The temperature--magnetic-field phase diagram of the frustrated magnet BaCuSi$_{2}$O$_{6}$. 
As a result of the reduction in dimensionality in the vicinity of QCP, the shading changes from dark to light.  The yellow, red and blue solid  circles are experimental data points obtained from different measurements that separate the quantum paramagnetic region (blue) from the XY antiferromagnetic long-range order (yellow). 
Adapted from~\cite{Sebastian2006} with permission from NPG.}{\label{phas}}
\end{figure}
Due to the large number of low-lying magnetic states in frustrated magnets, the effects of external stimuli, such as
an applied magnetic field, can often be quite dramatic and can tune such systems to completely new quantum states
with different (usually also exotic) magnetic or electronic properties. Such a situation can, for example, be found in low-dimensional systems near a QCP~\cite{Sachdev475,PhysRevLett.80.5627,Si2001}, which separates fundamentally different ground states via a quantum phase
transition. A clean realization is, however, often difficult to find in actual materials due to the often unavoidable presence
of additional 3D interactions. One promising candidate to realize the idea of a quantum phase transition through a QCP is the frustrated magnet BaCuSi$_{2}$O$_{6}$ known also as Han Purple.  It crystallizes in a body-centred tetragonal structure where Cu$^{2+}$ ($S = 1/2$) magnetic moments form a 2D plane of dimers oriented out-of-plane (shown in the inset in Fig.~\ref{phas}) with intra-dimer antiferromagnetic interactions $J$, in-plane inter-dimer interaction $J'$, and a frustrated inter-layer interaction $J_f$. 
In zero applied magnetic field, the ground-state of this system is direct product of singlets ($S = 0$) on each dimer with a finite energy gap to degenerate triplet excitations ($S_{z} = 1$, $0$, and $-1$) (Fig.~\ref{phas}).
\textcolor{black}{This is an example of a quantum paramagnet, a quantum entangled state with only short-range correlations even at zero temperature.} 
An applied external magnetic field $H$ Zeeman splits the energies of the excited triplet states, eventually closing the energy gap between the singlet state
and the lowest-energy triplet state ($S_{z} = -1$) for $H$ above a critical value $H_{c}$. 
This leads to antiferromagnetic order above a critical field for $H>$  $H_c$ at low $T$. 
%
\textcolor{black}{While for a typical antiferromagnet, the magnetic transition temperature $T_N$ vanishes above a critical magnetic field, in BaCuSi$_{2}$O$_{6}$ $T_N$ is instead enhanced by an applied magnetic field, at least up to a certain strength of the external applied magnetic field (Fig.~\ref{phas}), where it attains a maximum value, before it starts decreasing again under even larger applied magnetic fields. 
Moreover, from fitting a critical field dependence $T_{N} \sim (H - H_{c})^{\beta}$ to the phase boundary separating the field-induced antiferromagnetic state from the quantum paramagnetic region for $H$ close to $H_c$ one obtains a critical exponent of $\beta = 2/3$.
This suggests the realization of a field-induced magnetic Bose–Einstein-condensed state in this quantum magnet~\cite{Sebastian2006}.}

\textcolor{black}{Another interesting example of a magnetic field-induced transition was observed in the kagome antiferromagnet Zn-brochantite, ZnCu$_3$(OH)$_6$SO$_4$, an intriguing compound that will be more closely presented in Subsection \ref{kago}. 
In this compound, an applied magnetic field induces a pairing instability of intrinsically gapless magnetic excitations of its low-field quantum spin-liquid ground state, turning the system into a gapped spin liquid under a large applied magnetic fields at low $T$~\cite{gomilsek2017field}. Interestingly, this transition from a gapless to a gapped quantum spin liquid ground state can be
understood as a fully-magnetic analogue of the exotic triplet superconducting transition found in electronic systems (see Subsection~\ref{pairing}).}
 
\section{\textbf{Spin ice in 3D pyrochlores}}
\label{spinicc}
\begin{figure*}
	\begin{center}
		\includegraphics[height=6cm,width=14cm]{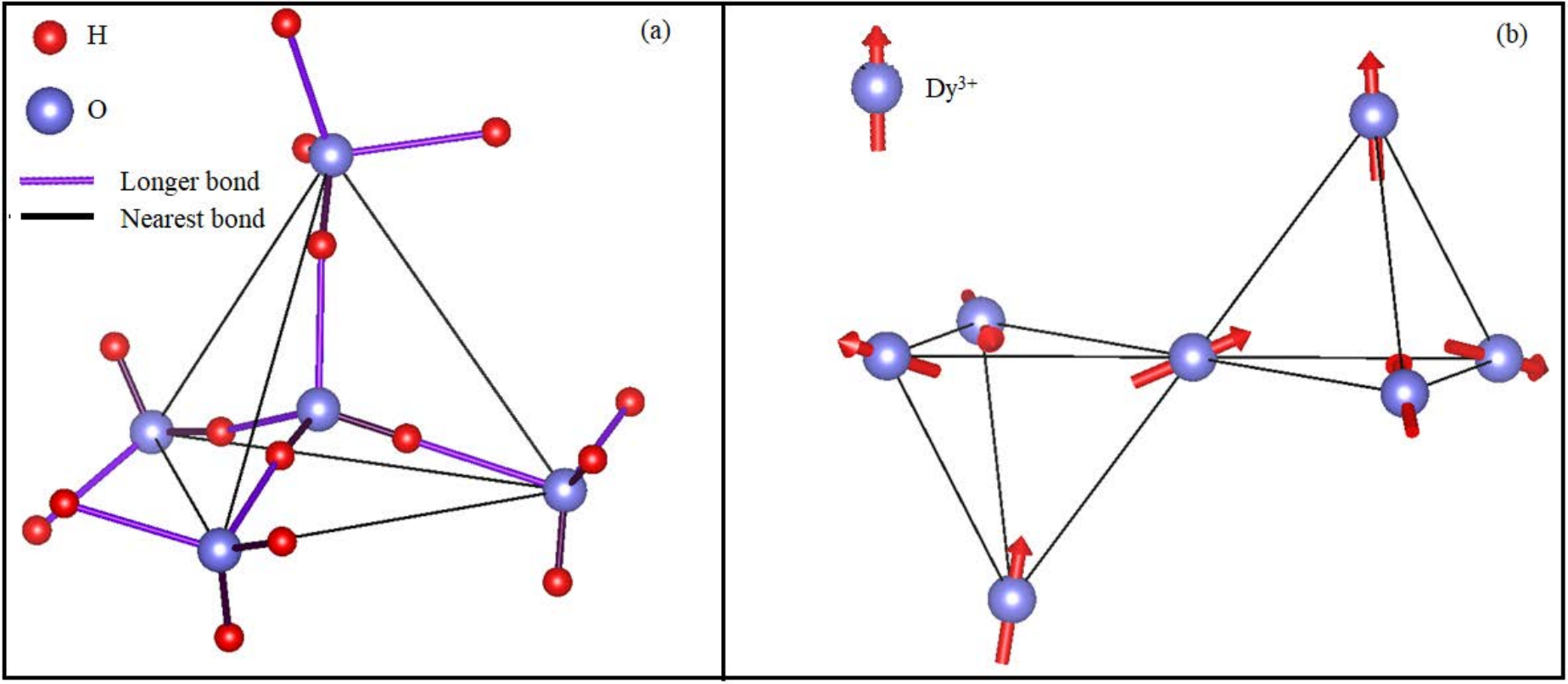}
		\caption{(a) Schematics of the water-ice structure and (b) the spin-ice structure on the pyrochlore lattice of Dy$_2$Ti$_2$O$_7$. In water ice, each oxygen atom is surrounded by two closer hydrogen atoms and two that are farther away, whereas in spin ice two spins always point towards the center of a tetrahedron while the other two spins point away from its center.}{\label{spin_water_ice}}
	\end{center}
\end{figure*}
\textcolor{black}{We start our review of unconventional magnetism of frustrated magnets on the pyrochlore lattice, for which the variety of unanticipated magnetic phenomena that were discovered is truly exceptional. Firstly, we highlight the spin-ice ground state that can be realized on this lattice.}
The term spin ice originates from an analogy between classical magnetic ground states with the structure of water ice~\cite{doi:10.1021/ja01315a102}. 
In water ice, each oxygen atom is surrounded by four neighbouring oxygen atoms forming the corners of a tetrahedron (see Fig.~\ref{spin_water_ice}(a)). 
The distance from the central oxygen atom to the four neighbouring oxygen atoms is 2.76~\r{A} with a hydrogen atom in-between, either 0.95~\r{A} or 1.81~\r{A} away from the central oxygen. Neglecting
thermal effects, each oxygen in water ice is surrounded by two hydrogen atoms that are closer to it, and two that are
farther away from it. This is called the ice rule of water ice. 
Crucially, there is no energetic preference for which hydrogen will be farther away or closer to a particular oxygen atom, so long as all the local ice rules are obeyed. 
Considering the two hydrogen atoms that are closer to a given oxygen atom as belonging to the same molecule, the ice rule corresponds to saying that water ice is made from H$_2$O molecules, while the energetic equivalence of different ice structures obeying the ice rules corresponds to saying that individual H$_2$O molecules can be arbitrarily rotated in water ice, so long as all
four possible hydrogen bonds between  H$_2$O molecules are optimally satisfied (which restricts the individual rotations
to 6 possibilities). 
It turns out that obeying ice rules on the lattice of usual water ice is a weak-enough constraint that
the number (degeneracy) of energetically-equivalent structural states of water ice becomes a macroscopic (extensive)
quantity, i.e., it grows exponentially with the size of the system~\cite{Ramirez1999,doi:10.1021/ja01315a102,Giauque1936,pauling1992nature}. 
A completely analogous scenario can be
realized in a magnetic analog of water ice, appropriately called spin ice~\cite{Bramwell1495}, where a spin analogue of the ice rule also
leads to macroscopic ground-state degeneracy, as we shall see next. 

\subsection{Ground state\label{SI}}
Spin-ice ground states are most often realized in magnets with the chemical formula A$_2$B$_2$O$_7$ (A = Yb, Pr, Dy, Ho, Bi etc.~and B = Ti, Ir, Zr, Mo, Ru etc.). 
Here, the positions of hydrogen atoms in water ice are replaced by orientations of
spins of trivalent magnetic ions that are located on a face-centered-cubic lattice of corner-sharing tetrahedra, forming the
pyrochlore lattice~(\ref{spin_water_ice} (b)). 
Due to strong Ising anisotropy in these systems, spins align themselves along a local $\langle 111 \rangle$ quantization axis, which corresponds to the direction from the center of the tetrahedron to one of the corners of that tetrahedron. 
\textcolor{black}{For either ferromagnetic or antiferromagnetic nearest-neighbor spin interactions, the system is frustrated.
Considering a single tetrahedron, its ground-state with ferromagnetic interactions is six fold degenerate due to six different 2-in--2-out configurations (this is the spin ice rule, which requires that two spins point into the tetrahedron while the other two spins have to point out of the tetrahedron), whereas the ground state with antiferromagnetic interactions is two-fold degenerate, because only the all-in and all-out spin configurations have the lowest possible energy~\cite{PhysRevLett.115.197202,Ramirez1999,PhysRevLett.79.2554}}. 
\textcolor{black}{As the tetrahedra are connected only through single corners, the spin ground states of such a ferromagnetic pyrochlore lattice are precisely those that obey the 2-in--2-out spin ice rule~\cite{Ramirez1999,GARLEA2015203,PhysRevB.77.214310}, in perfect analogy with water ice with its ice rules.
Spin ices thus have a macroscopically large number of degenerate ground states, resulting in a substantial residual (zero-point) entropy at $T = 0$, in apparent violation of the third law of thermodynamics~\cite{Ramirez1999,doi:10.1063/1.2186278}.
This emergent phase is also known as a Coulomb phase~\cite{henley2010coulomb}.
Note that the existence of extensively many possible ground-state configurations is in stark contrast to magnetically ordered states, where typically only a handful of privileged spin configurations are
ground states.}

In general, the entropy due to spin degrees of freedom of a magnetic system has a maximum value of $R\ln{(2S+1)}$ per mole, where $S$ is the spin. For example, if the system consists of spins $S_{\rm eff} = 1/2$, the maximum entropy is $R\ln{2}$. 
Any deviation of the high-$T$ entropy release towards values lower than this indicates that there is a zero-point entropy
present at temperatures close to absolute zero.
\begin{figure}
\includegraphics[height=10cm,width=9.3cm]{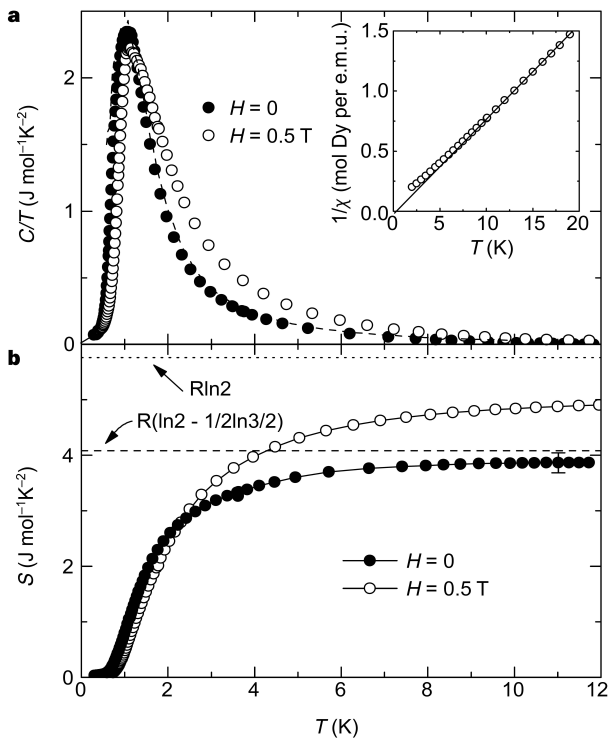} 
\caption{(a) Temperature dependent specific heat divided by temperature in zero-field and in 0.5~T applied field for Dy$_2$Ti$_2$O$_7$. Inset shows the inverse susceptibility vs. temperature with the Curie Weiss fit ($\theta_{CW} = 0.5$~K) indicating weak ferromagnetic interaction. (b) Estimated entropy from specific heat data indicates the presence of zero-point entropy 0.5$R$ln(3/2). Adapted from~\cite{Ramirez1999} with permission from NPG.}{\label{spin-ice entropy}}
\end{figure}
An example of this is the spin-ice material  Dy$_{2}$Ti$_{2}$O$_{7}$, where
the total spin entropy release of Dy$^{3+}$ ($S_\mathrm{eff} = 1/2$) magnetic ions measured in the temperature range from 0.2~K to 12~K  (Fig.~\ref{spin-ice entropy}(a)) is substantially less than the full spin entropy, expected if there was no zero-point (residual) entropy. 
The zero-point entropy of this spin-ice material was thus estimated to be $\Delta S$=(1 $-$ $(0.67\pm 0.04))R \ln 2$~\cite{Ramirez1999}, which is close to the theoretical zero-point entropy of water ice $R/2 \ln(3/2)$ (Fig.~\ref{spin-ice entropy}b)~\cite{doi:10.1021/ja01315a102}.
\textcolor{black}{The presence of the residual Pauling entropy in Dy$_2$Ti$_2$O$_7$ was, however, later questioned by more precise heat-capacity measurements, where special attention was paid to extremely long spin relaxation times, which affects thermal relaxation at low temperatures and could therefore have potentially caused an erroneous estimate of residual entropy in previous studies~\cite{pomaranski2013absence}. 
These measurements, taking into account long internal equilibration times, have rather suggested the absence of the low-temperature plateau in entropy, which might be a sign that higher-order processes (perhaps of quantum nature) could be important in this material.}

For more in-depth discussion, it is useful to look at the microscopic description of spin ices in more detail. In pyrochlore materials, $f$-electrons of their rare-earth ions behave as Kramers doublets with effective spin  1/2 spin--orbit entangled degrees of freedom due to the crystal electric field (CEF) splitting of the ion's energy levels. 
The nearest-neighbor exchange interaction are thus often strongly anisotropic, sometimes even Ising in character~\cite{PhysRevLett.84.3430, PhysRevLett.95.217201}. 
The emergent spin-ice behavior of these systems is attributed not just to short-range exchange interactions, but also to  long-range dipolar interactions between spins. The corresponding Hamiltonian is given by~\cite{PhysRevLett.84.3430}
	\begin{equation}
    \begin{split}
		\mathcal{H} = &-J\sum_{\langle ij \rangle}\mathbf{S}_{i}^{z_{i}}\cdot\mathbf{S}_{j}^{z_{j}} \\
        &+ Dr_{nn}^{3}\sum_{j>i}\left(\frac{\mathbf{S}_{i}^{z_{i}}\cdot\mathbf{S}_{j}^{z_{j}}}{|\mathbf{r}_{ij}|^{3}}-\frac{3(\mathbf{S}_{i}^{z_{i}}\cdot\mathbf{r}_{ij})(\mathbf{S}_{j}^{z_{j}}\cdot\textbf{r}_{ij})}{|\mathbf{r}_{ij}|^{5}}\right),
    \end{split}
	\end{equation}
where the Ising moment at site $i$ with a local Ising axis $z_i$ is given by the $\mathbf{S}_{i}^{z_{i}}$ vector, local Ising axes point along the set of $\langle 111 \rangle$ vectors and $\mathbf{r}_{ij}$ denotes the position vector between $i^{\rm th}$  and $j^{\rm th}$ sites. 
Nearest-neighbor exchange energy and dipolar energy between two neighbouring sites are given by $J_{nn} = J/3$ and  $D_{nn}=5D/3$, respectively.
For example, in the spin-ice materials Ho$_{2}$Ti$_{2}$O$_{7}$ and Dy$_{2}$Ti$_{2}$O$_{7}$, long-range dipolar interactions ($D_{nn} \approx 2.35$~K) are stronger than the nearest-neighbor magnetic exchange interaction ($J_{nn}\approx -0.52$~K and $J_{nn}\approx -1.24$~K, respectively)~\cite{Jaubert_2011,PhysRevLett.84.3430}. 

\subsection{Magnetic monopoles and Dirac strings\label{SI_monopoles}}
One remarkable emergent phenomenon in spin ices is the appearance of low-energy excitations that behave as effective
magnetic monopoles~\cite{Castelnovo2008}. 
Maxwell's equations in vacuum state that
	\begin{equation}
		\nabla \cdot \vec{E}=\frac{\rho}{\epsilon_{0}},
	\end{equation}
	\begin{equation}
		\nabla \cdot \vec{B}=0,
	\end{equation}
where $\vec{E} ,\vec{B}$, and $\rho$ are the strength of the electric field, the density of the magnetic field, and the electric charge density, respectively, while $\epsilon_{0}$ is the permittivity of free space. \textcolor{black}{Thus, while electric monopoles (electric charges) indeed exist there should be no analogous magnetic monopoles ~\cite{doi:10.1098/rstl.1865.0008,RevModPhys.71.863}}, according to Maxwell, as that would imply a non-zero divergence
of the magnetic flux density, $\nabla \cdot \vec{B}\neq 0$.

The concept of hypothetical magnetic monopoles was first introduced in a seminal papers by Dirac~\cite{10.1098/rspa.1931.0130}.
Castelnovo \textit{et al}. have subsequently shown that in the spin-ice state of the frustrated pyrochlore material Dy$_{2}$Ti$_{2}$O$_{7}$ the flipping of one
spin could stabilize an excited state with one 3-in–1-out and one 1-in–3-out defect, which, over long distances, behave
effectively as two mobile magnetic monopoles (sources and sinks of a discretized version of the magnetic field). In spin
ice, we thus naturally find an effective fractionalization of elementary magnetic dipoles (spins) into pairs of effective
magnetic monopoles~\cite{Castelnovo2008}.
It needs to be stressed that these monopoles are not true elementary (microscopic) particles, but rather emergent, long-wavelength phenomena.

\begin{figure}
		\includegraphics[width=1\linewidth]{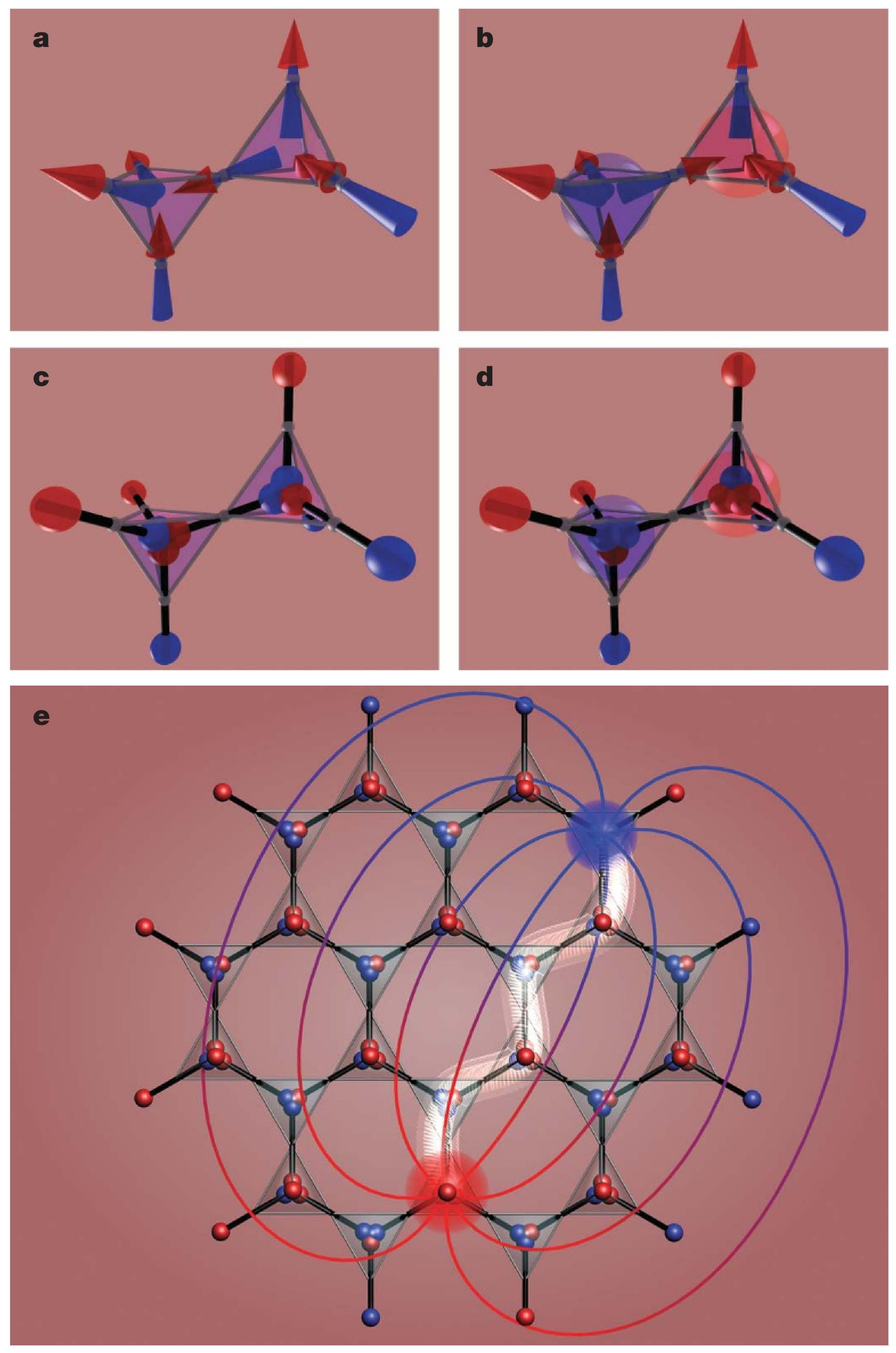} 
		\caption{(a) Two neighbouring tetrahedra are shown in the ground-state 2-in--2-out configurations (i.e., obeying the spin ice rule). (b) The common spin connecting two tetrahedra is flipped, resulting in the 3-in--1-out configuration (denoted by a red sphere) and a 1-in--3-out configuration (blue sphere) on them. (c, d). Every  spins in (a, b) is  replaced by two opposite emergent magnetic charges (the dumbbell picture). (e) The monopoles can move apart by local spin flips, forming an effective Dirac string of flipped spins that connects them.
			Adapted from~\cite{Castelnovo2008} with permission from NPG.}{\label{monopole}}
	\end{figure}

To see why they are called monopoles, we employ an illustrative way to represent each spin in a spin ice via a dumbbell, as presented in Fig.~\ref{monopole}, that consists of two opposite effective magnetic monopole charges located at each end of the spin. 
In the ground-state 2-in--2-out configuration, the net magnetic moment and hence the total magnetic charge, of each tetrahedron is zero. 
However, the excited 3-in--1-out and one 1-in--3-out configurations that arise from a spin flip violate the 2-in--2-out ground state spin ice rule, which leads to a net magnetic moment on the two neighbouring tetrahedra (Fig~\ref{monopole}(b)).
This can be visualised as a formation of an magnetic monopole--anti-monopole pair, with an effective magnetic monopole moment of $Q_{\rm eff}=\pm\sqrt{8/3}\mu_{\rm B} / r_{nn}$~\cite{Castelnovo2008}. 
%
By local spin flips that preserve the ice rules, and thus cost no exchange interaction energy, the monopole and anti-monopole can spatially separate and move around the lattice as two independent quasiparticles with only an effective Coulomb interaction  between them (potential energy proportional to
1/$r$) arising from dipolar magnetic interactions~\cite{Castelnovo2008}. 
The wake of flipped spins between the monopoles, as they move apart, represents an effective Dirac string connecting them (see Fig~\ref{monopole}(e)).
{This Dirac string is a topological object, as no local measurement can statistically distinguish it from a normal ice-rule obeying background, and yet it connects two distinguished monopole excitations and can thus be detected globally~\cite{Castelnovo_2012}.}
Topological features such as these ensure that spin ices, as non-local spin-textures, are robust again thermal and quantum fluctuations and can be realized in several classes of materials~\cite{qi2009inducing,milde2013unwinding,donnelly2017three,rana2023three}.
\textcolor{black}{Furthermore, it has recently been discovered that due to the the specific directions of local dipolar magnetic fields at spin positions in a spin ice, the emergent monopoles actually mostly move on a sublattice
	in real space that is effectively disordered with a well-developed dynamical fractal structure on short and intermediate
	length scales~\cite{hallen2022dynamical}. 
This results in an anomalous shape of the spectral density of magnetic noise, which was experimentally
observed in the paradigmatic Dy$_2$Ti$_2$O$_7$ spin-ice pyrochlore~\cite{dusad2019magnetic}, while leaving no signatures in its bulk thermodynamics.}

Remarkably, spin-ice behaviour arising from Ising-like spins on a frustrated 2D intermetallic-based kagome lattice, instead of the usual 3D pyrochlore lattice, has also been observed~\cite{Zhao1218}.
\textcolor{black}{Furthermore, artificial spin ices, where individual spins are replaced by ferromagnetic domains nanopatterned into arbitrary lattices, also show many of the same effects, at least on a classical level, while offering unprecedented possibilities for tunability, manipulation, and detailed observation of spin-ice phenomena~\cite{Wang2006, skjaervo2020advances,nisoli2013colloquium,goryca2021field,yue2022crystallizing}. 
}
 
\subsection{Spin fragmentation \label{frac}}
\begin{figure}[t]
	\includegraphics[width=1\linewidth]{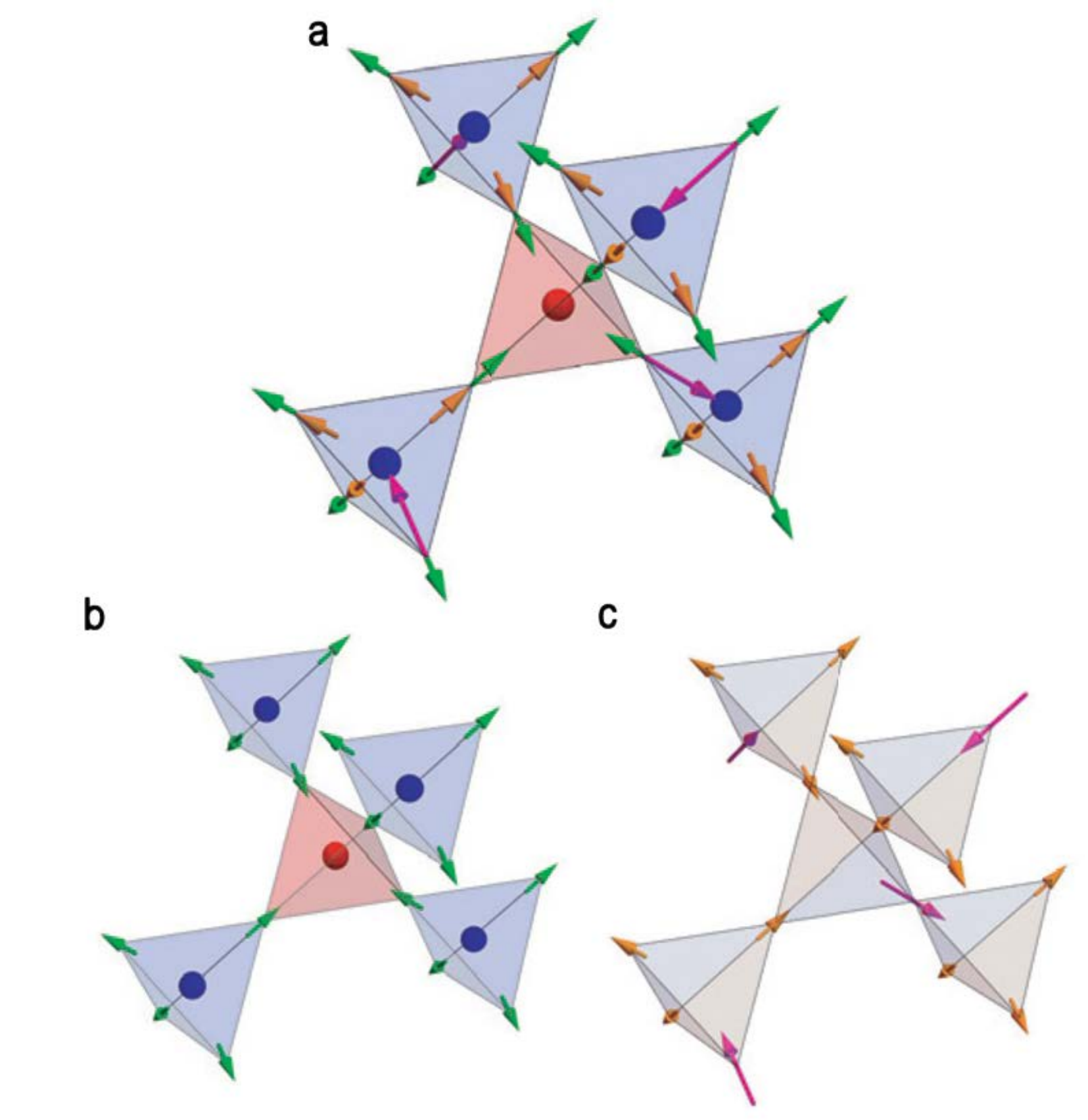} 
	\caption{\textcolor{black}{ (a) Crystallization of magnetic monopoles (red and blue spheres) due to dipolar interactions between magnetic moments leads to fragmentation of local moments 
into a sum of two parts, (b) a divergence-full contribution (all-in or all-out configurations) represented by green arrows and (c) a divergence-free contribution (emergent Coulomb phase) represented by orange and magenta arrows.
 Experimentally, this leads to  coexistence of magnetic Bragg peaks, arising from the divergence-full contribution, and pinch-point pattern characteristic of short-range correlations of the divergence-free phase. 
 Adapted from~\cite{petit2016observation} with permission from NPG.}}
	{\label{fragment}}
\end{figure}

Another intriguing phenomenon predicted for spin-ice states in pyrochlores is spin fragmentation~\cite{brooks2014magnetic}.
In certain cases, a single spin degree of freedom --the local magnetic moment --can be effectively decomposed into an ordered part, arising from magnetic monopole crystallization, and a persistently-fluctuating part, characteristic of the dynamically-fluctuating Coulomb (divergence-free) phase
(Fig.~\ref{fragment}).
The resulting magnetic state which is simultaneously ordered and fluctuating has indeed been observed experimentally, for the first time in the pyrochlore material Nd$_2$Zr$_2$O$_7$, where neutron scattering experiments revealed the coexistence of magnetic Bragg peaks, characteristic of long-range magnetic order, and a diffuse pinch-point pattern, characteristic of the magnetic structure factor of disordered spin ice~\cite{petit2016observation}.
Experimental realization of this effect on the pyrochlore lattice requires strong Ising anisotropy combined with effective ferromagnetic interactions, which is key to monopole crystallization, while the remaining fluctuating, divergence-free part is favoured entropically.

A similar state arising from the crystallization of emergent magnetic degrees of freedom has also been found in the Ising kagome spin-ice material Dy$_3$Mg$_2$Sb$_3$O$_{14}$~\cite{paddison2016emergent}.
Furthermore, spin fragmentation was also proposed as a possible mechanism for explaining the coexistence of magnetic order and persistent spin dynamics in the kagome antiferromagnet YCu$_3$(OH)$_6$Cl$_3$~\cite{zorko2019YCu3muon}, which is likewise characterized by strong magnetic anisotropy~\cite{arh2020origin}.
It may also be an intrinsic property of amplitude-modulated magnetically-ordered states, where persistent magnetic fluctuations can be attributed to the disordered part of the magnetic moment at each magnetic site~\cite{pregelj2012persistent}.

\subsection{Quantum spin ice}
\textcolor{black}{As discussed above, rare-earth magnets with large $f$-electron spins on the pyrochlore lattice obeying spin-ice rules are proposed to host macroscopically degenerate ground states.
This makes spin ices as examples of classical spin liquids, i.e., spin systems with a macroscopically degenerate ground state that thus do not form long-range magnetic order, and instead remain dynamical (fluctuating) at all $T > 0$. 
Specifically, in spin ices, thermal fluctuations at low $T$ can freely switch between different 2-in--2-out spin-ice configuration at almost no energy cost~\cite{Bramwell_2020}.
Although these effects are entirely classical, quantum effects can also play an important role in pyrochlores.
Materials such as Dy$_{2}$Ti$_{2}$O$_{7}$, Ho$_{2}$Ti$_{2}$O$_{7}$, and Ho$_{2}$Sn$_{2}$O$_{7}$ have been widely studied in search of collective quantum effects in spin-ice states~\cite{Ramirez1999,Bramwell_2020}, including some pyrochlore magnets with an effective pseudospin-1/2 moment, as well as Yb$_{2}$Ti$_{2}$O$_{7}$~\cite{PhysRevB.94.205107,PhysRevX.1.021002}.
Quantum fluctuations induced by transverse components of spins in certain pyrochlore materials can lead to a quantum spin-ice state, where competition between local interactions suppresses conventional long-range order and instead favour a long-range quantum entangled state supporting emergent electromagnetism associated with an emergent $\mathrm{U}(1)$ gauge field~\cite{RevModPhys.89.041004,gingras2014quantum,PhysRevB.69.064404,PhysRevB.86.075154}. 
This is an example of a quantum spin-liquid (QSL) state, where quantum effects enable spins to fluctuate between different classical spin configurations by being in highly entangled, quantum superpositions of them. 
Recently, a quantum spin ice has also been suggested for a breathing pyrochlore lattice Ba$_{3}$Yb$_{2}$Zn$_{5}$O$_{11}$~\cite{PhysRevB.94.075146,Dissanayake2022,Park2016,Rau_2018,PhysRevLett.116.257204,PhysRevB.93.220407}.}

\textcolor{black}{Interestingly, the recent signature of a quantum spin-ice state arising from the dipole--octupole nature of the ground-state Kramers doublet of certain rare-earth magnetic ion  on the pyrochlore lattice, in materials Ce$_{2}$Sn$_{2}$O$_{7}$, Ce$_{2}$Zr$_{2}$O$_{7}$, and Ce$_{2}$Hf$_{2}$O$_{7}$~\cite{Sibille2020, PhysRevLett.115.097202,Gao2019, PhysRevX.12.021015, PhysRevLett.122.187201,Bhardwaj2022, PhysRevMaterials.6.044406} sparked the search for quantum spin-ice states beyond the usual microscopic magnetic dipoles~\cite{PhysRevB.95.041106}.
A similar situation has been recently proposed for Pr$_2$Zr$_2$O$_7$ pyrochlore, where the Pr ions with a non-Kramers ground-state doublet behave as magnetic dipoles along one direction and as orbital quadrupoles along two other directions~\cite{tang2023spin}.
The dipolar moments couple to each other through regular exchange interactions, while the octupole degrees of freedom are coupled via lattice distortions. 
QSL behavior in this material is due to enhanced spin-–orbital quantum dynamics of the non-Kramers doublet. 
Although the pyrochlore magnets outlined above are suitable quantum spin-ice candidates, there are still uncertainties about the exact nature of their QSL ground state~\cite{https://doi.org/10.48550/arxiv.2211.15140,pnas.2008791117}. 
Despite the maturity of the initial proposal, the quest for a clear experimental manifestation of quantum spin ice states on the pyrochlore lattice remains
highly topical to this day~\cite{PhysRevLett.108.067204}.} 

\section{\textbf{Quantum Spin Liquids in 2D}}\label{qsll}
\begin{figure}[b]
	\includegraphics[height=5.2cm, width= 8.7cm]{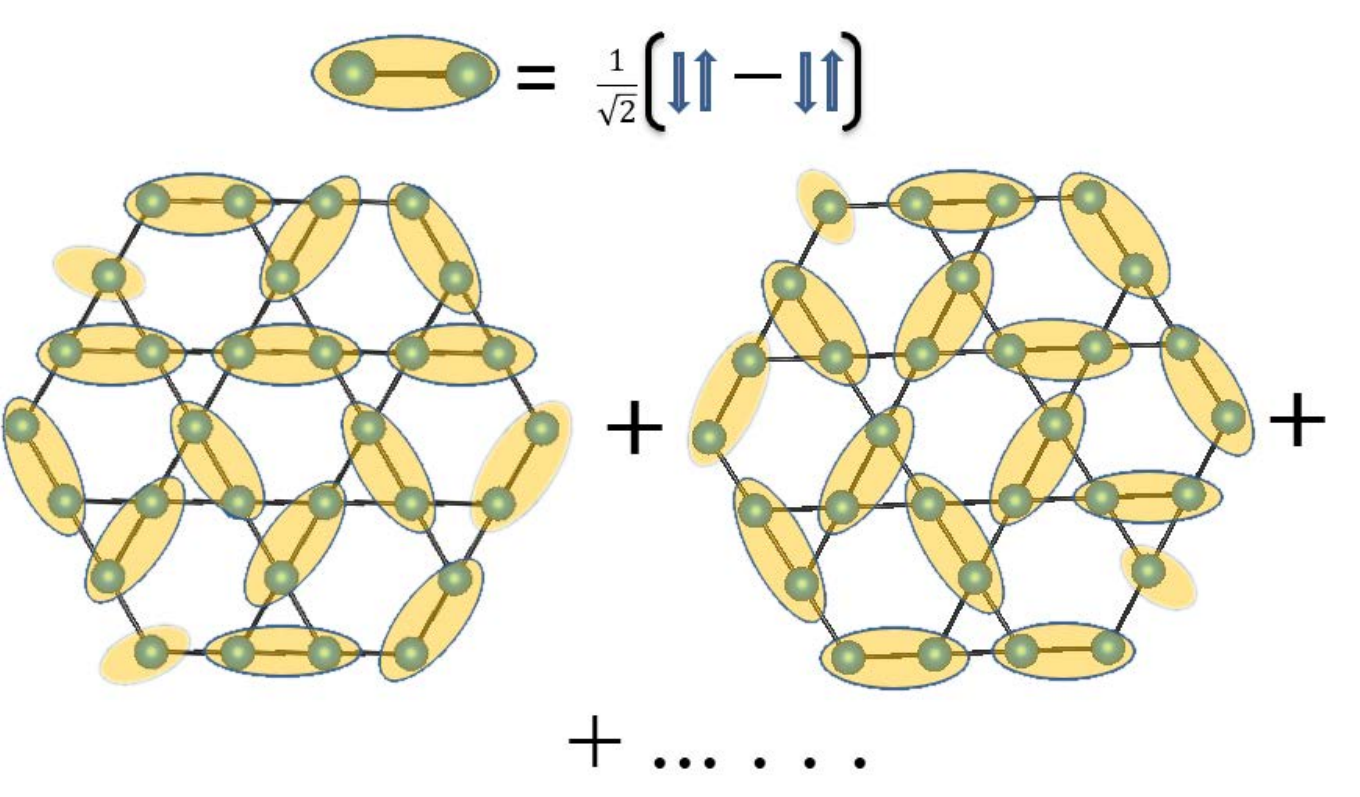}
	\caption{Schematic representation of a resonating valence bond state on the kagome lattice, if spins are short-range entangled only over the nearest-neighbor sites.}{\label{rvb}}
\end{figure}

Next, let us turn our attention to unconventional aspects of general QSLs, which represent some of the most exotic manifestation of magnetic frustration.
A QSL is an entangled quantum spin state, where strong frustration-induced
quantum fluctuations prevent a symmetry breaking phase transitions even at $T = 0$, despite strong pairwise exchange interactions between spins. 
Due to their quantum entangled nature, QSLs are ideal hosts of quasiparticles with fractionalized quantum numbers which coupled to emergent gauge fields~\cite{Balents2010,powell2020emergent}. 
The first theoretical description of a QSL is that of the resonating valence bond (RVB) state, which was introduced by P. W. Anderson and P. Fazekas already in the beginning of the 1970's~\cite{ANDERSON1973153,Fazekas1974-FAZOTG}. 
Due of its proposed significance in high-$T_{c}$ superconductivity~\cite{anderson1987resonating} and in quantum
topological states~\cite{wen2004quantum}, the RVB theory has remained a major research topic in contemporary condensed matter physics even
since.

We start by imagining two spin-1/2 moments located at sites $i$ and $j$ on a frustrated lattice. 
Their singlet state (valence bond) is given by the unique antisymmetric superposition of the two possible antiferromagnetic configurations of these two spins (Fig.~\ref{rvb})
\begin{equation}
	\Psi(i,j)= \frac{1}{\sqrt{2}}( \ket{\uparrow_{i} \downarrow_{j}}-\ket{\downarrow_{i} \uparrow_{j}} ).
\end{equation} 
Here $\uparrow_{i}$ denotes the spin at the site $\emph{i}$ is pointing up, while $\downarrow_i$ denotes it pointing down, with respect to an arbitrary quantization axis.  
We note that in the spin singlet state, the total spin of the spin pair is $S=0$, the state is fully isotropic in spin space, and the two spins are maximally entangled. 
This state is also the ground state of a single antiferromagnetic Heisenberg exchange interaction between the two spins, with excited states represented by spin triplets. 
So, if all the spins in a given system form such valence bonds, the ground-state of the entire system will have total spin $S_{\rm tot}=0$.
\begin{figure*}[t]
	\includegraphics[width=\textwidth]{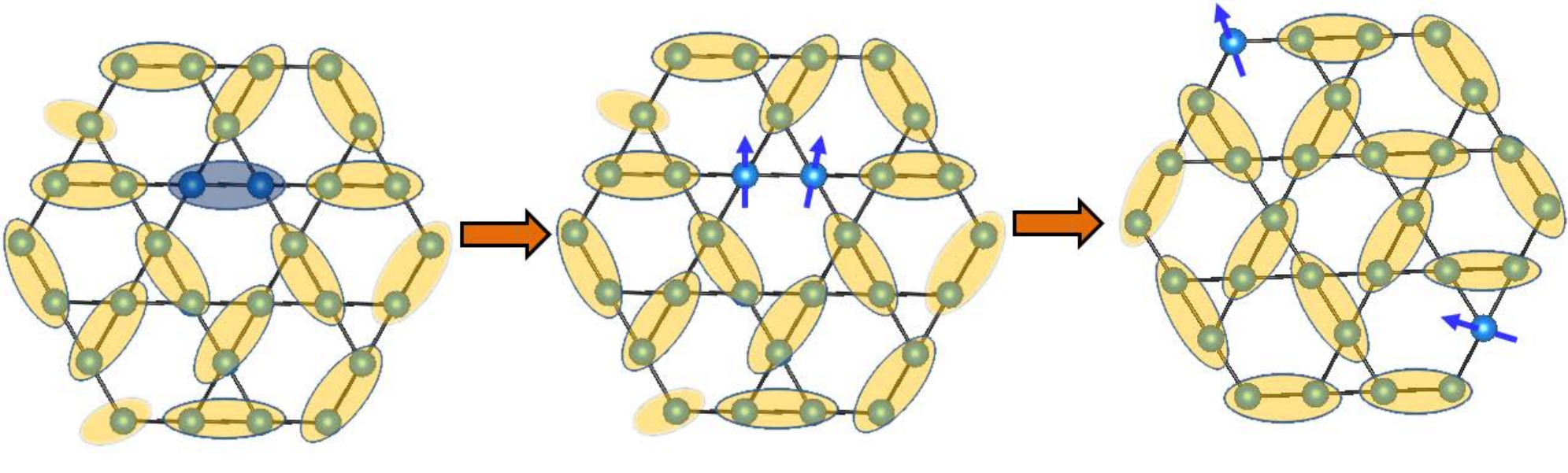}
	\caption{\textcolor{black}{ Spinons (blue arrows) are spin-1/2, charge-free quasiparticles created by breaking a valence bond (singlet pair; blue oval). Independent
			movement of the two created spinons is possible by rearranging unbroken valence bonds (orange ovals). In this illustration, the underlying spin
			lattice is a 2D kagome lattice with short-ranged antiferromagnetic spin correlations.}}{\label{spe}}
\end{figure*}
If a spin state is constructed where every spin is only entangled with its own unique (usually nearest-neighbour) partner spin the total system wave function can be written as a direct product of all the valence bonds present. 
Such a state, which only exhibits short-range entanglement, is called a valence bond crystal and breaks lattice translational symmetry. 
\textcolor{black}{It is worth noting that, in contrast to the valence bond crystal state, the creation of short-range entangled singlet bonds that do not break any lattice symmetries is possible on spin-dimer based lattices, such as the Shastry–Sutherland lattice in SrCu$_{2}$(BO$_{3}$)$_{2}$~\cite{Zayed2017,PhysRevLett.124.206602,PhysRevLett.82.3168,ShinMiyahara_2003,Shi2022}
and the square lattice of dimers in BaCuSi$_{2}$O$_{6}$ (Fig.\ref{phas})~\cite{PhysRevLett.93.087203,PhysRevLett.124.177205,Sebastian2006}.}

\textcolor{black}{According to the RVB theory, a quantum-entangled state can be stabilized as the ground state of low-dimensional, low-spin, highly-frustrated antiferromagnets due to frustration-driven quantum fluctuations that suppress classical long-range Néel order.  
As shown in Fig.~\ref{rvb}, the wave function of an RVB state is a superposition of many possible spin-pair valence-bond configurations that ultimately behaves like a disordered valence-bond liquid~\cite{10.1143/PTPS.145.37}. 
In such a state, the spins can dynamically change their spin singlet partners~\cite{Savary_2016,Balents2010}. 
The energy spectrum of a QSL state formed from only short-range valence bonds~\cite{PhysRevB.79.064405}, as depicted in Fig.~\ref{rvb}, is gapped, leading to exponential temperature dependence of magnetic susceptibility, specific heat, and NMR spin--lattice relaxation rate~\cite{PhysRevLett.86.1881,10.1143/PTPS.145.37}. 
To realize long-range quantum entanglement, not only nearest-neighbor spins must form valence bonds, but also spins that are farther apart~\cite{Savary_2016,RevModPhys.89.025003}.
As the valence bonds become weaker with increasing distance, this can lead to a QSL state with a gapless continuum of excitations
and consequently a power-law temperature dependence of thermodynamic quantities and the NMR spin–lattice relaxation
rate~\cite{PhysRevLett.58.2790,PhysRevLett.99.266403,PhysRevLett.98.067006}.}  

In an RVB state, both translational and rotational lattice symmetries can be preserved. 
Its can be represented as~\cite{RevModPhys.89.025003}
\begin{equation}
\ket{\Psi}=\sum_{i_{1}j_{1}....i_{n}j_{n}}a(i_{1}j_{1}\dots i_{n}j_{n})\ket{\psi(i_{1},j_{1}).....\psi(i_{n},j_{n})},
\end{equation}
where $(i_{1},j_{1})....(i_{n},j_{n})$ are  individual valence-bond (dimer) configurations that cover the entire lattice (Fig.~\ref{rvb}), $a(i_1 j_1 \dots i_n j_n)$ are their complex amplitudes in the superposition, and the sum is over all possible dimer configurations. 
In principle, the amplitudes $a(i_{1}j_{1}....i_{n}j_{n})$ can be determined variationally by minimizing the ground-state energy of the associated  Hamiltonian. 
As valence bonds minimize local antiferromagnetic Heisenberg exchange interactions between a pair of spins, spins that are far apart in a valence bond are usually only weakly connected to each other in the superposition (i.e., they have a small amplitudes $a$), and only a tiny amount of energy is usually sufficient to break them (i.e., to change them to a spin triplet state). 
\begin{figure}[b]
	\includegraphics[width=8.4cm]{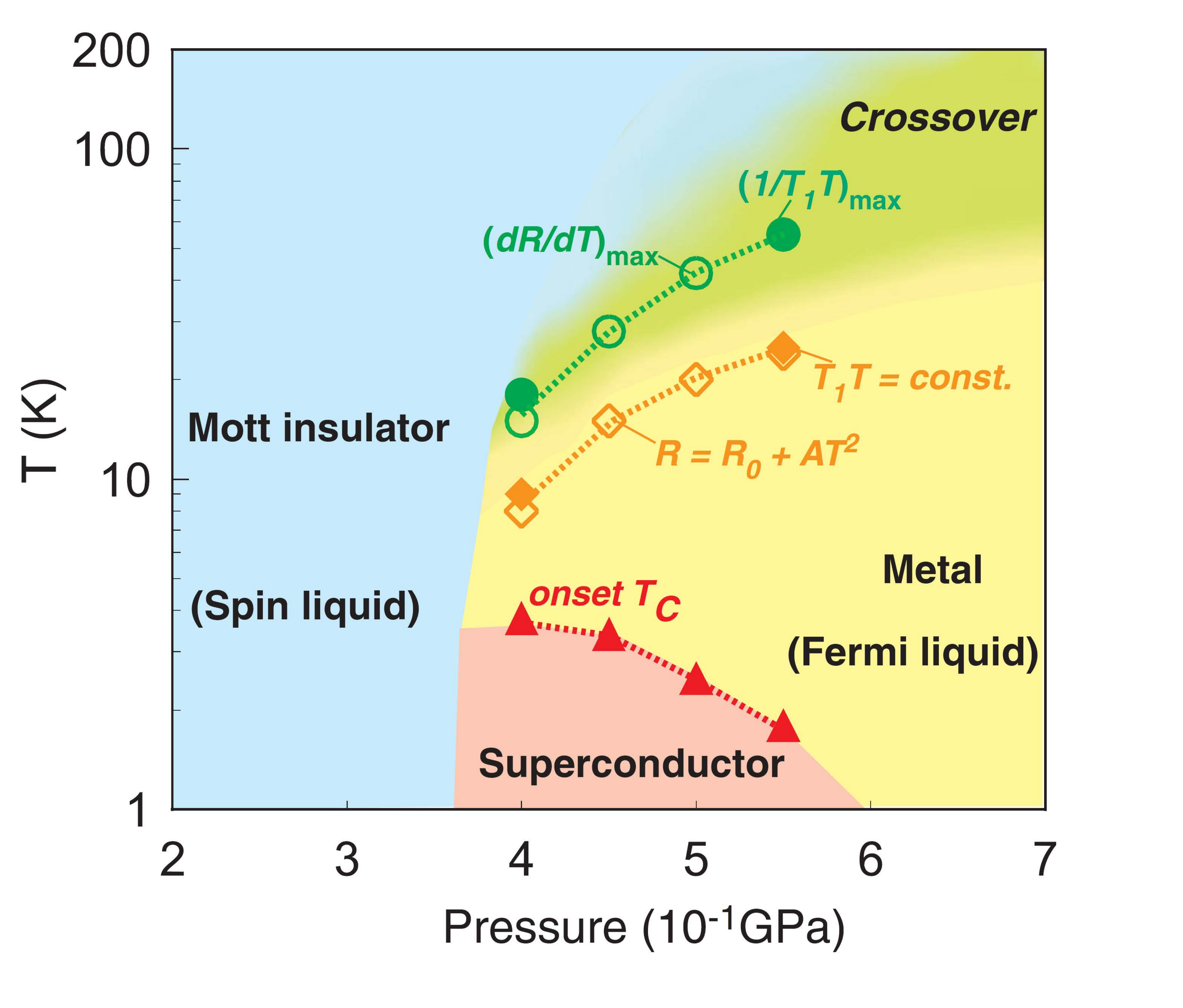} 
	\caption{Temperature vs.~pressure phase diagram of the triangular lattice $\kappa$--(BEDT-TTF)$_{2}$Cu$_{2}$(CN)$_{3}$, based on the NMR and resistivity measurement under hydrostatic pressures. Resistivity ($R$) measurements reveal the onset of superconducting transition. Upper boundary of the Fermi liquid region was determined from the deviation of $\frac{1}{T_1T}$ ($T_{1}$ is the NMR spin--lattice relaxation time) from the Korringa relation and $R$ from the $R_0+ AT^2$ behavior. The crossover lines and Mott transition temperature were defined from the maxima of $\frac{1}{TT_1}$ and $\frac{dR}{dT}$. Adapted from~\cite{PhysRevLett.95.177001} with permission from APS. }{\label{phasedia}}
\end{figure}
Therefore, states that consist of long-range valence bonds are expected to have spin excitations with lower energy than those in short-range RVB states or in valence bond crystals. 
 
\textcolor{black}{One of the most fundamental predictions of QSL theory is the existence of fractionalized $S = 1/2$ spinon excitations in these states, making these central to their experimental identification. 
The transformation of a singlet to a triplet state on a dimer, i.e., a change in spin by 1 (a spin flip), is the elementary magnetic excitation of the simplest dimer system (Fig. \ref{spe}) and leads to two neighbouring spins pointing in the same direction. 
In an RVB state, the two spins in the excited triplet
state are called spinons. They each carry spin-1/2 and can freely and independently propagate through the system via local rearrangements of  unbroken  valence bonds in the RVB superposition. 
The phenomenon where two independent spin-1/2
spinon quasiparticles arise from one spin-1 spin-flip excitation is known as spin fractionalization (not to be confused with
spin fragmentation from Subsection \ref{frac}), and resembles the appearance of two monopole excitations from a single spin flip
in classical spin ices discussed in Subsection \ref{SI_monopoles}. 
This gives rise to a continuum response in inelastic scattering experiments which is markedly different from sharp spin-1 spin wave (magnon) dispersions found in conventional long-range ordered magnets  where spin fractionalization is not present (Subsection~\ref{INS}). 
Furthermore, in RVB states, multiple other fractionalized quasiparticles besides spinons can also emerge, leading to even more complex spin dynamics. 
Besides the breaking of a singlet bond, spinons can also  spontaneously form around spin lattice defects.} 

\textcolor{black}{Experimental realization of spinon excitation has been known for a long time, and has also been well documented, in quasi-1D Heisenberg chains~\cite{PhysRevB.52.13368,Lake2005,Vasiliev2018}. 
In recent years, however, there has been enormous efforts devoted to the experimental observation of these exotic spinon excitations also in higher dimensional quantum magnets, where they are
accompanied by novel phenomena~\cite{han12fractionalized,PhysRevX.11.021044}.}
\textcolor{black}{Low-energy fractional spinon excitations in the RVB-type QSLs are expected to arise under spin--charge separation and could also result in low-loss transmission of spin current through electrical
	insulators~\cite{Hirobe2017}. 
}
  
There are several theoretical predictions concerning the realization of QSLs and associated exotic excitations in quantum materials beyond 1D. 
The paradigmatic 2D kagome and triangular lattice antiferromagnets are thought to be ideal hosts of QSL states~\cite{Carrasquilla2015,Meng2010,PhysRevLett.86.1881,PhysRevB.60.1064}. 
Most common theoretical models related to QSL states are based on isotropic pairwise spin interactions. But QSL Hamiltonians can also contain anisotropic terms, such as anisotropic exchange interactions
and dipolar interactions. These perturbing interactions can play an important role in the magnetism of candidate QSL
materials as they can often easily lift the near-degeneracy of the isotropic QSL ground state and destabilize it. It is thus
very challenging to find good model QSL materials where pertubing interactions are small enough that isotropic QSL
physics survives.

\begin{table*}
	\caption{Some representative frustrated materials with an exotic ground state on the 3D pyrochlore spin lattice and 2D triangular, kagome and honeycomb (Kitaev) lattices.} {\label{table}}\centering
 	\begin{tabular}{ | c | c | c | c | c |c|}
		\hline
		\textbf{Material name} &   \textbf{Lattice }   &\textbf{$ |\theta_{\rm CW}|$} (K)  & \textit{f}  & Ground state  & Ref. \\[2 ex ] \hline
		Ho$_{2}$Ti$_{2}$O$_{7}$ & pyrochlore & 1.9   & ${>}$38  & spin ice & \cite{Bramwell1495,Bramwell2009} \\ [2 ex] \hline 
		Dy$_{2}$Ti$_{2}$O$_{7}$ & pyrochlore & 1.2   & ${>}$20 & spin ice &  \cite{Bramwell1495,Bramwell2009} \\ [2 ex] \hline 
		Ce$_{2}$Zr$_{2}$O$_{7}$ & pyrochlore & 0.57  & ${>}$28  &   QSL & \cite{Gao2019} \\ [2 ex ] \hline
		ZnCu$_{3}$(OH)$_{6}$Cl$_{2}$ (herbertsmithite) & kagome & 314   & ${>}10^{3}$  & QSL& \cite{Shores2005} \\[2 ex] \hline  
		ZnCu$_{3}$(OH)$_{6}$FBr (Zn-barlowite) & kagome & 200   & ${>}10^{3}$  & QSL & \cite{Feng_2017} \\[2 ex] \hline 
		ZnCu$_{3}$(OH)$_{6}$SO$_{4}$ (Zn-brochantite) & kagome & 79   & ${>}$3700  & QSL& \cite{li2014gapless,gomilsek2016instabilities} \\[2 ex] \hline
		$\kappa$--(BEDT-TTF)$_{2}$Cu$_{2}$(CN)$_{3}$  & triangular &  375  & ${>}10^{3}$ & QSL & \cite{PhysRevLett.91.107001}   \\[2 ex] \hline
		EtMe$_{3}$Sb[Pd(dmit)$_{2}$]$_{2}$ & triangular  & 350   & ${>}10^{3}$ &  QSL & \cite{PhysRevB.77.104413} \\[2 ex] \hline 
		YbMgGaO$_{4}$ & triangular & 4  & ${>}$80 & QSL &  \cite{Li2015} \\ [2 ex ]\hline 
		NaYbS$_{2}$ & triangular & 7  & ${>}$140 & QSL &  \cite{PhysRevB.100.241116} \\ [2 ex ]\hline 
		NdTa$_{7}$O$_{19}$ & triangular & 0.46 & ${>}$ 11 & QSL & \cite{Arh2021} \\ [2 ex ]\hline 
		TbInO$_{3}$ & triangular & 1.13  & ${>}$170 & QSL &  \cite{Clark2019} \\ [2 ex ]\hline 
		$\alpha$--RuCl$_{3}$ & honeycomb & 40  & 6  & zigzag AFM order  &  \cite{PhysRevB.91.180401,PhysRevB.93.134423} \\ [2 ex] \hline 
		H$_{3}$LiIr$_{2}$O$_{6}$ & honeycomb & 105  &  $ {>}1000$   & QSL & \cite{Kitagawa2018}\\ [2 ex] \hline	 
	\end{tabular}
\end{table*}

The first discovered candidate QSL material was the organic Mott insulator $\kappa$--(BEDT-TTF)$_{2}$Cu$_{2}$(CN)$_{3}$~\cite{PhysRevLett.91.107001}, where (BEDT-TTF)$_{2}$ is a molecular dimer that carries $S = 1/2$ spin and forms an isotropic triangular lattice. 
Long-range magnetic order was found to be absent in this material down to 32~mK despite the presence of strong nearest-neighbor exchange interactions ($J/k_{\rm B} \approx 250$~K). 
Remarkably, various novel states such as superconductivity, QSL, and Fermi-liquid states (see Fig.~\ref{phasedia}) can be induced in this compound by applying external pressure, resulting in a very rich phase diagram~\cite{PhysRevLett.95.177001}. 
The next QSL material discovered was also an organic Mott insulator, EtMe$_{3}$Sb{[Pd(dmit)$_{2}$]}$_2$~\cite{PhysRevB.77.104413}, where [Pd(dmit)$_2$]$_2$ dimers carry localized spin $S = 1/2$ and form an anisotropic triangular lattice~\cite{Kato2004}. 
There are several other promising QSL candidates on various 2D spin lattices, which we will present in this section, while in-depth experimental studies of various aspects of select candidate materials will be presented in the Section\ref{expsig}. 

\textcolor{black}{In the context of QSLs, spin-1/2 quantum magnetic materials (see Table~\ref{table} for a list of select materials) are highly
	sought-after, since quantum fluctuations are most strongest when spin is the lowest. 
Examples with spin 1/2 can be found in material with molecular building blocks, such as (BEDT-TTF)$_{2}$ or [Pd(dmit)$_2$]$_2$, mentioned above.
A further example are 3$d$ transition metals with quenched orbital momentum, the most typical being Cu$^{2+}$. 
This magnetic ion is the basis of the majority of QSL candidates on the kagome lattice, such as herbertsmithite~\cite{Shores2005}, Zn-brochantite~\cite{li2014gapless} and Zn-barlowite~\cite{Feng_2017} (Subsection~\ref{kago}). 
Next, Kramers rare-earth ions can also provide an effective $J_{\rm eff} = 1/2$ magnetic moment at low $T$ due to crystal electric field (CEF) splitting of strongly spin--orbit-coupled energy levels. 
This is the situation encountered in novel triangular-lattice QSL candidates, such as  YbMgGaO$_4$~\cite{PhysRevLett.115.167203}, NaYbS$_2$~\cite{PhysRevB.98.220409}, and NdTa$_{7}$O$_{19}$~\cite{Arh2021} (Subsection~\ref{REtri}).
Moreover, a similar situation may be found it 4$d$- and 5$d$- systems, where spin--orbit coupling is also large.
Prominent examples of the latter class of materials include the honeycomb lattice found in Na$_{2}$IrO$_{3}$~\cite{PhysRevLett.108.127204} and $\alpha$--RuCl$_{3}$~\cite{PhysRevB.90.041112}, which are proximate Kitaev QSL candidates (Subsection~\ref{majorana}).}

\textcolor{black}{In addition to these examples, several other QSL candidates exist in 2D. As this review by no means aims to provide
	an exhaustive list of all of these, we will rather focus on the material classes mentioned above, which are particularly
	illustrative and demonstrate the richness of the field.
Additionally, in the experimental Section~\ref{expsig} some interesting results are presented for a couple of other 2D frustrated materials that fall outside the three selected classes of materials mentioned above. 
The first example (Subsection~\ref{Tcond}) is highly topical $5d$ transition-metal dichalcogenide 1T-TaS$_{2}$~\cite{Klan2017,law20171t,Valero2021, PhysRevLett.129.017202}, where a polaron-spin QSL state was discovered~\cite{Klan2017}. 
The second example (Subsection~\ref{TIO}) is the triangular--honeycomb antiferromagnet TbInO$_3$ with persistent spin dynamics, which features a spin lattice crossover from a triangular to a honeycomb lattice at low $T$ due to CEF effects~\cite{Clark2019}.
}

\begin{figure}	
\includegraphics[width=7.5cm]{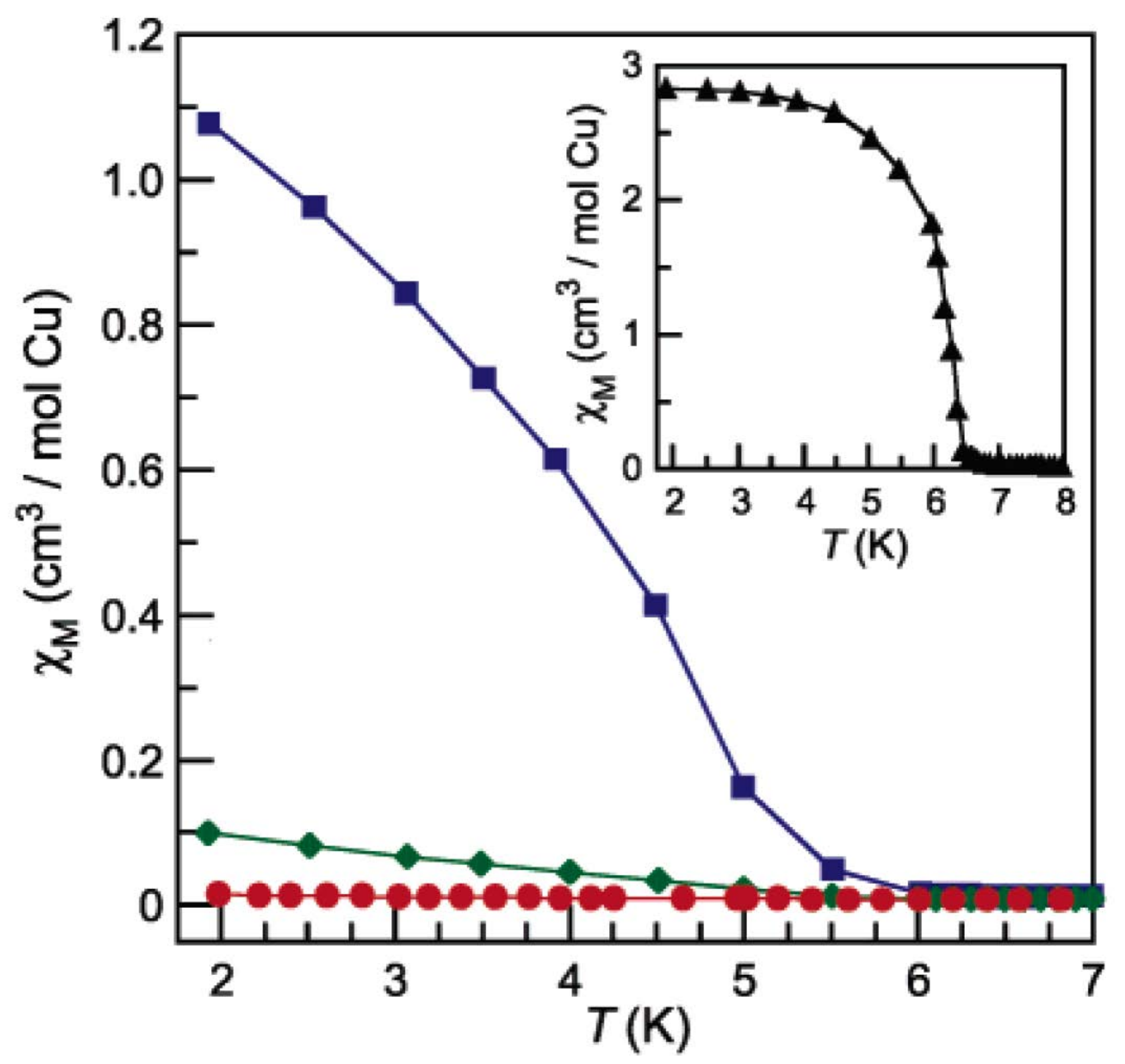} 
\caption{The low-temperature magnetic susceptibility ($\chi_M = M/H$) of the atacamite family of compounds Zn$_x$Cu$_{4-x}$(OH)$_6$Cl$_2$ for $x = 0$ ($\blacktriangle$ in inset), $x = 0.5$ ($\blacksquare$) $x = 0.66$ ($\blacklozenge$), and $x = 1$ ($\newmoon$).  Adapted from~\cite{Shores2005} with permission from ACS Publications.}{\label{ZnCu3(OH)6Cl2}}
\end{figure}

\subsection{Kagome antiferromagnet\label{kago}} 
\textcolor{black}{To date, the most studied QSL candidate has been the kagome-lattice antiferromagnet herbertsmithite, ZnCu$_3$(OH)$_6$Cl$_2$~\cite{mendels2010quantum, mendels2016quantum, RevModPhys.88.041002}. 
Therefore, we start our discussion of 2D QSL candidates on the kagome lattice and highlight the impacts of crystal structure, various perturbing interactions, and the presence of defects on QSL states on this lattice.}

\textcolor{black}{The 2D kagome lattice which consists of corner sharing triangles of spins (Fig.~\ref{frustrated_lattices}(b)), and offers a unique setting for realizing novel magnetic properties. 
The kagome geometry has a lower coordination number than those in the triangular lattice (Fig.~\ref{frustrated_lattices}(a)), and is thus less constrained.
This leads to enhanced quantum fluctuations, which result in a plethora of interesting quantum phenomena including QSLs state, massive Dirac fermions in ferromagnetic kagome metals, and the anomalous Hall effect~\cite{Ye2018,PhysRevB.97.134411}.
In the case of the antiferromagnetic model with isotropic Heisenberg interactions between spins located on nearest-neighboring kagome sites, a consensus of a QSL ground state has been reached. 
Nevertheless, the central question of whether this state is gapped or gapless still remains unanswered, even theoretically~\cite{clark2017closing, changlani2018macroscopically}. 
}

\textcolor{black}{Experimentally, kagome-lattice representatives were first found in 1990’s, yet the real breakthrough came only in 2005 with the discovery of herbertsmithite~\cite{Shores2005}, a member of the paratacamite family Cu$_3$$M$(OH)$_6$Cl$_2$~\cite{Roberts1996,braithwaite2004herbertsmithite} where
	$M$ = Zn.}
Here, Cu$^{2+}$ sites form a kagome lattice, while the $M$ site sits between the kagome layers.  Under substitution where
the inter-layer site becomes partially occupied by magnetic Cu$^{2+}$ ions ($S$ = 1/2), in place of non-magnetic Zn$^{2+}$ ions,
there appear additional Cu–O–Cu bonds to this site, which in turn leads to an effective weak ferromagnetic interaction
between the kagome layers, which, ultimately, induces long-range magnetic order at low-$T$ when the concentrations of
inter-layer Cu$^{2+}$ ions is large (see Fig.\ref{ZnCu3(OH)6Cl2}).
Herbertsmithite, ZnCu$_3$(OH)$_6$Cl$_2$, is the end member of the paratacamite family wherein  $M$ = Zn$^{2+}$ stabilizes a rhombohedral structure~\cite{Shores2005}. 
The Cu$^{2+}$ site is surrounded by four OH$^-$ (hydroxide) ligands and two chloride ligands that are attached to it axially but are farther apart. 
The three equivalent Cu$^{2+}$ ions in the unit cell are connected though hydroxide bonds and form an equilateral triangle.
The triangles are corner sharing, forming a perfect kagome lattice in the crystallographic $ab$-plane~\cite{Shores2005}. 
In this kagome antiferromagnet, the nearest-neighbor spins interact antiferromagnetically with a strong dominant Heisenberg-type exchange of $J/k_{\rm B} = 190$~K.
Despite this strong interaction, herbertsmithite does not exhibit any symmetry-breaking magnetic phase transition down to at least 50~mK and the spins instead remain dynamical, forming a liquid-like ground state down to the lowest accessible temperatures~\cite{PhysRevLett.98.077204,PhysRevLett.98.107204}. In other words, herbertsmithite is a QSL with a very large frustration parameter $f$ (Table~\ref{table}).
Furthermore, an inelastic scattering continuum of excitations was evidenced in this material~\cite{RevModPhys.88.041002,han12fractionalized}. 
However, despite huge experimental efforts
the central question of whether this state is gapped or gapless is still a subject of ongoing debate~\cite{Fu655,Khuntia2020,Wang2021,PhysRevX.12.011014}. The main
reason for the ambiguity is the relatively large percentage ($\sim$15 \text{\%}) of Cu/Zn site mixing defects (also called impurities),
which are unfortunately unavodiable and arise during synthesis due to the similar ionic radii of the two ions.

The discovery of herbertsmithite generated a flurry of activity in the search for QSL candidates in other variants of this family of compounds, as well as in other, structurally closely-related, kagome materials obtained either by a change of anions, cations, or by charge doping~\cite{PhysRevLett.113.227203,C6TC02399A,li2014gapless,PhysRevMaterials.3.074401,PhysRevX.6.041007,PhysRevLett.110.207208,PhysRevLett.118.237203,Tustain2020}. 
An interesting example of a related kagome antiferromagnet is Zn-brochantite, ZnCu$_{3}$(OH)$_{6}$SO$_{4}$, where Cu$^{2+}$ ions with $S = 1/2$ moments form a distorted kagome lattice~\cite{li2014gapless}. 
Magnetic susceptibility and specific heat measurements suggested the absence of long-range ordering down to 50~mK despite a Weiss temperature of $-79$~K, again indicating a highly frustrated system (Table~\ref{table}). 
Furthermore, temperature-independent magnetic susceptibility and linear temperature dependence of magnetic specific heat at low $T$ indicate the presence of a gapless QSL  with a spinon Fermi surface (a so-called spinon metal) in this material~\cite{li2014gapless}. 
This suggestion was confirmed by local-probe NMR and $\mu$SR measurements, which showed no magnetic order down to at least 21~mK~\cite{gomilsek2016instabilities,gomilsek2016muSR,Gomilsek2019}, as well as other exotic phenomena, including a magnetic field-induced instability of this gapless QSL state under spinon pairing, producing a gapped QSL at high applied fields and low $T$~\cite{gomilsek2017field}.
Furthermore, the long-standing theoretical prediction of a charge-free, spinon Kondo effect of impurities embedded in a QSL~\cite{PhysRevB.74.165114,florens2006kondo,kim2008kondo,ribeiro2011magnetic,vojta2016kondo} was first experimentally confirmed in this compound~\cite{Gomilsek2019} via local-probe techniques, mainly $\mu$SR. 

\textcolor{black}{The third structurally related kagome QSL candidate, which is of considerable recent focus~\cite{Tustain2020, smaha2020materializing, Fu655, Wang2021, PhysRevLett.128.157202, Yuan2022}, is Zn-barlowite, ZnCu$_{3}$(OH)$_{6}$FBr (Table~\ref{table}).
Like herbertsmithite and Zn-brochantite, Zn-barlowite is also plagued by magnetic impurities, which arise due to intrinsic ion mixing between the Cu site and the Zn site in the crystal structure~\cite{RevModPhys.88.041002}.
However, while the minimal amount of magnetic Cu$^{2+}$ ions on the Zn site in the former two compounds is well established to be around 15\% even in the best available crystals, the situation is less clear in Zn-barlowite. 
Here, impurity concentrations of only a few percents have been reported~\cite{Tustain2020,Fu655,Yuan2022}. 
It is important to stress that this compound was first reported to have a gapped QSL ground state based on NMR results~\cite{Feng_2017}. 
However, more refined NMR studies have later suggested that its non-ordered ground state contains spin singlets with a spatially inhomogeneous spin gap~\cite{Wang2021, Yuan2022}.
} 

\subsection{Rare-earth based triangular lattice\label{REtri}}
\textcolor{black}{The simplest frustrated geometry, the 2D triangular spin lattice, was the first one proposed as possibly hosting a magnetically disordered ground state by G. H. Wannier already in 1950~\cite{wannier1950antiferromagnetism}.
He showed that the ground state of this lattice under highly anisotropic Ising exchange interactions between nearest neighbours features extensive degeneracy and is thus an example of a classical spin liquid (in close similarity to the the spin ices discussed in Subsection~\ref{SI}). 
Later, in 1973, P. W. Anderson studied the same lattice but with isotropic Heisenberg interactions and proposed it could host the quantum RVB state i.e., a QSL~\cite{ANDERSON1973153}. 
Although it was demonstrated theoretically that the ground state of the nearest-neighbor Heisenberg model on the triangular lattice exhibits the conventional 120$^{\circ}$ order in the 1990's~\cite{PhysRevLett.69.2590,PhysRevB.60.1064,PhysRevLett.60.2531}, Anderson's seminal paper nevertheless ignited the field of frustration and quantum spin liquids. 
These can be stabilized on the triangular lattice by next-neighbor exchange intearctions,~\cite{PhysRevB.93.144411, PhysRevB.92.041105, PhysRevB.91.014426} spatially-anisotropic exchange interactions~\cite{PhysRevB.74.014408, PhysRevB.94.035107, PhysRevB.95.165110}, magnetic anisotropy~\cite{PhysRevLett.112.127203, PhysRevX.9.021017}, or higher order exchange terms~\cite{PhysRevB.60.1064}, such as ring exchange, which is present in weak Mott insulators (i.e., close to the metal--insulator transition) like certain organic QSL candidates~\cite{PhysRevB.72.045105, PhysRevB.81.245121}.
}

\begin{figure}[t]
	\includegraphics[width=0.9\linewidth]{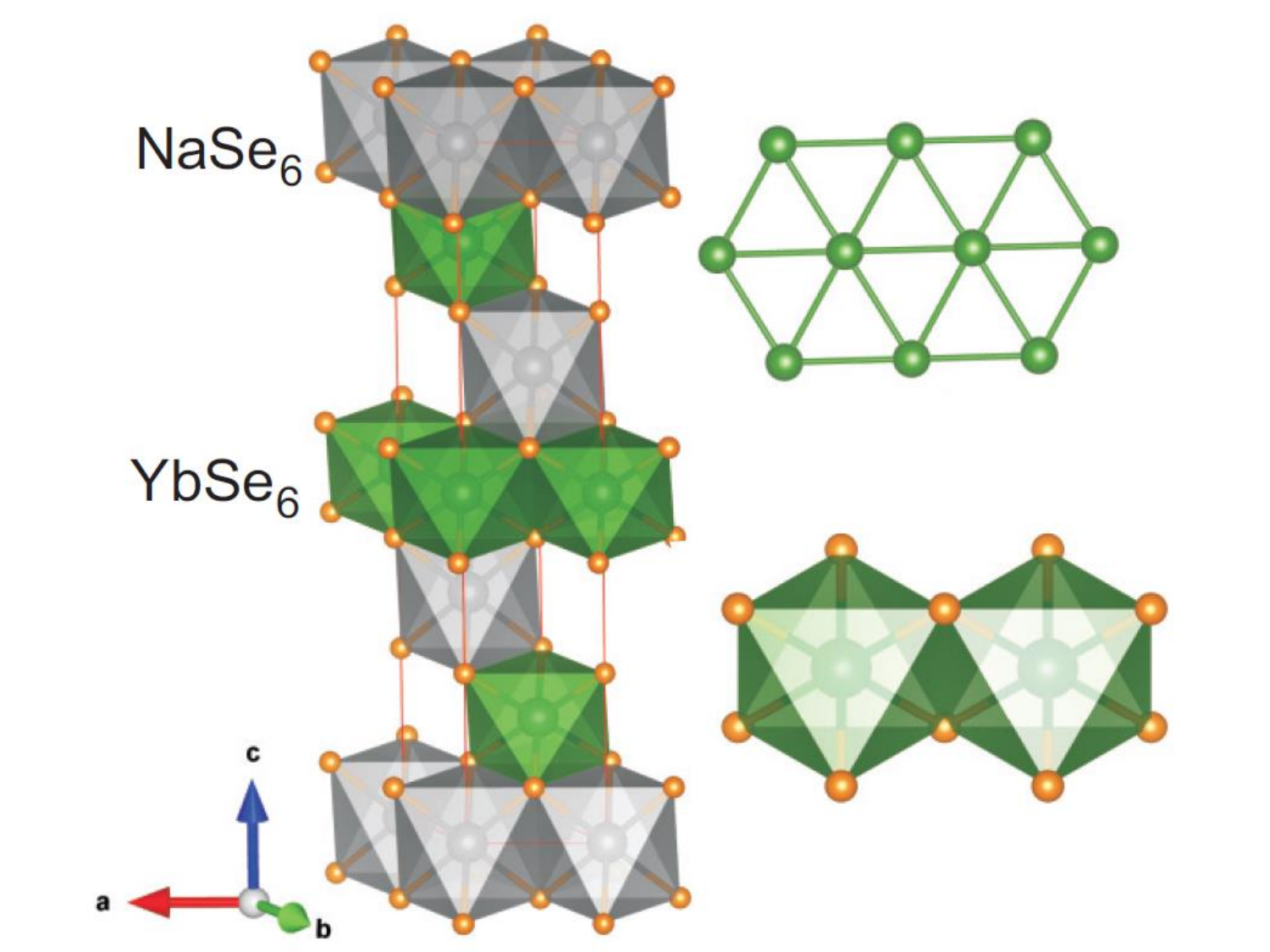} 
	\caption{Lattice structure of the Yb delafossite NaYbSe$_2$. Yb forms perfect triangular layers (green) by edge-shared YbSe$_6$ octahedra in the $ab$ crystallographic plane. Adapted from~\cite{PhysRevB.100.224417} with permission from APS.}{\label{Tri_Nayb}}
\end{figure}
\begin{figure}[t]
	\includegraphics[width=0.45\textwidth]{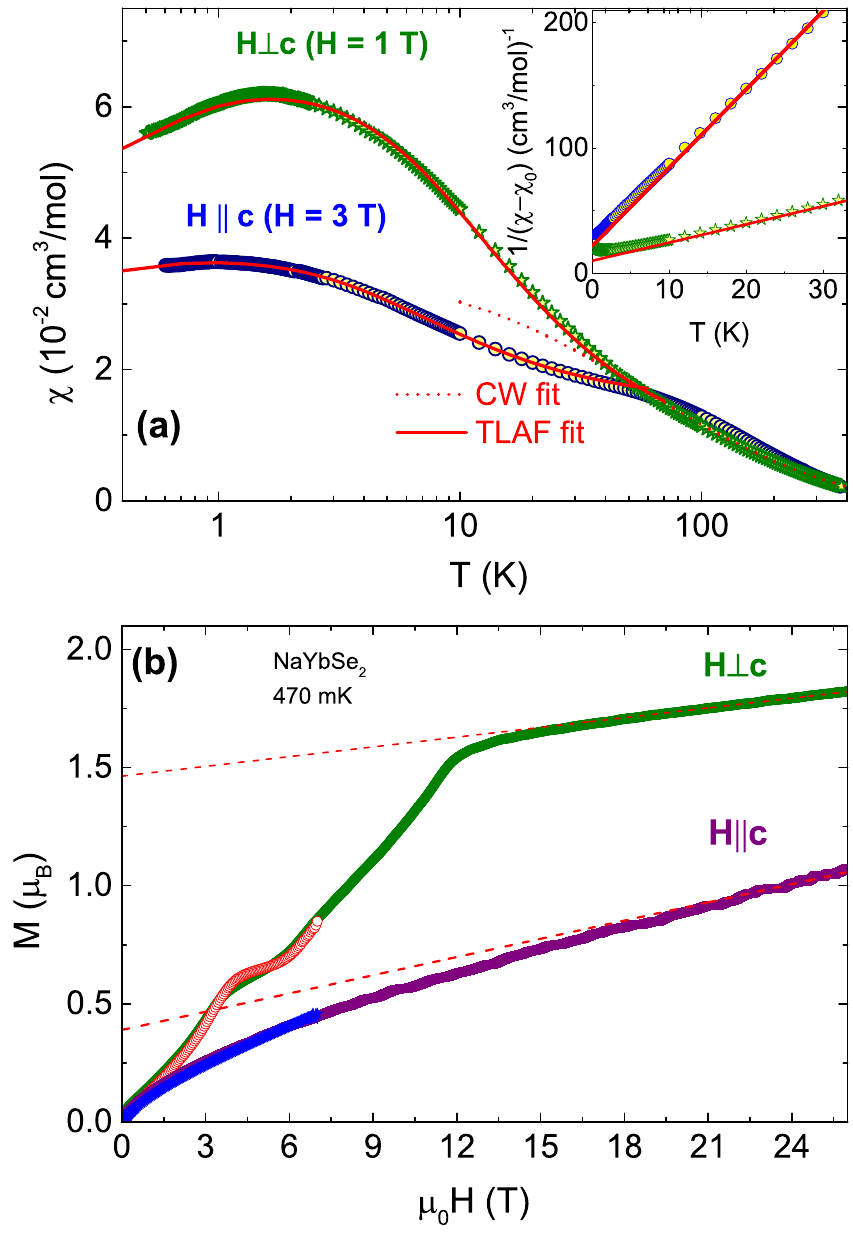} 
	\caption{
    (a) Magnetic susceptibility of NaYbSe$_{2}$ single crystals, where $c$ denotes the direction orthogonal to the triangular planes. 
    Red dotted and solid lines correspond to a fit with a high-$T$ Curie--Weiss (CW) law and  a triangular-lattice antiferromagnet (TLAF) spin model, respectively. 
    The inset shows the low-$T$ inverse susceptibility after subtracting the Van-Vleck		contribution along with CW fits for both directions. 
    (b) Magnetization isotherms at 470~mK with dashed lines showing the Van-Vleck contribution.
    Adapted from~\cite{PhysRevB.100.224417} with permission from APS.
	}{\label{mag_NaYb}}
\end{figure} 

\textcolor{black}{
In the field of triangular-lattice QSLs, the focus has recently shifted from organic compounds, such as $\kappa$--(BEDT-TTF)$_{2}$Cu$_{2}$(CN)$_{3}$~\cite{PhysRevLett.91.107001} and EtMe$_{3}$Sb{[Pd(dmit)$_{2}$]}$_2$~\cite{PhysRevB.77.104413}, to rare-earth-based compounds, which we will focus in this review.}
The strong impact of the spin--orbit coupling and bond-dependent anisotropic exchange interactions due to the CEF on the triangular lattice has become evident in recent years and has significantly boosted the search for new QSL materials~\cite{Balents2010,https://doi.org/10.1002/qute.201900089,Wen2019,RevModPhys.89.025003}. 
These materials nowadays represent a major avenue in the search for novel enigmatic QSLs (see Section~\ref{qsll}). 
Here, 4$f$ magnets are at the center of interest because the half-integer Kramers ions ordinarily exhibit a ground-state doublet with an effective spin-1/2 at low temperatures due to CEF splitting. 
This is a pseudospin state, because it involves strongly intertwined spin and orbital momenta.
As the CEF breaks rotational symmetry, this spin--orbit entangled ground state is usually characterized by strong magnetic anisotropy.

\textcolor{black}{To date the most intensively studied rare-earth-based QSL candidate with a triangular spin lattice has thus far been YbMgGaO$_{4}$~\cite{Li2015}. 
In this material, extensive INS studies demonstrated a continuum of excitations and suggested a gapless QSL state with a spinon Fermi surface~\cite{paddison2017continuous, shen2016evidence}. 
Various attempts to determine the exchange interactions in this compound all agree on the main interaction being of the order of ${\sim}2$~K and on the presence of large easy-plane magnetic anisotropy~\cite{Li_2020}. 
However, the strengths of off-diagonal exchange terms and further-neighbor interactions, which might crucially impact the spin state in this material, remain unresolved. 
Furthermore, like in kagome QSL candidates (Subsection~\ref{kago}), disorder seems to play an important role in this material.
It arises from Mg/Ga intersite mixing, which also distorts the local environment of Yb$^{3+}$ magnetic moments~\cite{PhysRevLett.118.107202}.
This is believed to significantly influence the magnetic behavior of this compound, leading to a QSL-like ground state~\cite{PhysRevLett.119.157201, kimchi2018valence}. 
A similar situation is encountered in the sibling compound TmMgGaO$_{4}$, where site-mixing disorder merges the two lowest singlet of the non-Kramers Tm$^{3+}$ ions into a quasidoublet~\cite{PhysRevX.10.011007}. 
However, in that case, small inhomogeneous inner doublet gaps are present, and these lead to partial magnetic ordering at low $T$.}

\textcolor{black}{Another triangular-lattice rare-earth-based family of compounds, which also contains several QSL candidates, is that of rare-earth chalcogenides Na$ReCh_2$, where $Re$ is a rare earth and $Ch$~=~O, S, or Se~\cite{Liu_2018}. 
The most studied representatives of this family are NaYbO$_2$, NaYbS$_2$, and NaYbSe$_2$, which, contrary to YbMgGaO$_{4}$, lack any inherent structural randomness~\cite{PhysRevB.100.144432,PhysRevB.100.241116,PhysRevB.98.220409}.} 
Fig.~\ref{Tri_Nayb} shows their common crystal structure, which consists of edge-sharing Yb$Ch_{6}$ and Na$Ch_{6}$ octahedra. 
In this lattice, Yb$Ch_{6}$ octahedra are weakly distorted with respect to the intra-plane Yb--Yb bond by an angle of 53.33$^\circ$ for NaYbSe$_{2}$, 48.60$^{\circ}$ for NaYbS$_{2}$, and 47.70$^{\circ}$ for NaYbO$_{2}$. 
The tilting angle for an undistorted octahedron is ${\sim}54.74^\circ$. 
The local spin anisotropy (determined by CEF splitting) and exchange anisotropy (arising from a complex superexchange pathway via chalcogenide $p$ states) depend strongly on this octahedral tilt~\cite{PhysRevB.98.220409,PhysRevB.99.180401,PhysRevB.100.224417}. 

Magnetic susceptibility, $\mu$SR and NMR measurements demonstrated the absence of long-range magnetic ordering down to at least 300~mK in single crystals of NaYbSe$_{2}$, as well as in  NaYbS$_{2}$ and NaYbO$_{2}$~\cite{PhysRevB.100.224417,PhysRevB.100.144432}. 
In addition, specific heat results obtained down to 100~mK for all three systems evidenced the absence of magnetic order and revealed the gapless nature of low-energy excitations from the constancy of $C_{m}/T$ at $T\rightarrow 0$. 
\textcolor{black}{These materials are thus candidates for a QSL ground state, which, however, undergoes a magnetic instability under the applied magnetic field leading to the 1/3 magnetization plateau (Fig.~\ref{mag_NaYb}) compatible with so-called $uud$ magnetic order~\cite{Bordelon2019,PhysRevB.99.180401,PhysRevB.100.224417}. 
In NaYbO$_2$, the spectral weight of the excitation continuum is peaked at the K point of the Brillouin zone~\cite{Bordelon2019,PhysRevB.100.241116}, suggesting a gapless Dirac QSL~\cite{PhysRevLett.120.207203}. 
Whether this state and its field-induced transition are stabilized primary due to easy-plane magnetic anisotropy, which has been observed in these systems, or whether next-neighbor exchange interactions also play an important role remains to be clarified.
}

In NaYbSe$_{2}$, the application of high pressures tunes the Mott insulator state into a metallic state with strong Kondo-like correlations among the Yb moments~\cite{PhysRevLett.117.107203}. 
Like in organic salts, electrical resistivity shows signs of non-Fermi-liquid behaviour, and surprisingly, even superconductivity at about 10~K~\cite{Jia_2020}. 
The observation of pressure-induced superconductivity in this rare-earth dichalcogenide delafossite is quite striking in the context of perturbation-induced novel states in quantum matter~\cite{RevModPhys.78.17,Lee_2007,PhysRevB.103.184419}. 
This offers a new route to realizing superconductivity in QSL candidates and is a step forward in exploring its possible origins.

\begin{figure*}[t]
	\includegraphics[width=\textwidth]{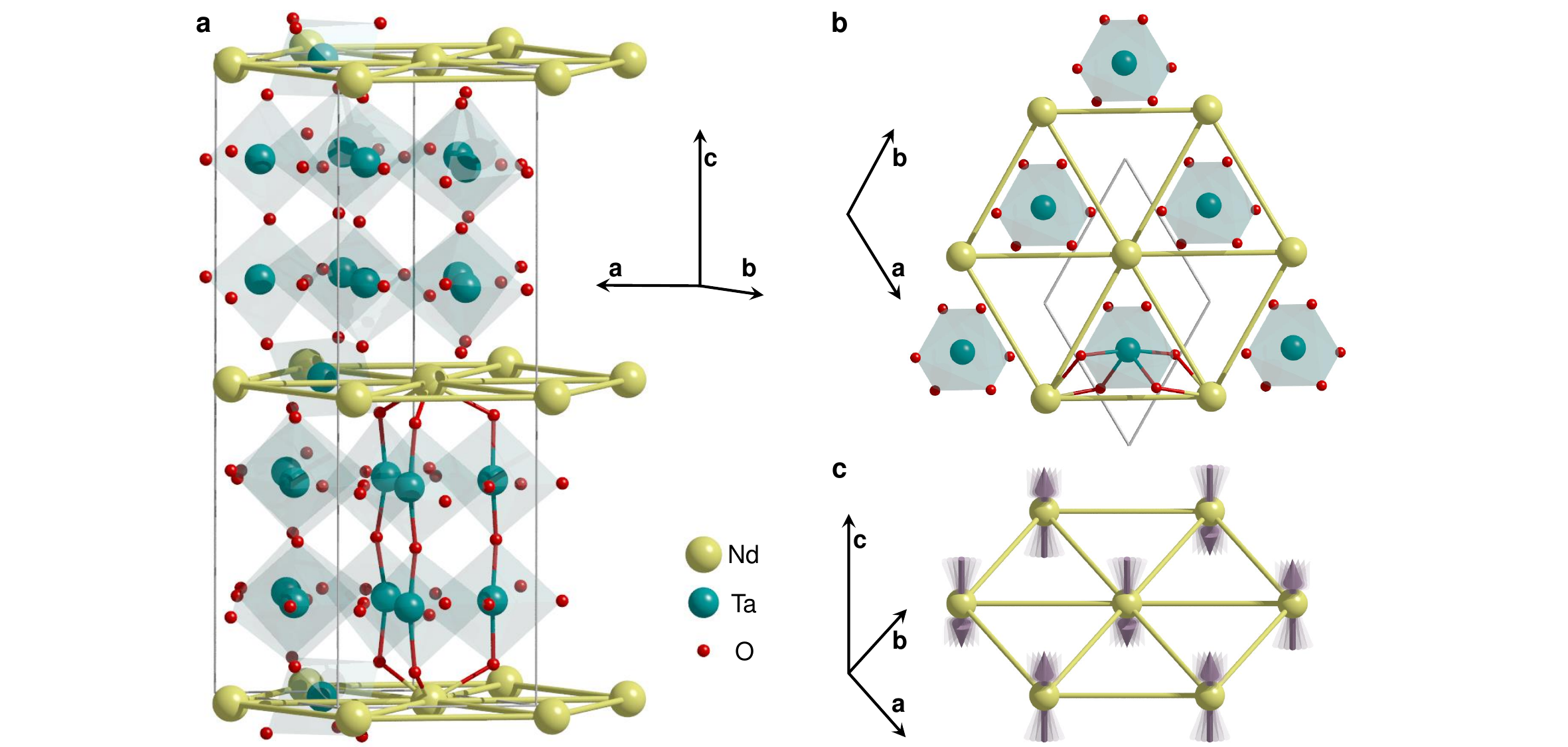} 
	\caption{(a) Layered crystal structure of NdTa$_7$O$_{19}$ with (b) a perfect triangular lattice of Nd$^{3+}$ magnetic ions.
    (c) The corresponding spins adopt a dynamical QSL ground state with strong Ising correlations in the direction perpendicular to the triangular planes. 
    Adapted from~\cite{Arh2021} with permission from NPG.
	}{\label{NTOstuct}}
\end{figure*} 
\textcolor{black}{Recently, a novel family of rare-earth-based disorder-free antiferromagnets on the triangular lattice has been introduced, namely rare-earth heptatantalates, with NdTa$_7$O$_{19}$ being their first extensively studied representative~\cite{Arh2021}.
This compound has well-separated layers of magnetic Nd$^{3+}$ ions, features an equilateral triangular lattice within the $ab$ crystallographic plane (Fig.~\ref{NTOstuct}), and lacks any detectable site disorder or nonstoichiometry. 
Just like the previously presented triangular-lattice QSL candidates, it also displays a disordered and fluctuating ground state. 
However, what makes it unique is that it displays predominantly Ising-like spin correlations (see Subsection~\ref{NTOIN} for details) instead of easy-plane or isotropic ones, making it the first QSL candidate of this kind~\cite{Arh2021}. 
This unusual state was argued to be due to the predominantly Ising nature of the nearest-neighbor exchange interactions, but a precise microscopic characterization of the corresponding spin Hamiltonian remains an open task. 
Importantly, other members of the family, which share the same crystal structure but have a different rare-earth ion, have also recently been successfully synthesized~\cite{wang2023synthesis}. 
As the properties of the newly discovered QSL states in these systems can be traced back to large magnetic anisotropy, which is ion-specific, it is quite intriguing to imagine various other exotic magnetic states and phenomena that could emerge in these isostructural compounds, in analogy to the richness of exotic phenomena found in various pyrochlore compounds (Section~\ref{spinicc}). 
}

\subsection{Kitaev QSL on the honeycomb lattice}\label{majorana}
\begin{figure}
	\includegraphics[width=8cm]{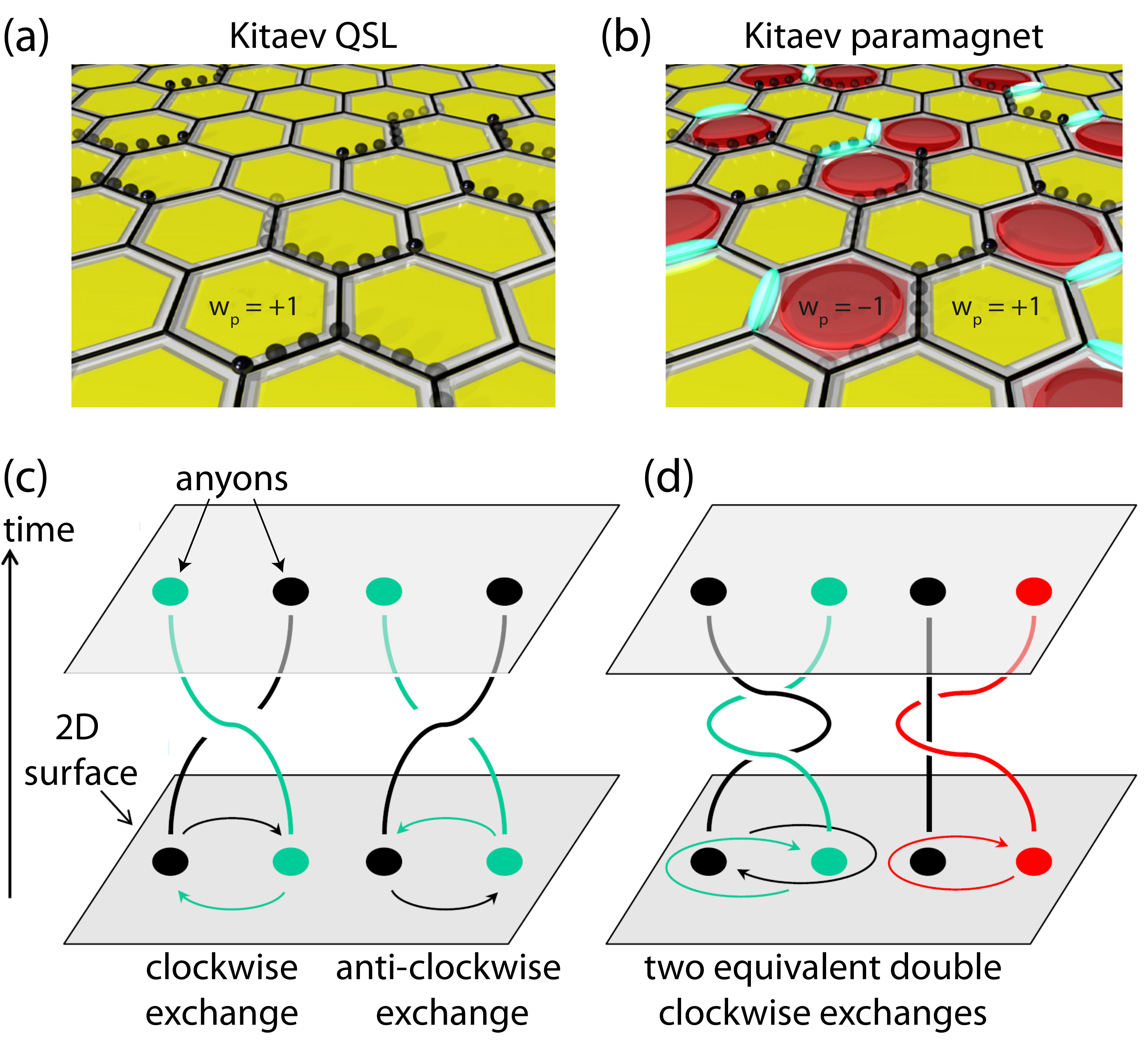}
	\caption{%
    (a) QSL state on the Kitaev honeycomb model at low $T < T_L$. Only low-energy Majorana fermions (black balls) are thermally activated, while $\mathbb{Z}_2$ gauge fluxes (yellow hexagons) are almost frozen in the zero-flux state. 
    (b) At intermediate temperatures $T_L < T < T_H$, $\pi$ gauge fluxes also become thermally activated (red hexagons).
    (c) Anyons distinguish between clock-wise and anti-clockwise exchange. For Abelian anyons the difference between the two  resulting system wavefunctions is just an overall phase factor, while for non-Abelian anyons the two wavefunctions
    become linearly independent.
    (d) The change in the system wavefunction under anyon exchange is topological, as it only depends on the sequence and number of exchanges. The relevant topological data is described by the braid group, which is why anyon exchange is often called anyon braiding.
    (a,b) Adapted from~\cite{Do2017} with permission from NPG, (c,d) adapted from~\cite{zeuch2016entangling} with permission.
    }{\label{kitaev_anyon_braiding}}
\end{figure}

\begin{figure}[b]
	\includegraphics[width=0.5\textwidth]{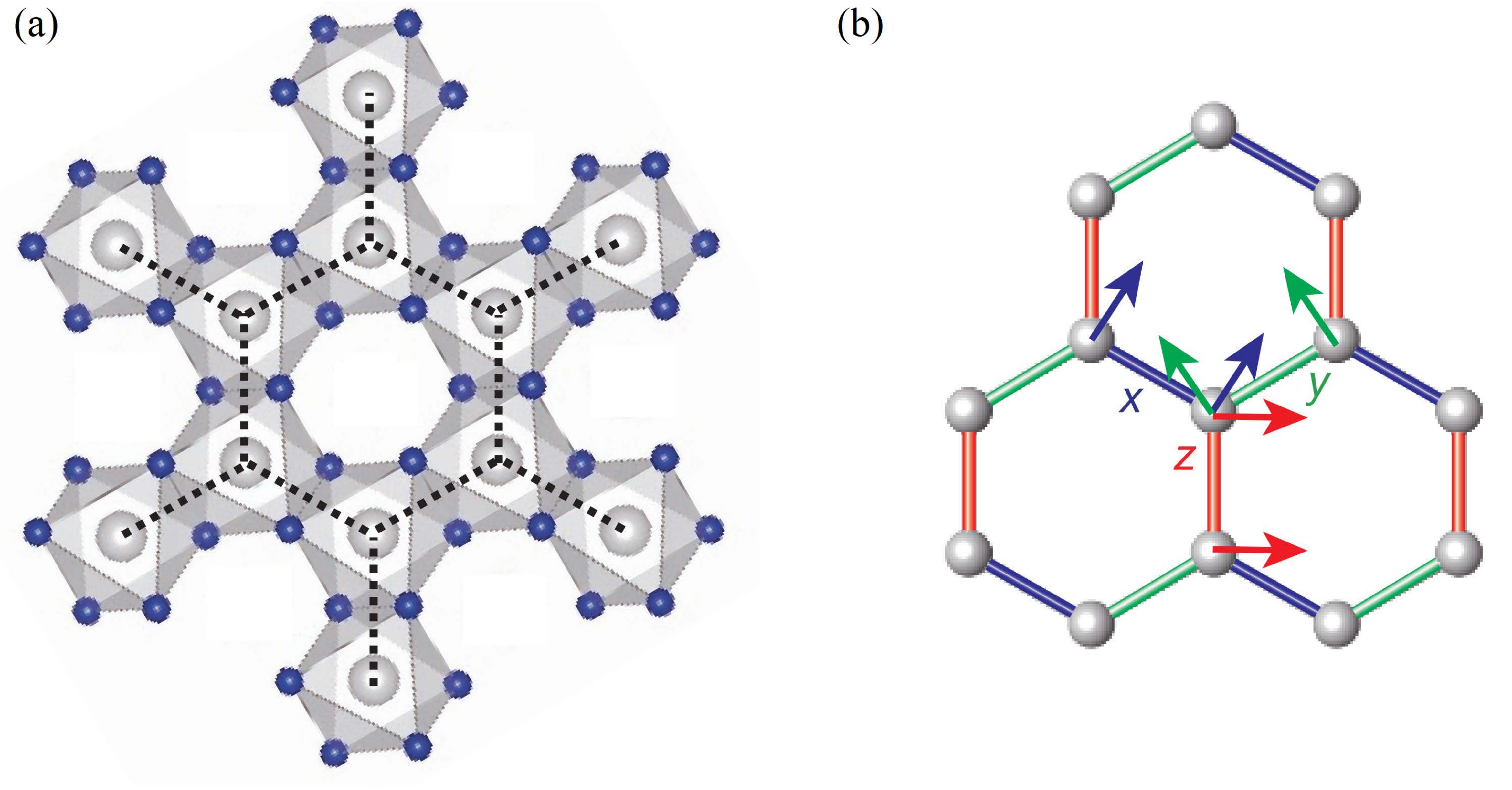} 
	\caption{(a) Honeycomb lattice of the nearest-neighboring Ru$^{3+}$ magnetic ions ($J_{\rm eff} = 1/2$; 4$d$$^{5}$) in $\alpha$--RuCl$_{3}$ formed by edge-sharing RuCl$_{6}$ octahedra. Here, Ru$^{3+}$ and Cl$^-$ ions are represented by gray and blue spheres. 
    (b) \textcolor{black}{Sketch of bond-dependent ferromagnetic Ising interactions between effective spin-1/2 magnetic moments (arrows) in the Kitaev honeycomb model where competing interaction between three 120$^{\circ}$ bonds with orthogonal Ising axes $\gamma = x, y, z$ provides a novel route to spin frustration.} 
    Adapted from~\cite{Takagi2019} with permission from NPG.}{\label{Kitaevin}}
\end{figure}

\textcolor{black}{The 2D honeycomb lattice is another 2D lattice that can support QSL states.
As it is a bipartite lattice, isotropic Heisenberg exchange between nearest-neighbor spins simply leads to long-range magnetic order at sufficiently low $T$.
However, anisotropic magnetic interactions, especially bond-dependent Ising-type exchange interactions arranged in specific patterns, which we then call Kitaev interactions (Fig.~\ref{Kitaevin}(b)), can stabilize a QSL. 
The Kitaev model on the honeycomb lattice is particularly appealing as its ground state can be calculated exactly and can be shown to be a topological QSL~\cite{KITAEV20062}. 
Furthermore, excitations from this state are highly unusual because they behave as two types of anyons --- quasiparticles whose exchange statistics are neither fermionic nor bosonic but somewhere in-between~\cite{wilczek1982quantum}.
Namely, a spin flip, which is a spin-1 excitation arising, e.g., due to an experimental probe interacting with the spin system, in the case of a Kitaev QSL fractionalizes into one emergent Majorana fermion  (Fig.~\ref{kitaev_anyon_braiding}(a)), which is its own antiparticle, and a pair of emergent $\mathbb{Z}_2$ bosonic gauge fluxes (Fig.~\ref{kitaev_anyon_braiding}(b))~\cite{Jansa2018,Do2017,KITAEV20062}. 
The mutual statistics of these two types of quasiparticle excitations non-Abelian anyons of Ising type~\cite{KITAEV20062} and thus, in principle, suitable for robust,
topologically-protected quantum computation via anyon braiding and fusion (Fig.~\ref{kitaev_anyon_braiding}(c,d))~\cite{KITAEV20032, PhysRevLett.98.087204, Chen_2008,PhysRevLett.98.247201}. 
}

\textcolor{black}{The notion of antiparticles was introduced in 1928 by Paul Dirac~\cite{dirac1928quantum}, who 
solved the relativistic wave equation for spin-1/2 particles, where the positive energy solution describes the properties of an electron and the negative energy solution defines its antiparticle positron, which has the same mass but opposite electric charge. 
In 1937, Ettore Majorana proposed an alternative relativistic equation that could be represented in a form similar to the Dirac equation but instead described particles, now called Majorana fermions, that were their own antiparticles. 
%
In materials, a similar situation can arise with zero-energy (zero-mode) ``partiholes'', which are bound states of electron and a hole that are different from conventional excitons in semiconductors~\cite{Wilczek2009}. 
The latter are bosonic (with integer spins) whereas partiholes are spin-1/2 particles.
}
These quasiparticles can be found in some superconductors~\cite{PhysRevB.73.220502,Volovik1999}, where flux vortices trap zero-mode spin-1/2 ``excitons'', and in the fractional quantum Hall effect~\cite{PhysRevB.61.10267,PhysRevLett.94.166802}. 
They usually coexists with other emergent low-energy excitations~\cite{PhysRevB.82.020509}.
Interestingly, as mentioned above, Majorana fermions also emerge as purely magnetic quasiparticle excitations in Kitaev-model QSLs, which we describe next.

%
The spin Hamiltonian of the Kitaev honeycomb model has the form
\begin{equation}
\mathcal{H} = -\sum_{<jk>_{\gamma}}J_K S_{j}^\gamma S_{k}^\gamma,
\end{equation}
where $\gamma = x, y, z$ denotes the direction of the Ising-type exchange interaction (the Kitaev interaction), between the two spin components of spins at sites $j$ and $k$, while $J_K$ gives the strength of this Kitaev interaction~\cite{KITAEV20062,PhysRevLett.117.157203}. 
Importantly, in the model, each spin is coupled to its three honeycomb nearest-neighbours via Kitaev exchanges with three orthogonal Ising axes $\gamma$ (Fig.~\ref{Kitaevin}(b)) that compete, which provides a novel route to spin frustration and stabilizes a disordered QSL state. 
Though this model was initially thought to be just a toy model (as strong Ising coupling is not expected for pure $S = 1/2$ spins), a theoretical proposal~\cite{PhysRevLett.102.017205} was later made on how Kitaev physics could be realized in materials with $J_{\rm eff} = 1/2$ pseudo-spins after the discovery of the Ir$^{4+}$-based $J_{\rm eff} = 1/2$ spin--orbit driven Mott-insulator Sr$_2$IrO$_4$~\cite{Kim1329}. 
\textcolor{black}{This inspired researchers to look for 4$d$/5$d$-transition-metal-based 2D honeycomb lattices that could exhibit bond-directional Ising-type exchange interactions found in the ideal Kitaev honeycomb model.} 
As a result, spin--orbit coupled 4$d$/5$d$ materials with a honeycomb spin lattice, such as (Na,Li)$_2$IrO$_3$~\cite{Winter_2017} and
$\alpha$--RuCl$_{3}$~\cite{PhysRevLett.108.127203,PhysRevB.93.195158,PhysRevResearch.2.042007,versteeg2020nonequilibrium,wagner2022magnetooptical}, with pseudospin $J_{\rm eff} = 1/2$~\cite{PhysRevB.93.134423,PhysRevB.98.100403,Do2017} have been thoroughly investigated as potential Kitaev QSL candidates over the last decade. 

\textcolor{black}{The local easy axes $\gamma$ of Ising exchange interactions in Kitaev candidate materials relies on the spatial orientation of a particular exchange bonds~\cite{10.1143/PTPS.160.155}. 
Therefore, the geometric orientation of magnetic ions (e.g., Ir$^{4+}$/Ru$^{3+}$) with surrounding ligands (O$^{2-}$/Cl$^{-}$) plays a crucial role in establishing Kitaev interactions~\cite{PhysRevLett.102.017205}. 
Corner-sharing octahedra lead to single exchange path with a 180$^{\circ}$ metal--ligand--metal bond that produces symmetric Heisenberg exchange between the spin–orbit entangled $J_{\rm eff} = 1/2$ moments.
On the other hand, edge-sharing octahedra (Fig.~\ref{Kitaevin}) may form two exchange paths with 90$^{\circ}$ metal--ligand--metal bonds, where destructive quantum interference between isotropic Heisenberg interactions on the two superexchange paths can lead to the dominance of strong anisotropic Kitaev interactions~\cite{PhysRevLett.102.017205,RevModPhys.87.1}. 
Layered iridium-based honeycomb magnets deviate from the ideal Kitaev case due to strong Heisenberg interactions that stabilize long-range antiferromagnetic order at low $T$~\cite{PhysRevB.95.144406,PhysRevLett.108.127204,PhysRevB.93.195158,PhysRevLett.112.077204}. 
However, in $\alpha$--RuCl$_{3}$, Ru--O--Ru bond angles are closer to the ideal 90$^{\circ}$ and lead to the dominance of Kitaev interactions over Heisenberg interactions and to lower magnetic phase transition temperatures than those observer in iridates, relative to the strength of the involved magnetic exchange interactions~\cite{PhysRevB.90.041112,PhysRevB.93.134423}.} 

\textcolor{black}{ 
Specifically, even though the ideal Kitaev model hosts an exact 2D QSL ground state, the most intensively studied Kitaev candidate $\alpha$--RuCl$_3$, like the majority of Kitaev candidate materials, shows zigzag long-range antiferromagnetic order below $T_N=6.5$~K. 
This arises from a combination of additional non-Kitaev interactions within heneycomb planes and inter-layer interactions~\cite{TREBST20221,PhysRevLett.112.077204,doi:10.1146/annurev-conmatphys-031115-011319}.
In this material,} interactions between Ru$^{3+}$ ($J_{\rm eff}= 1/2$) magnetic ions are still dominated by the anisotropic Kitaev term ($J_K$)~\cite{Yadav2016,PhysRevLett.117.157203}, unlike in Na$_2$IrO$_3$~\cite{PhysRevLett.108.127203}, where additional Heisenberg interactions are comparable in size to Kitaev interactions. 
Quantum Monte Carlo simulations at finite $T$ have predicted two well-separated energy scales for spin correlations and the appearance of spin fractionalization in $\alpha$--RuCl$_3$~\cite{PhysRevB.92.115122,PhysRevLett.117.157203}. 
The lower crossover temperature, $T_L$, is dominated by thermal fluctuations of emergent localized $\mathbb{Z}_2$ gauge fluxes, which for $T < T_L$ are frozen (Fig.~\ref{kitaev_anyon_braiding}(a)) and at $T > T_L$ become thermally activated (Fig.~\ref{kitaev_anyon_braiding}(b)), the higher crossover temperature, $T_H$, arises from a build-up of spin correlations between neighboring spins, which give rise to emergent, spin-fractionalized itinerant Majorana fermions at $T < T_H$, while at $T > T_H$ the system is in an ordinary (effectively non-interacting) paramagnetic state.  
The intermediate regime, $T_L < T < T_H$, with itinerant Majorana fermions and thermally activated $\mathbb{Z}_2$ gauge fluxes, is called a Kitaev paramagnet. 
This scenario was also verified experimentally~\cite{Do2017}. 
\begin{figure}
	\includegraphics[width=8cm]{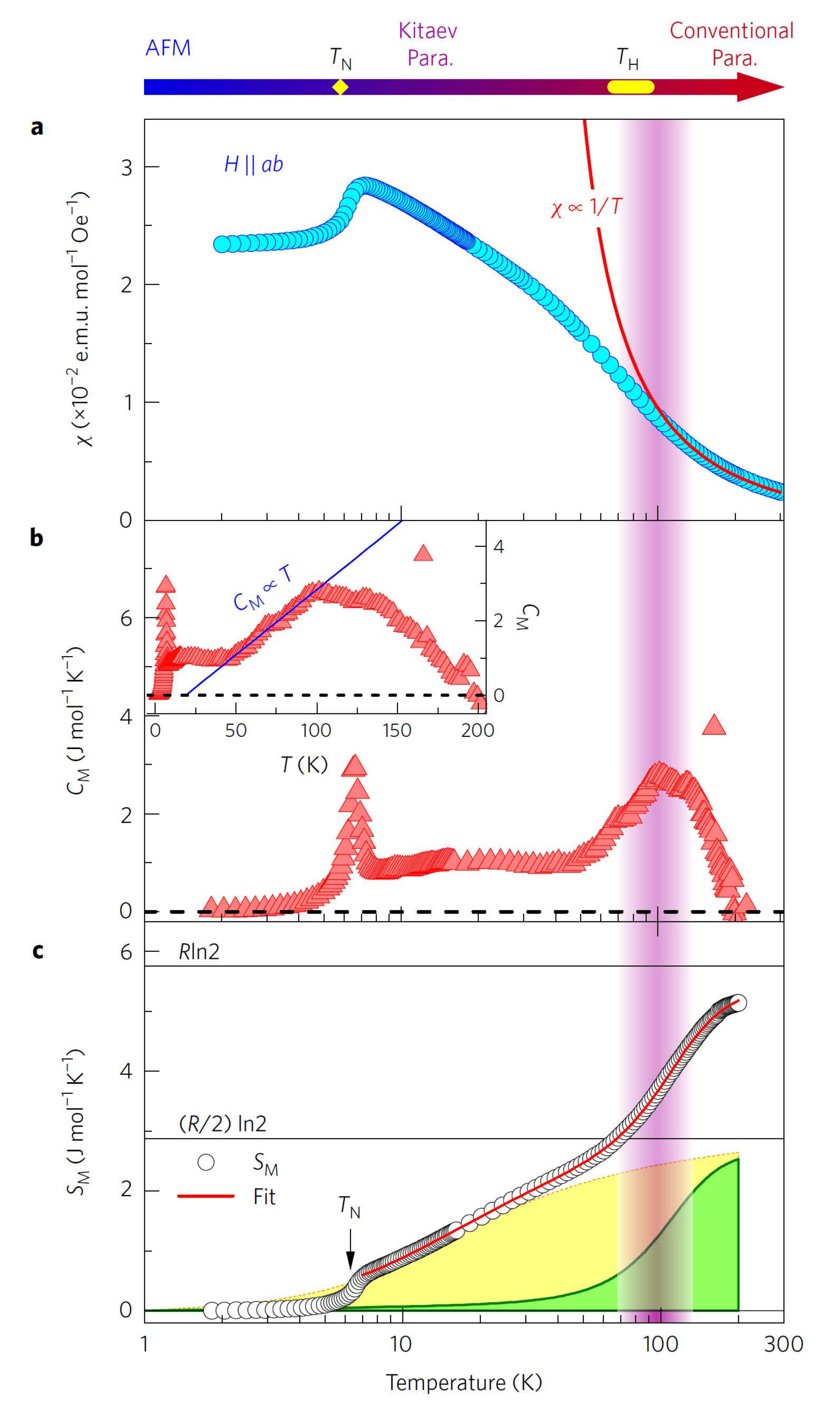} 
	\caption{(a) The $T$-dependence of static magnetic susceptibility of $\alpha$--RuCl$_3$ taken an applied field within the $ab$ plane. The Curie--Weiss law (red line) is fitted to the high-temperature data. (b) Magnetic specific heat (C$_M$) shows a kink at 6.5~K associated with an AFM transition. In addition, two broad bumps are observed, one extending from $T_N$ up to almost 100~K and another narrower one at 100~K correspond to the excitation of localized and itinerant Majorana fermions, respectively. The inset shows the linear dependence of  $C_M$ in the intermediate temperature region $50 \leq T \leq 100$~K, which is ascribed to a quasiparticle density of states akin to metals originating from itinerant Majorana fermions. (c) Magnetic entropy change versus $T$. The red line represents a fit taking into account both Majorana fermion contributions (green and yellow shadings). Adapted from~\cite{Do2017} with permission from NPG.}{\label{RUCl3_Specific_heat}}
\end{figure}
Namely, in $\alpha$--RuCl$_3$, magnetic specific heat $C_M$ shows two broad maxima, one at around $T_H \approx 100~K$, and another one extending from the antiferromagnetic transition temperature $T_N$ (Fig.~\ref{RUCl3_Specific_heat}) to roughly $T_H$. 
Interestingly, the $C_M$ data show a linear $T$ dependence in the intermediate region, $T_N < T < T_H$, in agreement with Kitaev QSL theory, which attributes this to an effective metallic-like density of states associated with itinerant Majorana fermions~\cite{PhysRevB.92.115122,PhysRevB.93.174425,Do2017}. 
The two-stage entropy release can be well reproduced by considering the contributions from localized low-$T$ gauge-flux excitations and itinerant high-$T$ Majorana fermion excitations~\cite{Do2017}.

In recent years, numerous approaches have been proposed to realize more ideal Kitaev spin-liquid candidates. 
One of these is via the modifications of inter-layer interactions, which can induce unwanted long-range ordering. 
In this respect, the observation of random-bond disorder within honeycomb layers in $\alpha$--Li$_{2}$IrO$_{3}$ led to discovery of H$_{3}$LiIr$_{2}$O$_{6}$ with a honeycomb lattice, where the inter-layer H$^{+}$ are positionally disordered~\cite{Kitagawa2018}. 
It was found that despite a strong antiferromagnetic interaction between Ir$^{4+}$ moments, this honeycomb magnetic material shows no evidence of long-range magnetic ordering down to at least 50~mK, consistent with a QSL ground state. 
Unlike the idealized, clean scenario considered by Kitaev, recent reports suggest that the presence of significant disorder in H$_{3}$LiIr$_{2}$O$_{6}$ appears to mimic a spin-liquid-like state associated with the bond-disordered Kitaev model~\cite{PhysRevLett.122.047202,PhysRevX.11.011034}.
In a similar spirit, the honeycomb lattice Ag$_{3}$LiIr$_{2}$O$_{6}$ was prepared by replacing Li atoms between the layers of $\alpha$--Li$_{2}$IrO$_{3}$ by Ag atoms to modify the inter-layer bonds~\cite{PhysRevLett.123.237203}. 
Contrary to H$_{3}$LiIr$_{2}$O$_{6}$, this material exhibits long-range antiferromagnetic order at low $T$~\cite{PhysRevB.104.115106,PhysRevB.103.094427,PhysRevB.103.214405} despite evidence of strong Kitaev interactions~\cite{PhysRevResearch.4.033025,PhysRevB.105.235125}.

\textcolor{black}{
Recently, 3$d$ Co$^{2+}$-based honeycomb lattices, such as Na$_{2}$Co$_{2}$TeO$_{6}$ and BaCo$_{2}$(AsO$_{4}$)$_{2}$, where $J_{\rm eff} = 1/2$ moments of magnetic ions originate from the interplay between CEF and spin--orbit coupling~\cite{PhysRevMaterials.6.064405,PhysRevB.97.014407,Motome_2020,PhysRevB.106.224416,PhysRevB.106.195136}, have also been proposed as Kitaev QSL candidates.
Despite the fact that these cobaltates display long-range magnetic order at low $T$, it has been found that a dominant Kitaev exchange interaction is needed to describe their observed magnetic excitation spectrum~\cite{PhysRevB.106.014413,PhysRevB.104.144408,Zhang2023}. 
Additionally, the suppression of magnetic order under an applied magnetic field suggests the possibility of a field-induced QSL state, similar to the one proposed for $\alpha$--RuCl$_{3}$~\cite{PhysRevB.106.224416,Lin2021,doi:10.1126/sciadv.aay6953}. 
However, a direct link to Kitaev physics has yet to be established in these cobaltate materials. 
Finally, apart from the $d$-electron based honeycomb lattices, Kitaev interactions can also be realized among spin--orbit entangled $J_{\rm eff} = 1/2$ moments in the rare-earth-based honeycomb lattice materials, such as SmI$_{3}$~\cite{TREBST20221,doi:10.1126/sciadv.abl5671}. 
}

\section{\textbf{Experimental signatures of QSLs}\label{expsig}}
\textcolor{black}{Understanding the microscopic origin of non-equilibrium quantum phenomena, novel quantum phases, electron correlation, emergent gauge fields, and fractional excitations in quantum materials is a pressing issue. 
For instance, to
detect and identify a quantum spin liquid, experiments must not only prove that long-range magnetic ordering is absent,
but also that signatures specific to QSLs are present~\cite{lacroix2011introduction}, such as quantum entanglement and unconventional fractional excitations.
These are non-local properties that, at the present stage of experiments, can not be measured directly.
Therefore, a single indirect experiment is usually insufficient to characterize the complex behavior of a QSL candidate material unambiguously.
Rather, a combination of state-of-the-art techniques is usually required~\cite{Wen2019}. 
Techniques such as muon spectroscopy ($\mu$SR), nuclear magnetic resonance (NMR), electron spin resonance (ESR), inelastic neutron scattering (INS) and thermal conductivity offer a comprehensive set of approaches for probing exotic quantum states and their associated quasiparticle excitations over wide range of time scales. 
Below, we present a brief overview of various complementary experimental techniques that can be used to study quantum spin liquids.
Several selected examples demonstrate the main advantages of each technique.}
	
\subsection{Muon spin spectroscopy ($\mu$SR)}	
Bulk magnetization measurement techniques are not sensitive enough to study a low percentage of defects that could otherwise lead to magnetic freezing or ordering, while specific heat is in many cases affected by unavoidable disorder and defect nuclear Schottky and phonon contributions. 
NMR measurements are not possible for certain frustrated magnetic materials where NMR-active nuclei are absent.
Unavailability of single crystals and requirements of large amounts of polycrystalline samples can pose strong constraints on the feasibility of neutron scattering experiments. 
In all of these cases, muon spin spectroscopy ($\mu$SR) stands out as a unique local-probe technique for exploring ground-state magnetic properties of frustrated spin systems~\cite{blundell2021muon, yaouanc2011muon, lacroix2011introduction}. 

$\mu$SR is highly sensitive to small internal fields, both static and dynamical, due to unique particle-physics properties of the muon production process via weak, parity-violating interactions, which results in muons being almost 100\text{\%} polarized when entering a sample. 
It is possible to implant muons (when at rest, their mean life-time is $2.2 \mu$s) in materials to detect very small magnetic moments, as low as $10^{-4} \mu_{\rm B}$, and to probe local field distributions in real space~\cite{yaouanc2011muon,Nuccio_2014,Hillier2022}. 
$\mu$SR offers good sensitivity over a wide frequency window of electron spin fluctuations in the material under study  (10$^{4}$ to 10$^{12}$ Hz).
The experimental observable is the time decay of average muon spin polarization due to the magnetic moments
present in the studied material interacting with the muon spin via local fields at the muon site. This enables one to study
the local environments of a muon in a sample, which provides information regarding its static and dynamic magnetic
properties.

When a muon, implanted in a material under study, decays a positron is emitted preferentially in the instantaneous
direction of the muon spin due to the parity-violating properties of the weak interaction through which muon decay
proceeds. The spatial distribution of emitted positrons therefore provides information on the average spin polarization of
muons just before their decay. The experimental set up consists of very sensitive positron detectors. The normalized muon
spin polarization function $P_{\alpha}(t)$, where $\alpha$ indicates the direction ($x$, $y$, or $z$) along which the polarization is measured, and which is proportional to the projection of the ensemble average of the muon spin along this direction, thus provides a probe of local magnetic fields at the muon stopping site. 
As muons are initially fully polarized along the beam direction, one does not require any external magnetic field to polarize them in a $\mu$SR experiment, in contrast to other
local-probe techniques like NMR and ESR.
This fact offers a unique handle to probe the static and dynamic response of electron spins without any external field. 
This unique advantage of $\mu$SR makes it very powerful in cases where even very small applied magnetic fields could perturb or destroy the intrinsic properties of a material.
$\mu$SR is also well established for its unique sensitivity in studying diluted magnetic materials, like metallic spin-glasses, where even a few percent of magnetic ions can have a dramatic effect on their overall properties~\cite{doi:10.1002/pssb.2221440131}.

\textcolor{black}{By measuring the number of decay positrons versus time using forward ($N_{F}(t)$) and backward ($N_{ B}(t)$) detectors, one obtains the muon asymmetry 
	\begin{equation}
	A\left(t\right)=\frac{N_F\left(t\right)-\alpha N_B(t)}{N_F(t)+\alpha N_B(t)}= A_0P_z\left(t\right)+A_{\rm bg} ,
	\label{asss}
	\end{equation}
which is proportional to the muon polarization function $P_{z}(t)$ along the direction of the initial muon polarization (longitudinal or $z$-direction). Here,
$\alpha$ is a parameter used to compensate for possibly different sensitivities of the forward
and backward positron detectors as well as the position of the sample relative to these detectors~\cite{yaouanc2011muon,doi:10.7566/JPSJ.86.024705,Nuccio_2014}. 
$A_{0}$ is the initial asymmetry and $A_{\rm bg}$ is the contribution of background due to muons stopping outside the sample.
By using sets of detectors in directions perpendicular to the initial muon polarization information can be obtained about the transversal components of the muon polarization $P_{x,y}(t)$ in a similar manner.} 

\textcolor{black}{
The time dependence of muon polarization is due to the local magnetic field at the muon site $B_{loc}$, exerting a magnetic torque on a muon's magnetic moment.
The local magnetic field sensed by a muon is a combination of dipolar and hyperfine fields, but in magnetic insulators the former usually dominate~\cite{yaouanc2011muon}.
In the case of a static local field at an angle $\theta$ with respect to the initial muon polarization, the muon polarization will precess around the field direction and the time dependence of its longitudinal component will be given by~\cite{yaouanc2011muon}
\begin{equation}{\label{avmu}}
	\centering
	P_{z}(t)=\cos^{2}\theta + \sin^{2}\theta \cos( \omega_{\mu}t),
	\end{equation}
where $\omega_{\mu}=\gamma_\mu B_{loc}$ is the Larmor frequency of muon precession and $\gamma_\mu$ is the gyromagnetic ratio of the muon.
In a polycrystalline sample, though, the angle $\theta$ will be uniformly distributed, yielding the average muon polarization 
\begin{equation}
	\label{pwd}
 \centering
	P_{z}(t)=\frac{1}{3}+\frac{2}{3}\cos(\omega_{\mu}t).
	\end{equation}
The appearance of coherent oscillations below a certain temperature suggests the onset of magnetic ordering in a material under study \citep{R_otier_1997,Blundell:1999zz}.
This is often used as an undeniable fingerprint of a long-range ordering transition in magnets, including frustrated ones like kagome antiferromagnets~\cite{keren1996muon, zorko2019YCu3muon, Tustain2020}, triangular antiferromagnets~\cite{PhysRevLett.97.167203, Khatua2021,somesh2021universal}, and Kitaev honeycomb candidates~\cite{lang2016unconventional,majumder2018breakdown,bahrami2021effect}.
Furthermore, the constant 1/3 in Eq.~(\ref{pwd}) gives rise to a so-called ``1/3"-tail of the signal that survives to long times and is absent if local fields are dynamical (see below). 
}

\textcolor{black}{If not all muons encounter the same local field vector (e.g., due to magnetic disorder or incommensurate magetic
	order), a distribution of local fields must be considered. Narrow distributions $\Delta \ll \gamma_\mu B_{loc}$, where $\Delta$ denotes the width of the field distribution, lead to a damping of the oscillations at late times~\cite{zorko2014frustration}.
For a broad isotropic Gaussian  (random) distribution of local fields, on the other hand, the longitudinal muon polarization adopts the characteristic Kubo--Toyabe form~\cite{yaouanc2011muon}
	\begin{align}
	P_{z}(t) =\frac{1}{3}+\frac{2}{3}(1-\Delta^{2}t^{2})\exp\left(-\frac{1}{2}\Delta^{2}t^{2}\right).
	\label{ku}
	\end{align}
This function is characterized by a single dip and the lack of coherent oscillations indicates the presence of frozen disordered. 
This might either be of nuclear origin~\cite{yaouanc2011muon} but can also originate from electronic fields, e.g., as typical for conventional spin-glass materials such as CuMn and AuFe~\cite{PhysRevB.31.546}.
Frustrated systems also sometimes exhibit such fast damping of oscillations due to random static internal fields.
Prominent examples include the distorted kagome lattice material Pr$_{3}$Ga$_{5}$SiO$_{14}$~\cite{PhysRevLett.104.057202} and the honeycomb spin-liquid candidate Li$_{2}$RhO$_{3}$~\cite{PhysRevB.96.094432}.
}
\begin{figure}[t]
 	\includegraphics[width= 7.9 cm, height= 6.5 cm]{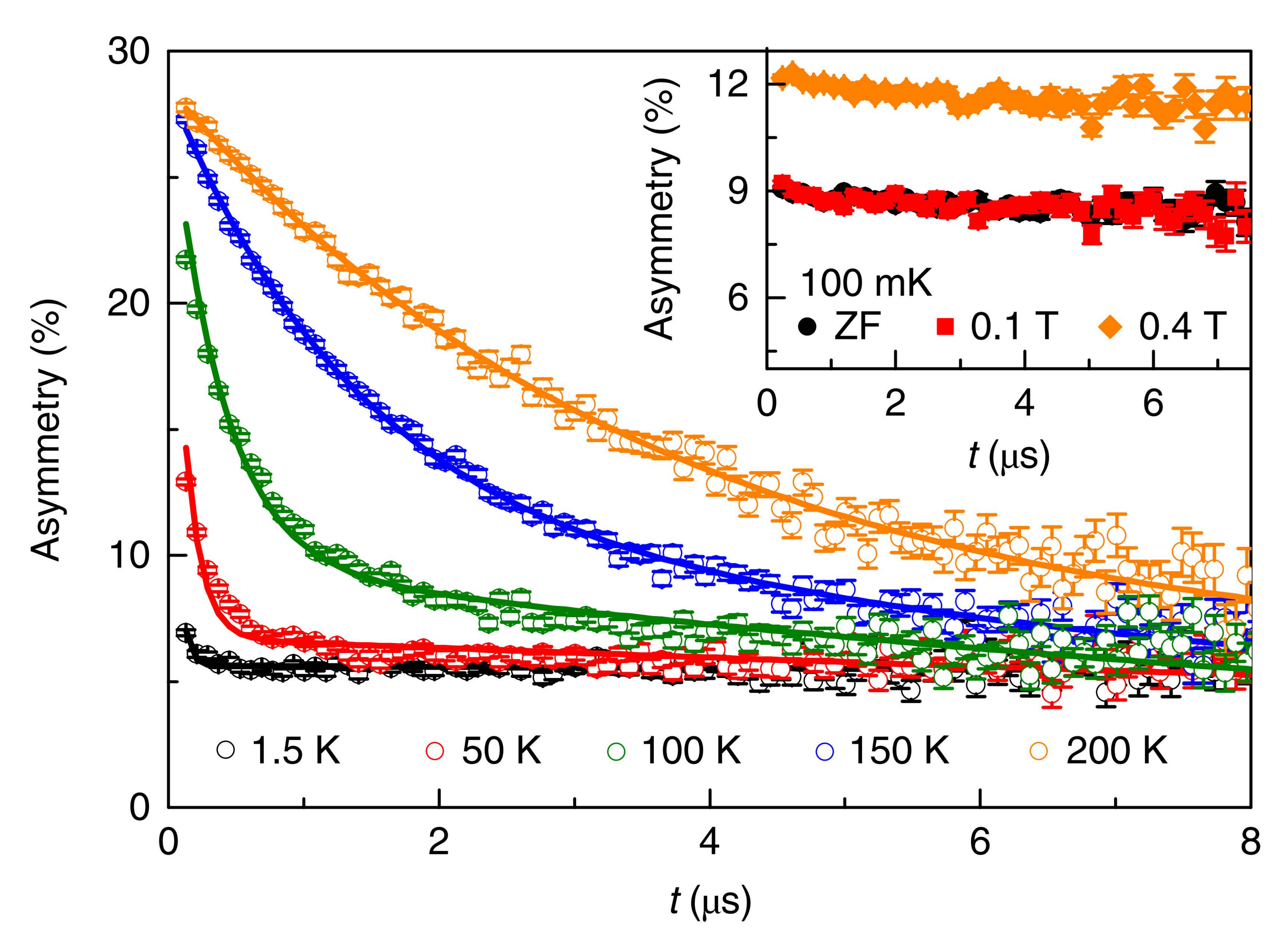} 
 	\caption{Time evolution of zero-field  muon spin asymmetry at various temperatures for the triangular lattice antiferromagnet TbInO$_{3}$, while the inset shows the 100~mK time dependent muon asymmetry in zero-field and under longitudinal fields. Adapted from~\cite{Clark2019} with permission from NPG. \label{fig:lamb}}
 \end{figure}

\textcolor{black}{
In sharp contrast to the cases of static local fields, where the muon polarization function exhibits oscillations or a single dip, in case of rapidly fluctuating local field ($\nu \gg \gamma_\mu B_{loc}$, where $\nu$ is the fluctuation frequency), muon polarization becomes a monotonically decaying function~\cite{yaouanc2011muon}
\begin{equation}
	P_{z}(t) = \exp\left[-(\lambda t)^{\beta}\right],   
	\end{equation}
where $\lambda$ denotes the average longitudinal muon spin relaxation rate and $\beta$ is the stretching exponent, which reflects the width of the distribution of muon spin relaxation rates \citep{PhysRevB.31.546,PhysRevB.20.850,R_otier_1997}.
Although a single local-field correlation time $\tau=1/\nu$ leads to single-exponential decay ($\beta = 1$), the stretched exponential behavior of the relaxation function is often observed experimentally in QSL candidates~\cite{PhysRevLett.117.097201,Gao2019,Arh2021}.
In the case of fast fluctuations of local fields ($\nu \gg \gamma_{\mu}B_{\rm loc}$), i.e., in the so-called motional narrowing regime, the muon spin relaxation rate can be related to the fluctuation rate of local magnetic fields via the Redfield relation~\cite{yaouanc2011muon,slichter2013principles}
 \begin{equation}
 \lambda=\frac{2\gamma_\mu^{2}B_{\rm loc}^{2}\nu}{\nu^{2}+(\gamma_{\mu}B_{L})^{2}}.
 \label{eq:mlamd}
 \end{equation}
The applied longitudinal magnetic field $B_{L}$ allows one to probe the local-field fluctuation density at a corresponding angular frequency $\gamma_{\mu}B_{L}$.
If local fields originate from electron spins, the muon spin relaxation rate $\lambda$ thus provides information concerning spin dynamics.
The $T$-independent behavior of $\lambda$, commonly referred to as a relaxation plateau due to persistent spin dynamics, is often observed in frustrated magnets at low $T$ and is  considered a characteristic
signature of a QSL ground state~\cite{PhysRevLett.104.057202,gomilsek2016muSR,PhysRevLett.117.097201,PhysRevB.100.241116,Clark2019,PhysRevLett.125.117206,PhysRevResearch.2.023191,PhysRevB.106.085115}. 
}

\begin{figure}[t]
	\includegraphics[width=6.5 cm]{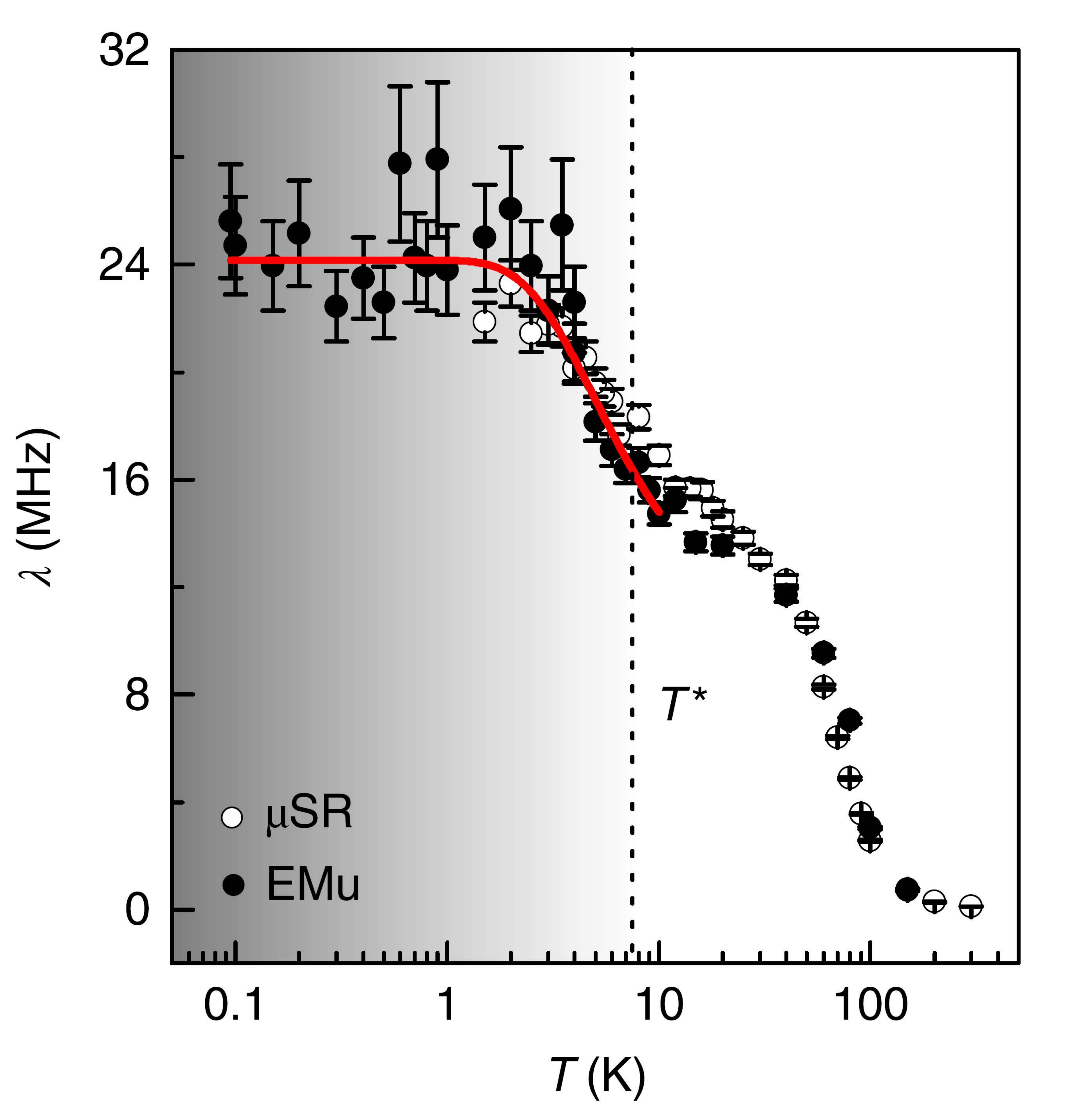} 
	\caption{The temperature evolution of muon relaxation rate $\lambda$ in zero field. The labels $\mu$SR and EMu  indicate the two spectrometers at the ISIS
		facility, Rutherford Appleton Laboratory in the UK that were used to collect the data. The red solid line is a fit to $\lambda(t) = \lambda_{0}/(1+A \exp(-T/T^{*}))$, which was used to obtain the relevant energy scale ($T^{*}$) namely  a thermally activated spin gap. Adapted from~\cite{Clark2019} with permission from NPG. \label{fig:zero}}
\end{figure}

\subsubsection{Persistent spin dynamics in TbInO$_3$\label{TIO}}
In Fig.~\ref{fig:lamb}, $\mu$SR asymmetry is shown for the rare-earth-based triangular-honeycomb antiferromagnet antiferromagnet TbInO$_{3}$, where 4\textit{f} magnetic ions Tb$^{3+}$ reside on a distorted triangular lattice, which effectivelly undergoes a transition to a honeycomb lattice due to crystal-electric-field effects~\cite{Clark2019}. 
Magnetic susceptibility data of this compound suggests the absence of long-range order and the CW temperature $\theta_{\rm CW} \approx -1.13$~K indicates very weak antiferromagnetic interactions between spins, as expected for rare-earth systems~\cite{PhysRevBtb,Clark2019}. 
As can be seen in Fig.~\ref{fig:lamb}, $\mu$SR spectra do not exhibit any oscillation, which would
correspond to muon spins precessing in a static internal field. 
Furthermore, the ``1/3"-tail that should be seen for a static but disordered ground-state is also absent, down to at least 100~mK. 
This rules out the presence of any static electronic magnetism at low $T$. 
Inset of the Fig.~\ref{fig:lamb} shows no difference in asymmetry between zero-field and applied field (0.1~T) data, which is also consistent with the absence of a static local field at a temperature of 100~mK.  
In order to describe a dynamic ground-state, Clark \textit{et al}. considered exponential relaxation of the signal and obtained the $T$-dependence of the muon spin relaxation rate $\lambda$ shown in Fig.~\ref{fig:zero}.
It was observed that $\lambda$ increases as $T$ decreases and exhibits a pleteau below 2~K~\cite{Clark2019}. 
The increase of $\lambda$ upon lowering $T$ was attributed to a slowing down of spin dynamics in the sample, which was due to the emergence of short-range electron spin correlations. 
The saturation of relaxation rate at very low temperature implies persistent spin dynamics which is a  characteristic feature of a QSL.

\begin{figure}[b]
	\includegraphics[width=1\linewidth]{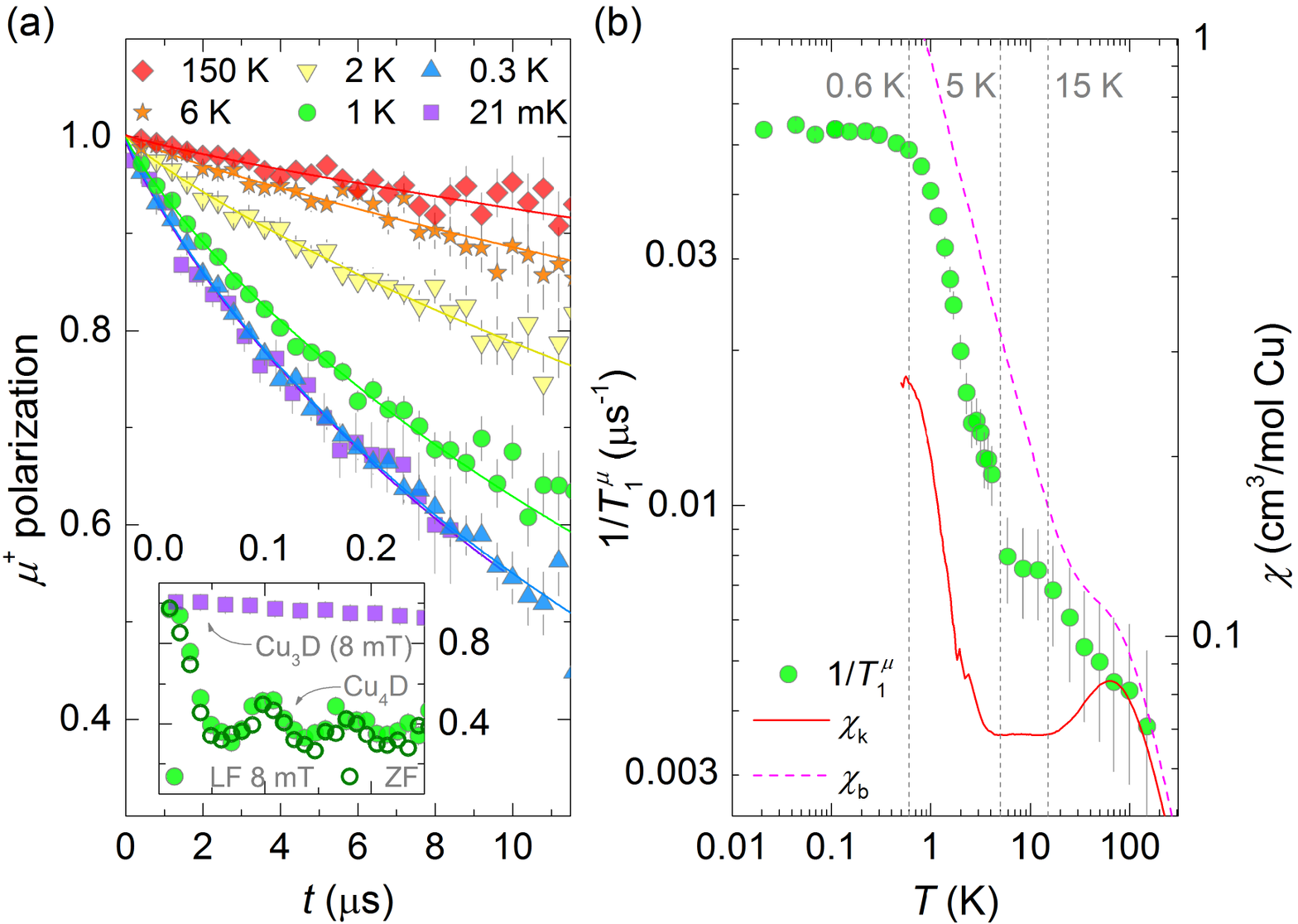} 
	\caption{(a) The monotonic decay of the muon polarization in Zn-brochantite, ZnCu$_3$(OD)$_6$SO$_4$, at various temperatures. The undoped brochantite, Cu$_4$(OD)$_6$SO$_4$, on the other hand shows clead oscillations (inset).  (b) The temperature dependence of the corresponding muon spin relaxation time and magnetic susceptibility.
		Adapted from~\cite{gomilsek2016instabilities} with permission from APS. }
	{\label{broch}}
\end{figure}

\subsubsection{Gapless QSL and spinon Kondo effect in Zn-brochantite}
\textcolor{black}{$\mu$SR was also used to determine the ground state properties of the kagome lattice antiferromagnet Zn-brochantite, ZnCu$_3$(OH)$_6$SO$_4$, featuring a QSL ground-state~\cite{li2014gapless,gomilsek2016instabilities}.
Like in TbInO$_{3}$, muon polarization curves were found to decay monotonically down to the lowest experimental temperature of 21~mK (Fig.~\ref{broch}(a)), compared to much larger average inter-kagome exchange interaction of 65~K~\cite{gomilsek2016instabilities}.
This is in sharp contrast to undoped brochantite, Cu$_4$(OH)$_6$SO$_4$, which orders magnetically and therefore exhibits clear oscillations in the muon polarization (inset in Fig.~\ref{broch}(a)).
A low-$T$ relaxation plateau was also observed in Zn-brochantite (Fig.~\ref{broch}(b)).}
\begin{figure}[t]
	\includegraphics[width=0.95\linewidth]{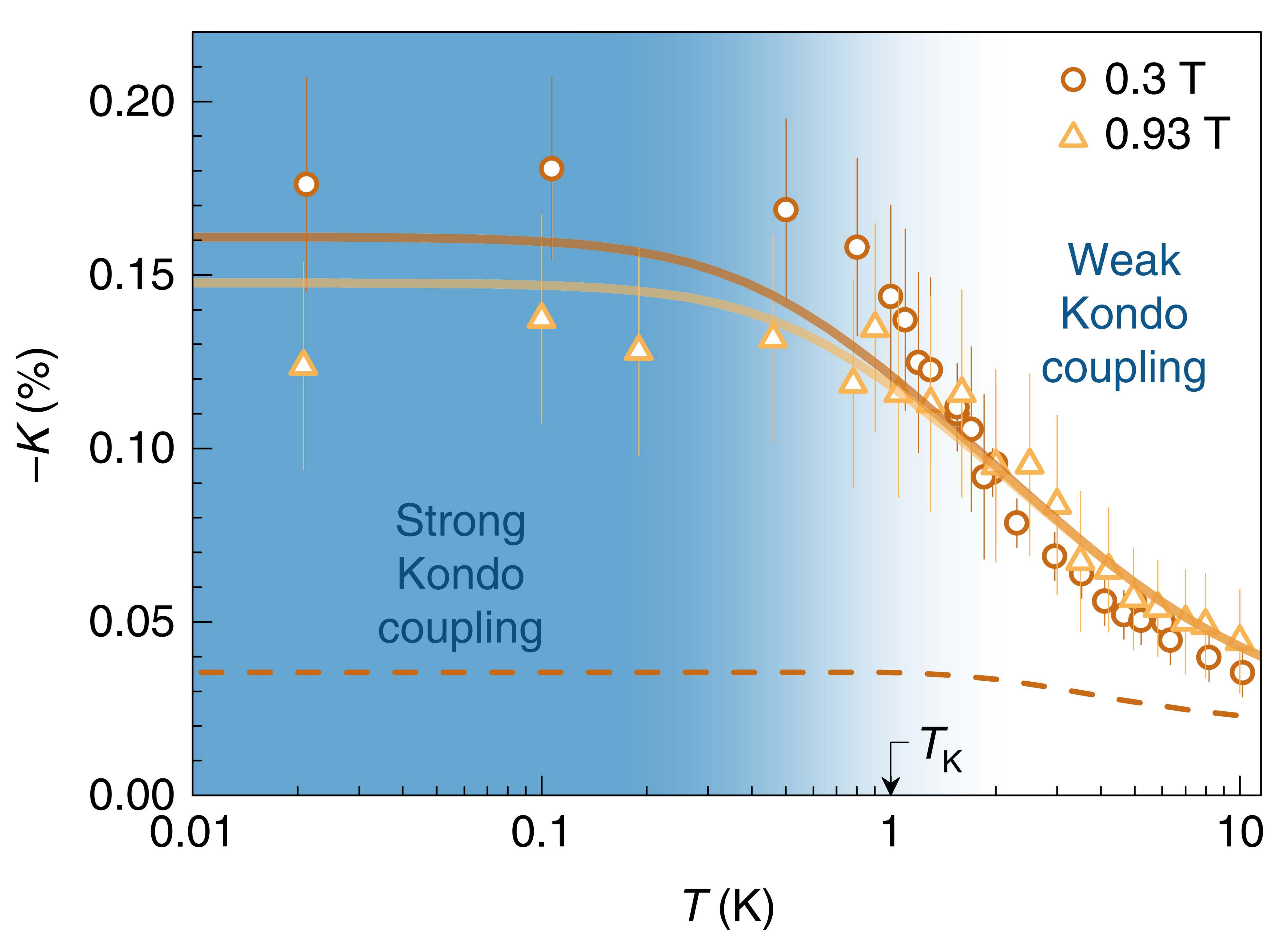} 
	\caption{The temperature dependence of the muon frequency shift $K$ in Zn-brochantite, ZnCu$_3$(OH)$_6$SO$_4$. The solid lines correspond to theoretical predictions for  Kondo screening of impurity spins with the Kondo temperature $T_K = 1.0$~K, while the dashed line is the subdominant additional, direct contributions to the total shift from intrinsic kagome spins.
		Adapted from~\cite{Gomilsek2019} with permission from NPG. }
	{\label{kondo}}
\end{figure}
\begin{figure*}[t]
	\includegraphics[width=1\linewidth]{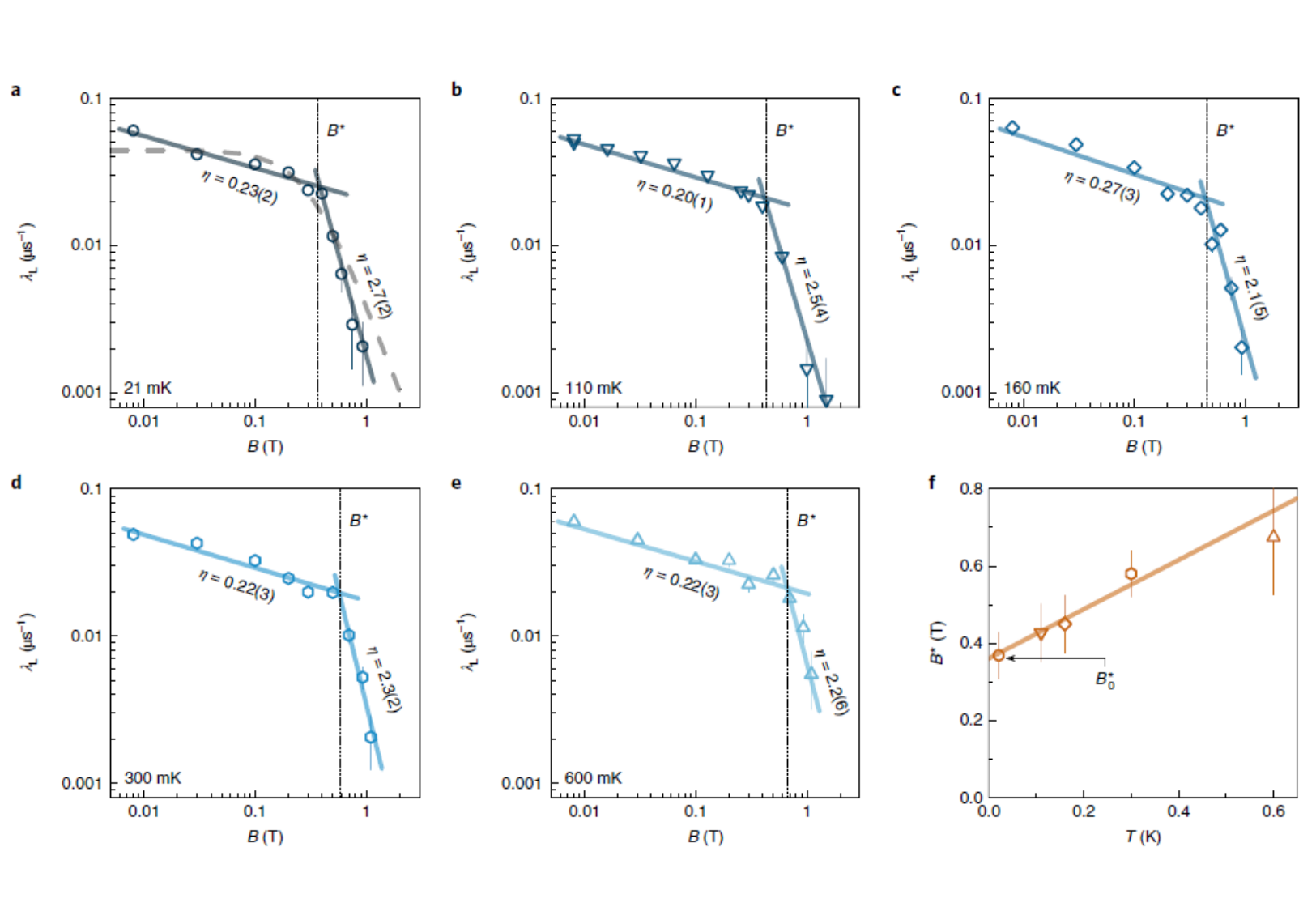} 
	\caption{(a--e) The field dependence of the longitudinal muon spin relaxation rate at different temperatures and (f) the corresponding temperature dependence of the critical field $B^*$, where the field dependences exhibit anomalous downturn due to the Kondo-ressonance splitting.
	Adapted from~\cite{Gomilsek2019} with permission from NPG. }
	{\label{kondorelax}}
\end{figure*}
\textcolor{black}{Like the iconic herbertsmithite~\cite{RevModPhys.88.041002} (see Subsection \ref{herbNMR}), Zn-brochantite is also hampered by sizable Cu/Zn site mixing, which induces defects in the spin lattice~\cite{li2014gapless}.
Ionic substitutions are a common problem of frustrated magnets, because when}
defects are present they usually dominate the low-temperature magnetic behavior of a QSL material, as they often behave as quasi-free spins.
Therefore, defects also determine the main decay of muon polarization, as muons usually couple to surrounding magnetic moments in electrical insulators via long-range dipolar interactions~\cite{yaouanc2011muon}.
However, When defects are coupled to host spins they can therefore act as effective in-situ probes of the magnetic state of the host~\cite{alloul2009defects}.
\textcolor{black}{Defects represent a local perturbation that induce a local response in the otherwise macroscopic QSL state. 
It is precisely this response that can shed light on the intrinsic properties of the host QSL state~\cite{PhysRevB.74.165114,florens2006kondo,kim2008kondo,ribeiro2011magnetic,vojta2016kondo,Gomilsek2019,chen2022evidence}.}

Using $\mu$SR, defects were employed as local-probes of the host material's electronic state in Zn-brochantite~\cite{gomilsek2016muSR,Gomilsek2019}.
The muon frequency shift $K=(\nu-\nu_0)/\nu_0$ was first used, where $\nu$ is the precession frequency in applied transverse magnetic field $B$ and $\nu_0=\gamma_\mu B$ is the reference precession frequency. 
$K$ is proportional to local defect susceptibility $\chi$ via $K=A\chi$, where $A$ is the magnetic coupling constant between the muon end electron spins. 
Measurements in Zn-brochantite showed that $K$ saturates below approximately 0.6~K (Fig.~\ref{kondo}).
This reveals that the QSL ground-state must be gapless and features a spinon Fermi surface, a state for which a constant defect susceptibility is predicted at zero temperature, in contrast to gapped QSL characterized by a gapped defect susceptibility, or gapless Dirac spin liquids which are predicted to have a divergent defect susceptibility at the lowest temperatures~\cite{gomilsek2016muSR}.
Such a state is also compatible with a linear term in the magnetic specific heat observed in this material at the lowest temperatures~\cite{li2014gapless}.
It should be emphasized that the saturation of the defect susceptibility in Zn-brochantite is not due to simple polarization effects in a strong magnetic field at low temperatures.
Namely, the magnitude of the saturated moment is only $0.13$--$0.27 \mu_{\rm B}$ (i.e., much less than
the full moment $\sim 1 \mu_{B}$) and the measured muon frequency shift, which is proportional to the defect magnetic moment divided by the applied field, does not scale with $1/B$ (Fig.~\ref{kondo}), as would be expected for a free saturated magnetic defect.
As also the defect-defect interactions were demonstrated to be incapable of explaining the experimental findings, Kondo
screening of the magnetic moments of defects by emergent itinerant fermionic spinon excitations of the kagome QSL
spins was suggested~\cite{Gomilsek2019}. While a reasonable fit to the experimental Knight-shift field-dependence was obtained already
with non-interacting spinons, the optimal fit indicated the possible presence of strong emergent gauge fields coupled to
spinons, which renormalized their effective g-factors. We note that such detailed information on the QSL state of this
material would be almost impossible to obtain via more direct measurements of its intrinsic magnetic state.

The spinon Kondo screening scenario was further experimentally confirmed by another fingerprint of the Kondo
effect; namely, Kondo resonance splitting in a finite applied field, which was detected via muon spin relaxation
measurements~\cite{Gomilsek2019}.
\textcolor{black}{In the QSL ground state the muon relaxation rate $\lambda$ was found to decrease with increasing strength
	of the longitudinal applied field, following a power law $\lambda \propto B_L^{-p}$ with a very small power $p\sim 0.2-0.3$ at low $B_{L}$ (Fig.~\ref{kondorelax}).
This sub-linear field dependence is unconventional and not in line with the usual exponential decay of the local-field
autocorrelation function that yields the Redfield relation with $p=2$ (Eq. (\ref{eq:mlamd})).
It rather suggests that the temporal correlations decay algebraically as $t^{-(1-p)}$~\cite{PhysRevLett.96.247203,gomilsek2016muSR},
as often observed in frustrated magnets~\cite{PhysRevLett.92.107204,PhysRevB.84.100401,PhysRevB.90.205103}.
However, at a temperature dependent critical field $B^*$ the power $p$ was found to change to a much higher value (Fig.~\ref{kondorelax}) that is in line with the Redfield relation.
This sudden change of behavior and the temperature dependence of $B^*$ (Fig.~\ref{kondorelax}(f)) could be explained within the Kondo picture as a manifestation of the Kondo-resonance splitting, which causes a dramatic decrease of the local density of states that are responsible for muon spin relaxation when the stength
of the Zeeman interaction overcomes the strength of the Kondo coupling.
}
Zn-brochantite thus appears to be the first
known insulator with so-called spinon Kondo screening, where the traditional role of itinerant electrons in screening
the localized magnetic impurity moments in metals is taken over by emergent itinerant (but charge-neutral) spinon
excitations arising from the electrically-insulating QSL ground state of this material.
\textcolor{black}{
Recently, experimental observation
of a characteristic Kondo resonance peak was reported also for defect magnetic ions in another gapless QSL candidate,
single-layered  1T-TaSe$_2$~\cite{Ruan2021, chen2022evidence}.
}Such non-perturbative Kondo coupling of defects to the host QSL state might, in fact,
be a generic feature of QSLs and could shed light on the ground-state properties of many other candidate QSL materials,
if it could be appropriately disentangled from other effects, like in the cases mentioned above.

\subsection{Nuclear magnetic resonance (NMR)\label{VB}}
Unavoidable disorder and defects in polycrystalline samples pose a strong constraint on revealing the true groundstate properties of frustrated magnets. Being a site-selective local-probe technique, nuclear magnetic resonance (NMR)
offers a powerful way of extracting the intrinsic magnetic susceptibility of the pure magnetic lattice and exploring its
spin dynamics without ambiguity. As a spectroscopic technique, NMR is usually able to single out any extrinsic defects
present in the material under study, in contrast to bulk magnetic susceptibility which sums both intrinsic and extrinsic
contributions equally. In NMR, a nuclear spin, whose levels are Zeeman split by an external magnetic field, is perturbed
by applying a radio frequency pulse and its spin polarization measured. In magnetic materials, electrons are coupled
to nuclear spins via hyperfine interactions with the corresponding Hamiltonian being~\cite{BERTHIER2017331,slichter2013principles,Abragam,KHUNTIA2019165435}
\begin{align}
	\mathcal{H} &= \frac{8\pi}{3}g \mu_{\rm B}\gamma_{n}\hbar \mathbf{I} \cdot \mathbf{S}(\mathbf{r})\delta(\mathbf{r})-g\mu_{\rm B}\gamma_{n}\mathbf{I} \cdot \left[\frac{\mathbf{S}}{r^{3}}-\frac{3\mathbf{r}(\mathbf{S} \cdot \mathbf{r})}{r^{5}}\right]\nonumber\\
	&- g\mu_{B}\gamma_{n} \hbar \left[\mathbf{I}\cdot\frac{\vec{L}}{r^{3}}\right],
	\label{hyper}
\end{align}
where $\mathbf{S}$ and $\mathbf{I}$ are the electron and the nuclear spin, respectively and \textbf{r}  is a vector connecting them, while $\mu_{\rm B}$ and $\gamma_n$ denote the Bohr magneton and the gyromagnetic ratio of the probing nucleus, respectively. \textcolor{black}{The first term in Eq.~(\ref{hyper})
is the Fermi contact interaction, accounting for situations when there is a finite probability of electron being at the
position of the nucleus.
Due to its isotropic form, the contact hyperfine interaction leads to finite
NMR shifts in magnetic materials even in polycrystalline samples. 
The second term
in Eq.~(\ref{hyper}) represents the dipolar interaction between the nuclear spin and the electron spin, which is usually responsible for broadening of the NMR spectra. The last term accounts for the interaction of the nuclear spin with the orbital momentum of electron in the studied material and leads to a $T$-independent NMR shift.}\\
In general, symmetry breaking phase transitions in magnetic materials, such as magnetic ordering or freezing, usually result in rectangularly-shaped NMR spectra in powder samples and/or a splitting of NMR spectra in single crystals.
A lack of such features, on the other hand, suggest the absence of such phase transitions, even in cases when the spin dynamics changes profoundly~\cite{itou2010instability,gomilsek2017field}. 
The intrinsic susceptibility from the spectra  can be calculated from the NMR shift, which is defined by
\begin{equation}
	\begin{split}
		K & = K_{\mathrm{chem}}+\frac{A_{\rm hf}}{N_{\rm A}\mu_{\rm B}}\chi(T) =  K_\mathrm{chem} + K_\mathrm{spin},
	\end{split}
\label{knifor}
\end{equation}
where $K_\mathrm{chem}$ is the temperature independent chemical shift due to the local chemical environment (local electronic orbitals), $A_{\rm hf}$ is hyperfine coupling constant characterizing the interaction between electron spins and the probing nucleus, and $\chi(T)$ is the temperature-dependent local magnetic susceptibility of electron spins.\\
Another important  observable probed by  NMR is the spin--lattice relaxation rate $1/T_{1}$. 
After the perturbation of the magnetization of nuclei by a radio frequency pulse, 
their magnetization returns to its thermal equilibrium with a characteristic time $T_1$, which is known as the NMR spin--lattice relaxation time.
 $T_{1}$ is determined by the interactions
of nuclear spins with their surroundings, including electron spins, which cause a loss of nuclear polarization over time.
The dynamical observable 1/$T_{1}$ for a magnetic field applied along the $z$ direction can be explicitly expressed in terms of
the electron spin–spin correlation function as~\cite{BERTHIER2017331,PhysRevLett.79.925}
\begin{equation}
\begin{split}   
	\frac{1}{T_1} &= \frac{1}{2}\gamma_n^2\sum_{\mathbf{q}}\sum_{\alpha=x,y,z}[A_{x\alpha}^2(\mathbf{q}) \\
    &+ A_{y\alpha}^2(\mathbf{q})] \left(\int_{- \infty}^{+\infty} \langle S_{\alpha}(\mathbf{q},t)S_{\alpha}(-\mathbf{q},0)\rangle e^{i\omega_nt}\dd t \right),
\end{split}
\label{NMRrela}
\end{equation}
\textcolor{black}{ where $A$ is the hyperfine coupling tensor, $\omega_n$ is NMR resonance frequency, and $\langle S_{\alpha}(\mathbf{q},t)S_{\alpha}(-\mathbf{q},0)\rangle$ is the spin--spin correlation function, where $S_{\alpha}(q,t)$ is the electron spin operator along $\alpha$ direction at time $t$ in \textbf{q}-space.
According to Eq.~(\ref{NMRrela}), in an NMR spin--lattice relaxation experiment, electron spin correlations (fluctuations) are effectively summed over all wave vectors as expected for a local-probe technique.
}

\subsubsection{Spin gap in herbertsmithite\label{herbNMR}}

\textcolor{black}{We start the NMR experimental examples with the paradigmatic quantum kagome antiferromagnet herbertsmithite~\cite{mendels2010quantum, mendels2016quantum, RevModPhys.88.041002}, ZnCu$_3$(OH)$_6$Cl$_2$, where various NMR experiments that have been performed over the years led to many fundamental findings~\cite{PhysRevLett.100.077203, PhysRevLett.100.087202, PhysRevLett.107.237201, Fu655, Khuntia2020, Wang2021}, but also to contradictory conclusions stemming from the complexity of the
	material.
In herbertsmithite, Cu$^{2+}$ ions on kagome planes are strongly exchange-coupled with one another via Cu–
O–Cu bridges. 
Although in a perfectly stoichiometric compound a single crystallographic site would exist, defects are
always present in real material samples due to unavoidable Cu/Zn site mixing, with some Cu$^{2+}$ ions occupying interlayer
Zn$^{2+}$ sites. However, these interlayer Cu$^{2+}$ defects are less strongly coupled to the kagome plane via oxygen atoms
than intra-layer Cu$^{2+}$ ions. NMR, being a spectroscopic technique, can effectively separate between the defect and
intrinsic contributions to local susceptibility in herbertsmithite by focusing on particular spectroscopic lines in the NMR
spectrum~\cite{Fu655, Khuntia2020}.}
\begin{figure}[t]
	\centering
	\includegraphics[width=\linewidth]{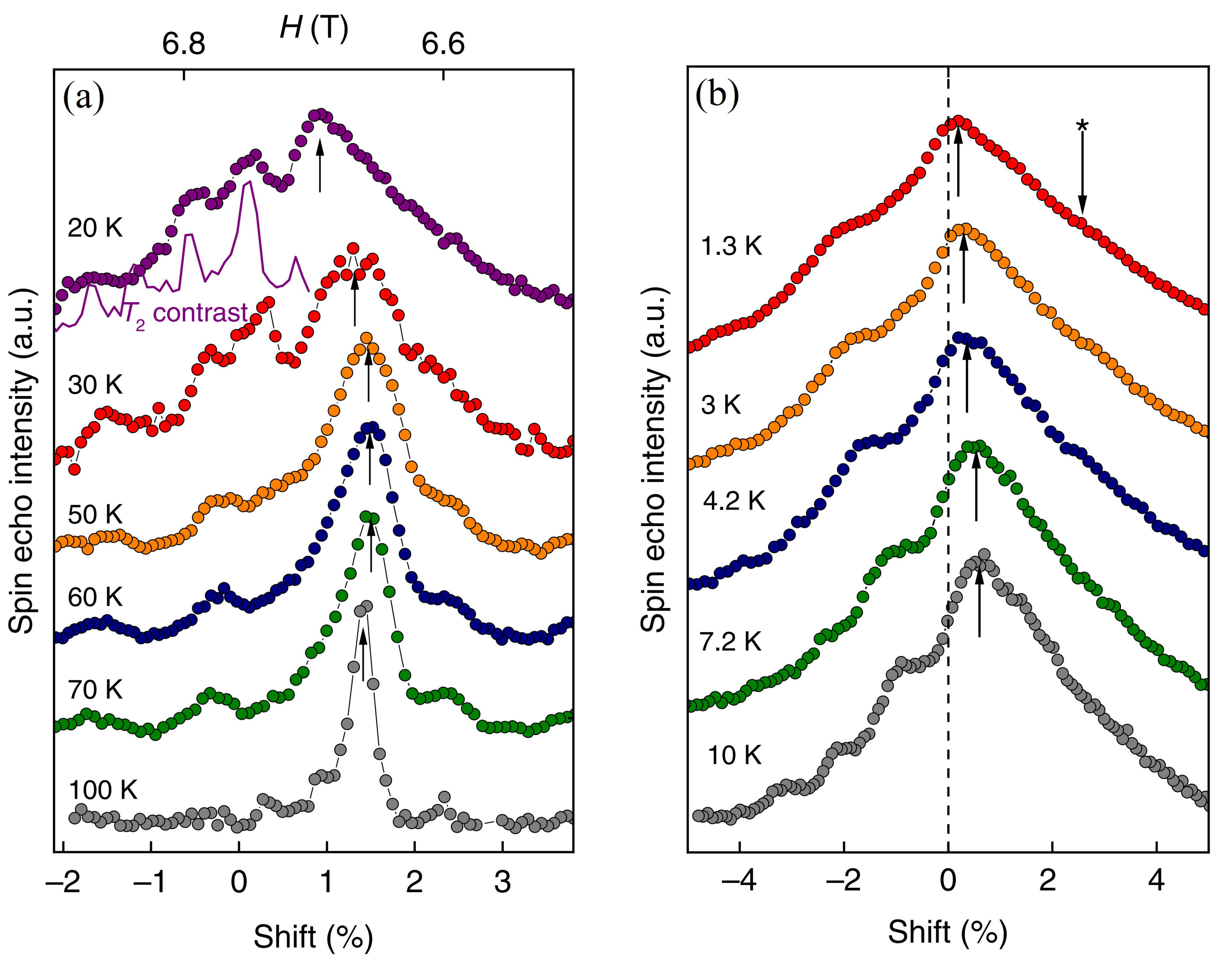} 
	\caption{ (a) The temperature evolution of the $^{17}$O NMR spectra of the quantum kagome antiferromagnet herbersmithite, ZnCu$_3$(OH)$_6$Cl$_2$. Adapted from~\cite{Khuntia2020} with permission from NPG.}{\label{Herberth}}
\end{figure}
\begin{figure*}[t]
	\centering
	\includegraphics[width=1\linewidth]{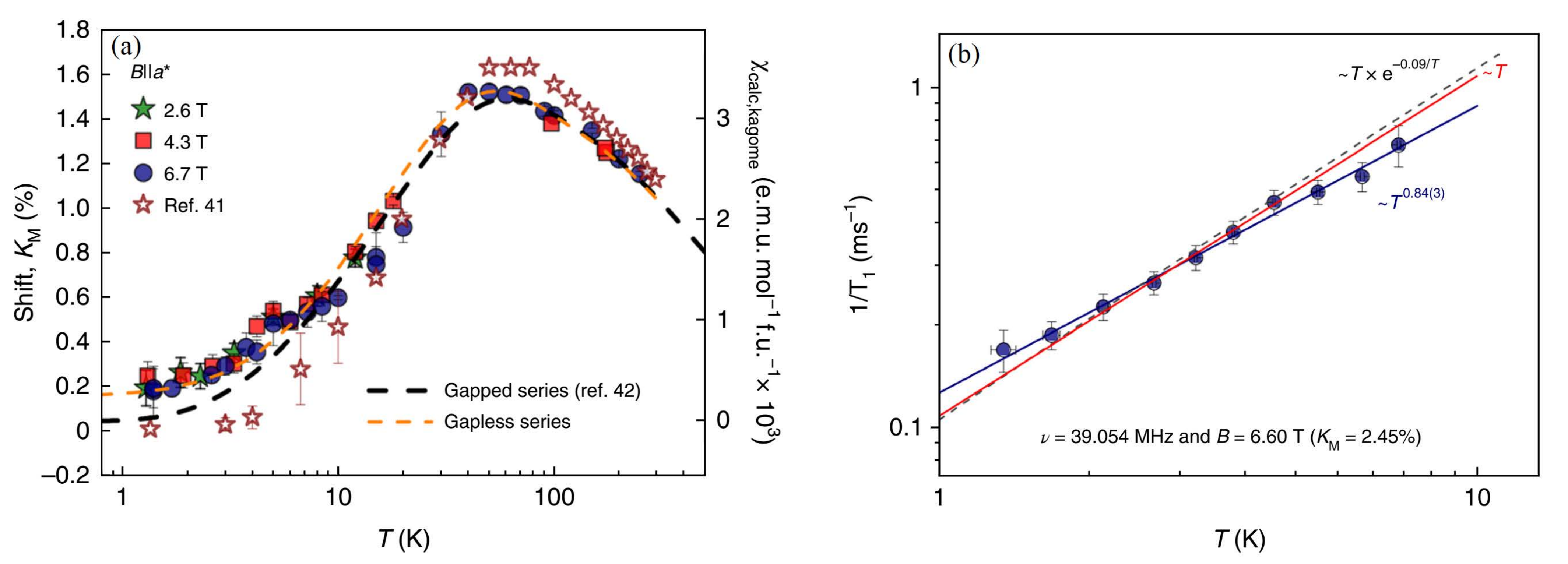} 
	\caption{(a) The NMR frequency shifts $K$ is plotted versus temperature for different applied magnetic fields. 
 Quasi-linear behavior and a finite value of susceptibility at zero temperature indicates gapless excitations. Previous data marked as ref.~41 is from Ref.~\cite{Fu655} and defers from other datasets due to different data analysis procedure.
 (b)  The spin--lattice relaxation rate $1/T_{1}$ follows a power law, which also indicates gapless excitations from the ground state. 
 Adapted from~\cite{Khuntia2020} with permission from NPG.}{\label{Herberth_kn}}
\end{figure*}

\textcolor{black}{Fig.~\ref{Herberth}(a) depicts the field-swept $^{17}$O NMR spectra on single crystal of ZnCu$_3$(OH)$_6$Cl$_2$ at $T\geq$ 20~K.
These spectra include the contributions of two oxygen sites, one of which is a defect site (O$_{2}$) and the other is a defect-free site (O$_{1}$). 
At 100~K, the spectrum is dominated by the O$_{1}$ site, which gives the main line, as indicated by arrows. 
Once $T$ is lowered, another set of well-resolved lines associated with defects becomes apparent at low frequencies, as shown by the purple line in the 20~K spectrum. 
These lines are nicely resolved in a so called $T_{2}$-contrast experiment, taking advantage of the fact that defect lines have a longer spin--spin relaxation time compared to the main line~\cite{Khuntia2020}.  
In order to determine the intrinsic kagome spin susceptibility from the NMR shift of the main line unambiguously, it is necessary to subtract the defect contribution from the NMR spectra.}

The shift of the  main line consists of two terms, $K_{\rm spin}$ and $K_{\rm chem}$ (Eq.~\ref{knifor}). 
The latter is constant while the variation of $K_{\rm spin}$ gives information about the intrinsic spin susceptibility.
This was obtained via contrast experiments on high-quality single crystals for the external magnetic fields applied along the reciprocal axis $a^{*}$, for which the spectra were not affected by the minute amount of anti-site disorder present in the sample.  
As depicted in Fig.~\ref{Herberth},  $^{17}$O NMR spectra do not show any signature of  magnetic ordering and yield intrinsic susceptibility that remains finite down to 1.3~K, which is two orders of magnitude smaller than the main exchange interaction between spins on the kagome lattice. 
The observation of finite NMR shift at low temperature suggested a gapless QSL, in agreement with existing theoretical conjectures~\cite{PhysRevLett.98.117205}. \textcolor{black}{
However, the NMR shift from a preceding NMR study (marked as Ref.~41 in Fig.~\ref{Herberth_kn}(a)) rather suggested zero shift and therefore a gapped QSL state~\cite{Fu655}, however this conclusion was obtained via a less refined analysis, not employing the $T_2$-contrast experiments.}
 
\textcolor{black}{The NMR spin--lattice relaxation rate $1/T_{1}$ can yield additional information on the potential presence gap in the QSL
	ground state of herbertsmithite.}
It essentially probes the spectral density  of the low-energy spin excitations at the NMR resonant frequency.
At low $T$, frustrated quantum materials with a spin gap in the excitation spectrum display thermally activated behavior, $1/T_{1}\sim \mathrm{exp}(-\Delta/(k_\mathrm{B} T))$, while a gapless spectrum  generally results in a power-law dependence of the spin--lattice relaxation rate, $1/T_{1}\sim T^n$.
In herbertsmithite, the nuclear spin--lattice relaxation rate exhibits the latter behavior, $1/T_{1} \sim T^{0.84}$~\cite{Khuntia2020}, (see Fig.~\ref{Herberth_kn}(b)), which is in agreement with theoretical predictions for a Dirac $\mathrm{U}(1)$ QSL state~\cite{PhysRevLett.98.117205,PhysRevX.7.031020,PhysRevLett.118.137202}.
\textcolor{black}{However, it has recently been found that the QSL state of herbertsmithite is highly inhomogeneous due to disorder and spatially varying excitation gaps have rather been suggested~\cite{Wang2021}.
Therefore, the debate about  whether the QSL ground state of herbertsmithite is gapped or gapless continues to this date~\cite{PhysRevX.12.011014}.}

\subsubsection{Temperature dependent spin correlations on the kagome lattice}
\begin{figure*}[t]
 	\includegraphics[width=0.8\linewidth]{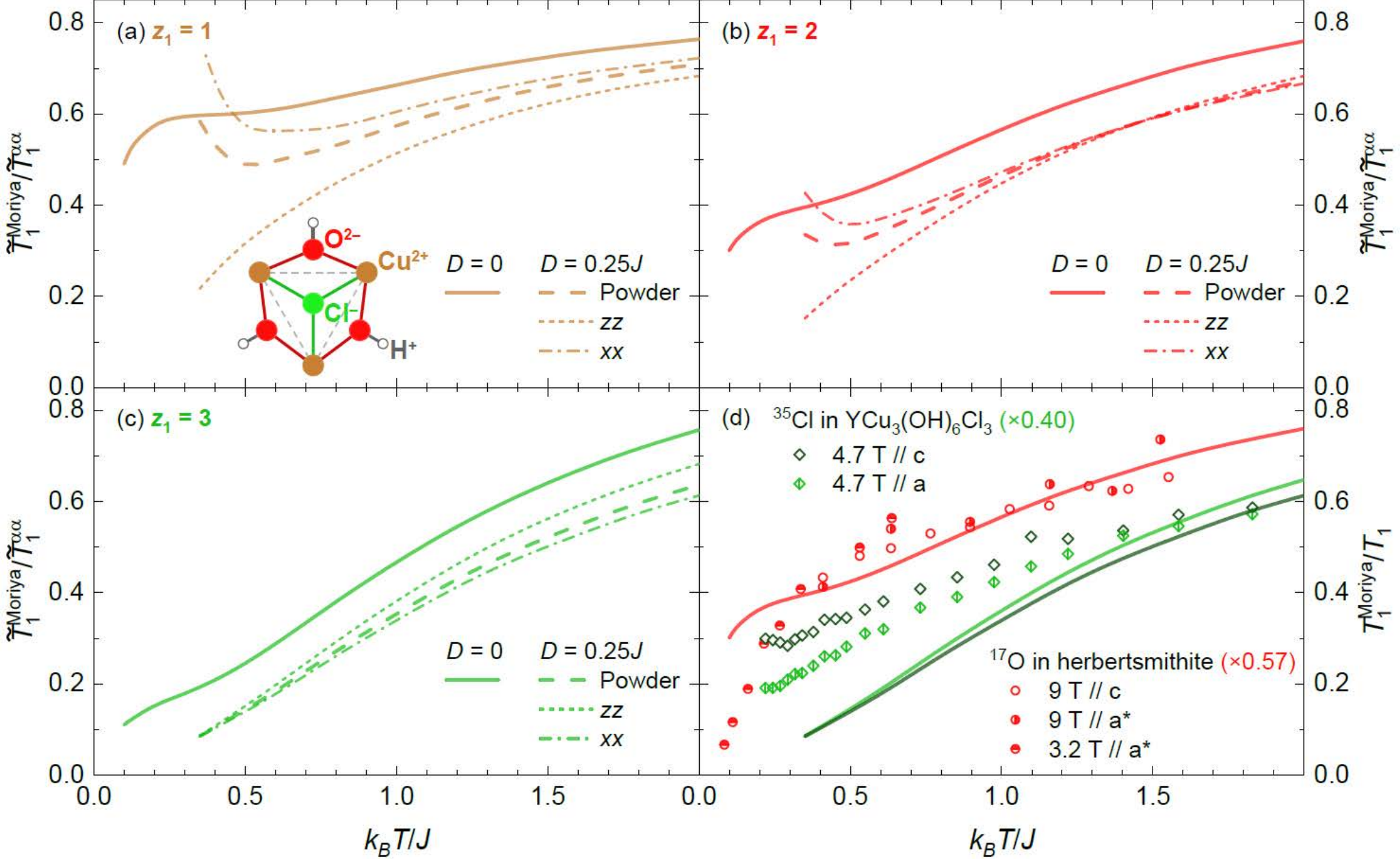} 
 	\caption{ (a)--(c) Directional spin--lattice relaxation rate contributions for coupling to a different number ($z_1$) of electron spins (see inset in (a)) normalized by the Moriya's high-temperature Gaussian approximation $1/T_1^{\rm Moriya}=\hbar z_1/(8 \sqrt{\pi}J)$. 
 		The numerical FTLM calculations were performed for DM interactions $D = 0$ and $D = 0.25J$ on a finite kagome lattice with $N$ = 30 sites.
 		(d) Comparison between theory and experiment on  single crystals of herbertsmithite~\cite{Fu655} and  for the kagome antiferromagnet YCu$_3$(OH)$_6$Cl$_3$.
 		Adapted from~\cite{prelovsek2021dynamical} with permission from APS.}
 	{\label{FTLM}}
 \end{figure*}

The expected temperature dependence of the nuclear spin–lattice relaxation in kagome antiferromagnets was recently
calculated by exact diagonalization techniques down to $T\simeq 0.1J$. 
Using the finite temperature Lanczos method (FTLM) Prelov\v{s}ek {\it et al.}~\cite{prelovsek2021dynamical} have calculated $1/T_1$ for different nuclear sites (see inset in Fig.~\ref{FTLM}) coupled to $z_1$ nearby electron spins for the case of perfectly isotropic Heisenberg Hamiltonian ($D=0$) and when additional out-of-plane Dzyaloshinskii-Moriya (DM) interactions up to $D/J=0.25$ are present.
Already for $D=0$ the temperature dependence of $1/T_1$ is strongly site dependent (compare panels (a)--(c) in Fig.~\ref{FTLM}), because coupling to multiple electron sites ($z_1>1$) at highly symmetric nuclear sites causes filtering of spin fluctuations for particular wave vectors ${\bf q}$ in Eq.~(\ref{NMRrela}).
Namely, for the oxygen site in herbertsmithite, which is symmetrically coupled to two copper spins (inset in Fig.~\ref{FTLM}), i.e., $z_1 = 2$, antiferromagnetic fluctuations of these two spins are effectively filtered out, while chiral correlations of all three spins of a fundamental kagome triangle are suppressed only to a certain degree~\cite{prelovsek2021dynamical}.
The latter are fully filtered out for the more symmetric chlorine site in herbertsmithite, which is positioned on the three-fold rotation symmetry axis of the fundamental kagome triangle (inset in Fig.~\ref{FTLM}), i.e., $z_1 = 3$.
In the case of a finite DM interaction $D>0$, spin fluctuations, and consequently $1/T_1$, become anisotropic (Fig.~\ref{FTLM}).  
For herbertsmithite, the nuclear spin--lattice relaxation results on $^{17}$O nuclei~\cite{Fu655} agree well with theoretical predictions for an isotropic Hamiltonian (Fig.~\ref{FTLM}(d)), as the DM interaction in this system is relatively small ($D/J =0.04-0.08$, see Subsection~\ref{DMherb})~\cite{zorko2008dzyaloshinsky,el2010electron} so that it does not substantially affect the development of spin correlations~\cite{prelovsek2021dynamical}.
The agreement gets somewhat worse for $T/J \lesssim 0.3$, where terms beyond the simplest Heisenberg Hamiltonian and DM interactions potentially become important.\\
Another example of how the $T$ dependence of $1/T_1$ can yield an informative insight into magnetism of a kagome antiferromagnet is provided by $^{35}$Cl measurements on Y-kapellasite, YCu$_3$(OH)$_6$Cl$_3$~\cite{prelovsek2021dynamical}.
This materials enters a negative-vector-chirality 120$^\circ$ long-range ordered ground-state below $T_N=0.15 J$~\cite{zorko2019negative}, which is attributed to a sizable DM interaction $D/J=0.25$~\cite{arh2020origin}.
Like in herbertsmithite, chlorine nuclei in this material are symmetrically coupled to three spins on the kagome triangle ($z_1 = 3$). 
The anisotropy of $1/T_1$ should be minimal at $T>T_N$ due to effective filtering out of the dominant and anisotropic chiral spin fluctuations (Fig.~\ref{FTLM}(c)), which are precursors to the ordered state.
However, the experimentally-observed decrease of $1/T_1$ with $T$ is less pronounced and anisotropic than predicted (Fig.~\ref{FTLM}(d)), which suggests a remnant chiral contribution to the spin--lattice relaxation rate. 
This could be a sign of broken three-fold rotation symmetry, at least on a local scale, perhaps in a similar manner as the three-fold rotational symmetry is globally broken in herbertsmithite~\cite{zorko2017symmetry} (see Subsection~\ref{DMherb}).
\begin{figure*}[t]
	\includegraphics[width=0.9\linewidth]{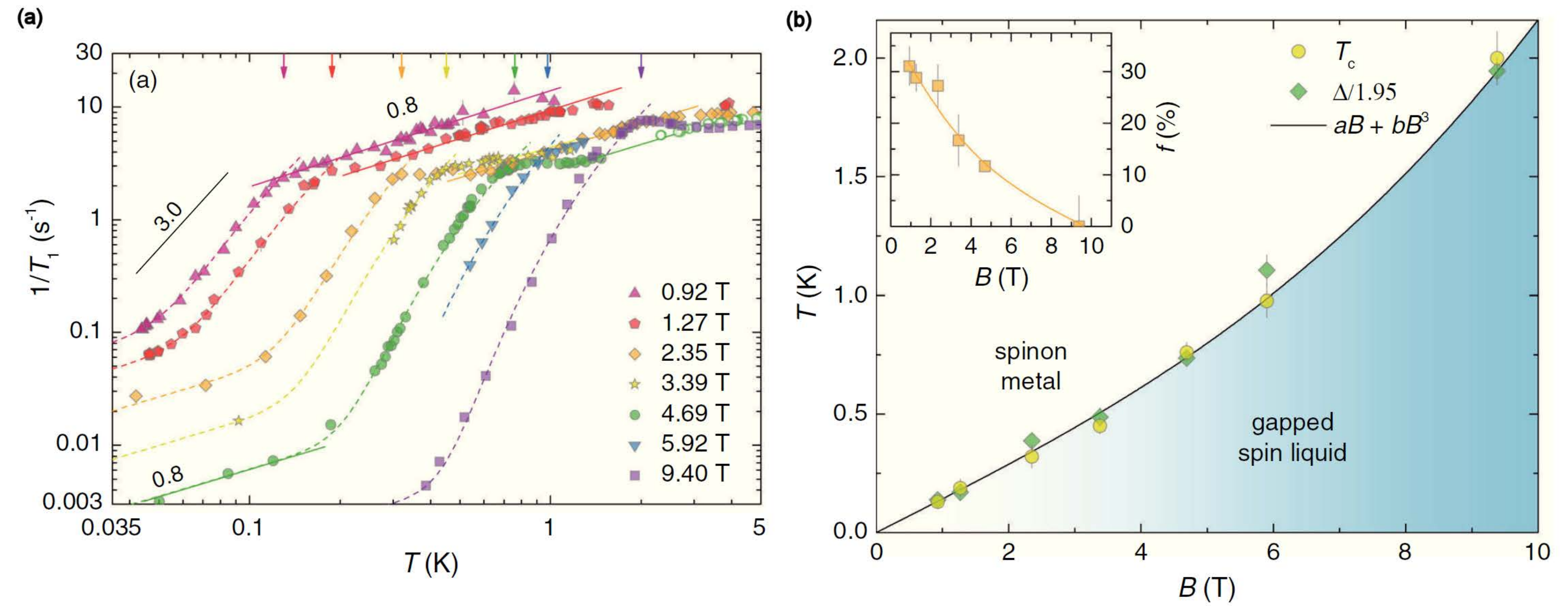} 
	\caption{(a) The temperature dependence of the $^2$D NMR spin--lattice relaxation rate of deuterated ZnCu$_3$(OH)$_6$SO$_4$ in several magnetic fields. 
		The solid lines indicate different power laws, while the dashed lines correspond to a thermally-activated (gapped) behavior.
		(b)Phase diagram of this compound a gapless spinon-metal QSL phase at low fields and a gapped QSL phase at high applied fields. The inset shows the fraction of the residual spinon density of states at the Fermi level.
		Adapted from~\cite{gomilsek2017field} with permission from APS.}
	{\label{pair}}
\end{figure*}

\subsubsection{Spinon pairing in Zn-brochantite\label{pairing}}
A sudden change in the temperature dependence of the nuclear spin--lattice relaxation is almost always a sign of a phase transition.
For instance, the opening of a magnon gap below the N\'eel temperature $T_N$ is by rule seen in a rapid decay of the relaxation rate $1/T_1$ below this temperature.
At first sight, a similar phenomenon was also experimentally observed in the kagome antiferromagnet Zn-brochantite, ZnCu$_3$(OH)$_6$SO$_4$ 
\cite{gomilsek2017field}, where a sudden change of the $^2$D NMR relaxation rate $1/T_1$ (Fig.~\ref{pair}(a)) was observed in a deuterated sample below a field-dependent critical temperature $T_c$ (Fig.~\ref{pair}(b))
Here, $T_c$ vanishes at $B = 0$ and very slowly increases approximately as $T_c \propto B$ in large applied fields.
Above $T_c$, the spin--lattice relaxation rate is proportional to $T^{0.8}$, which is close to the linear Korringa relation $1/T_1 \propto T$ expected for free fermions, e.g., conduction electrons in a metal or other itinerant excitations with a Fermi surface. 
\textcolor{black}{Deviations towards sublinear power-law behavior are expected~\cite{itou2010instability} in the case where spinon couple with emergent gauge field fluctuations~\cite{metlitski2015cooper}.}
Such behavior in insulating Zn-brochantite is thus consistent with a gapless QSL ground-state with a spinon Fermi surface~\cite{gomilsek2016muSR}.
This state is called a spinon metal state since itinerant spinons in many respects, e.g. in
relaxing nuclear spins, play effectively the same role as itinerant electrons do in metals. The decrease of the power from
1, expected for free fermions, to 0.8 is an important indication of the presence of spinon–spinon interactions mediated
by emergent gauge fields.
The kink in $1/T_1$ at $T_c$ does not correspond to a standard magnetic transition, which would manifest in divergent behavior originating from critical spin fluctuations at the low-symmetry deuterium sites in Zn-brochantite.
Moreover, no spectral broadening that would be indicative of spin freezing or ordering is observed below this temperature~\cite{gomilsek2017field}. Therefore, the system remain in a spin-liquid phase. 
The excitations in this low-$T$, high-field QSL phase are, however, very different than for $T > T_c$, as evidenced by much a stronger $T$ dependence of $1/T_1$ (Fig.~\ref{pair}(a)).
The nuclear spin--lattice relaxation rate follows thermal-activated behavior, $1/T_1\propto  T \mathrm{exp}(-\Delta/k_{\mathrm{B}}T)$, indicating the opening of a spin excitation gap $\Delta$, which increases with applied field and scales linearly with $T_c$ (Fig.~\ref{pair}(b)).
\textcolor{black}{This linear increase is much less steep than the temperature dependence of the critical field in the Kondo effect that also occurs in this material (see Fig.~\ref{kondorelax}f), so the two effects are clearly seperable~\cite{Gomilsek2019}.}

The instability of the spinon metal phase in Zn-brochantite at $T_c$ is believed to be due to spinon pairing arising from gauge-field mediated attractive interactions between spinons~\cite{gomilsek2017field}. 
This resembles the formation of Cooper pairs in superconductors~\cite{lee2023proposal} ($2\Delta/T_c = 3.9(1)$ is also quite close to the BCS value $2\Delta/T_c = 3.53$ of conventional superconductors~\cite{gomilsek2017field}), electron pairing in metals mediated by order-parameter fluctuations at quantum critical points, and composite-fermion pairing in quantum Hall
fluids~\cite{metlitski2015cooper}. 
The difference is that in this case the pairing affects purely magnetic excitations (spinons) with no charge degrees of freedom.

\begin{figure*}
	\centering
	\includegraphics[width=0.8\textwidth]{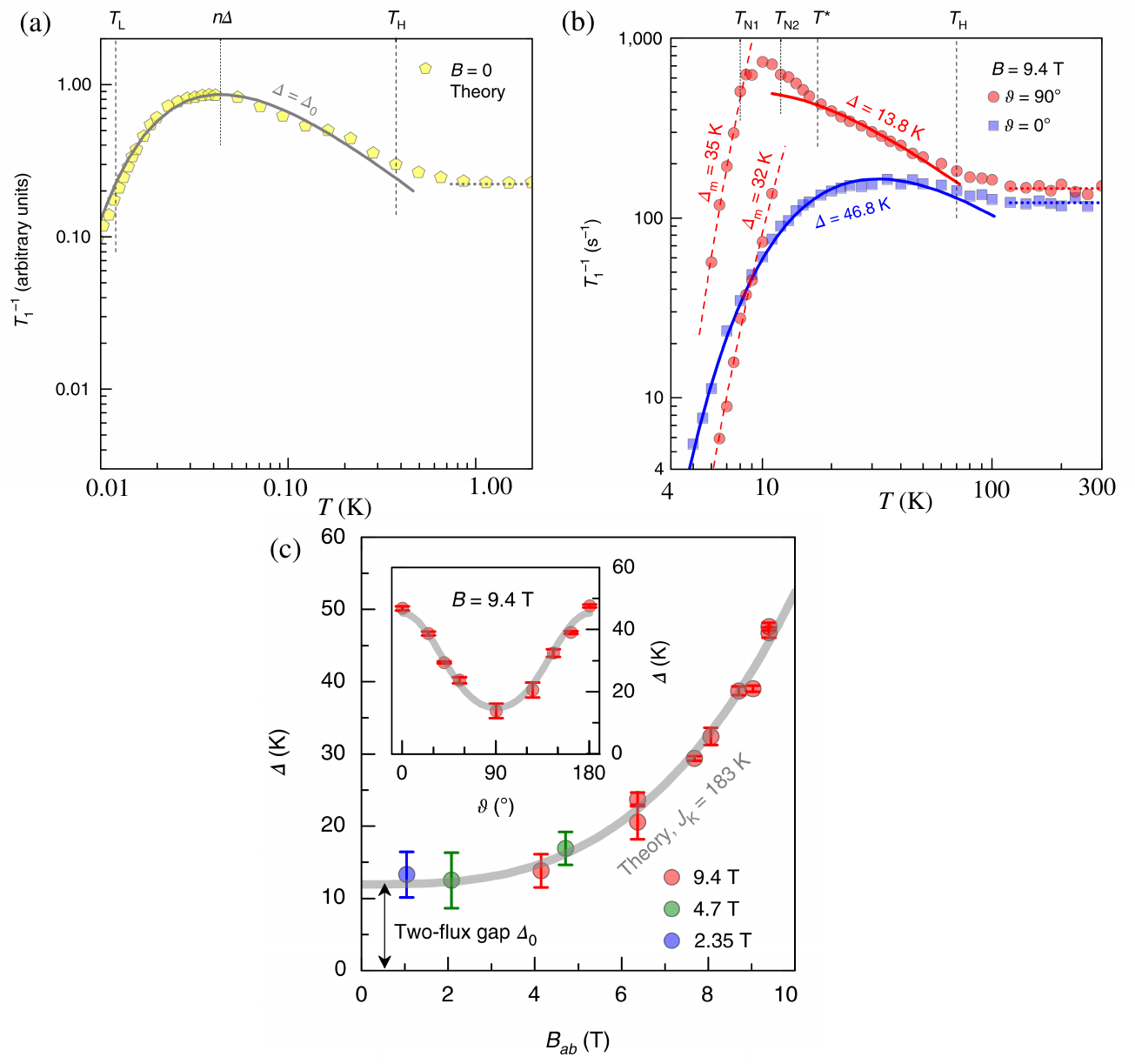} 
	\caption{(a) Theoretical NMR spin--lattice relaxation rate, $1/T_{1}$ vs. $T$ for ferromagnetic Kitaev model. 
 The solid gray line is the fit to the Eq.~(\ref{eqn42}) with $\Delta=0.065J_K$ which gives $n = 0.67$ in zero applied field. 
 The temperature independent classical paramagnetic value of $1/T_{1}$ is denoted by the dotted gray line at high $T$. 
 (b) Experimental NMR spin--lattice relaxation for two different directions of the applied field with respect to the honeycomb planes. The solid lines are fits with the model of Eq.~(\ref{eqn42}).
 (c) The excitations gap ($\Delta$) as a function of the applied effective in-plane magnetic field $B_{ab}$ where the solid grey line is the theoretically predicted excitations gap which is proportional to the cube of the field. 
 Inset shows $\Delta$ as a function of the orientation of the magnetic field $B_{\rm ab}$ = 9.4~T. 
 Adapted from~\cite{Jansa2018} with permission from NPG.}{\label{NMR_RUCl3}}
\end{figure*}
\subsubsection{Two types of fractional excitation in $\alpha$--RuCl$_3$}
Another illustrative example of the strength of
NMR in addressing the excitations of a QSL state is that of the Kitaev candidate $\alpha$--RuCl$_3$~\cite{Jansa2018}.  
In the absence of an applied magnetic field, Majorana fermions are gapless, while under weak magnetic fields, they become gapped with the gap predicted to vary as a peculiar third power of the applied field~\cite{Kasahara2018,KITAEV20062,PhysRevB.83.245104}.  
NMR measurements indeed confirmed the presence of field-dependent gapped excitations in this Kitaev material~\cite{Jansa2018}.
The $g$ factor of Ru$^{3+}$ ion is highly anisotropic with its values in the $ab$ plane and along the $c^*$ axis being $g_{ab}=2.5$ and $g_{c^*}= 1.1$~\cite{Yadav2016}, respectively.
If a magnetic field $\mathbf{B}$ is applied at an angle $\theta$ away from the $ab$ plane, the effective $g$ factor is given by~\cite{Jansa2018}
\begin{equation}
g(\theta)=\sqrt{g_{ab}^2 \cos^2(\theta) + g_{c^*}^2\sin^2(\theta)}.
\end{equation}
The effect is  similar as if a field of magnitude $B_{ab}= B g(\theta)/g_{ab}$ were applied in the $ab$ plane.
The theoretical NMR spin--lattice relaxation rate in $\alpha$--RuCl$_3$ is shown in Fig.~\ref{NMR_RUCl3}.
In the lower temperature range where a magnetic phase transition is observed, the data  are well fitted by the equation
\begin{equation}{\label{eqn41}}
\frac{1}{T_1}=T^2 \exp(\frac{-\Delta_m}{T}),
\end{equation} which is due to the gapped magnon excitations observed in the 3D magnetically ordered state (Fig.~\ref{NMR_RUCl3}(b)). 
In the Kitaev paramagnetic region ($17$~K~$\leq T \leq 70$~K), on the other hand, the data are well captured by the empirical expression
\begin{equation}{\label{eqn42}}
\frac{1}{T_{1}}=\frac{1}{T}\mathrm{exp} \left(\frac{-n\Delta}{T}\right),
\end{equation}
where $n = 0.67$ was  obtained for  the zero-field theoretical gap value.
The broad maximum signifies the excited pairs of gauge fluxes with a two-flux gap $\Delta _0=0.065 J_K$, where $J_K =190$~K is the strength of the Kitaev interaction~\cite{Do2017,PhysRevLett.112.207203}. 
Besides this zero-field two-flux gap, the rest of the spin gap precisely follows the prediction for Majorana fermion excitations, i.e., $\Delta - \Delta_0 \propto B^3$ (Fig.~\ref{NMR_RUCl3}(c)). 
The signature of both Majorana fermions and $\mathbb{Z}_2$ fluxes in the excitations gap suggests that the honeycomb lattice $\alpha$--RuCl$_3$ is a proximate Kitaev spin liquid.
\textcolor{black}{Another common signature of fractionalized quasiparticles is a continuum in the spin excitation spectrum. 
Its presence in $\alpha$--RuCl$_3$ was observed by Raman scattering experiments~\cite{PhysRevLett.114.147201, Nasu2016}.}

\textcolor{black}{At the end of this subsection, we note that it is also possible to carry out NMR measurements in the absence of an external magnetic field by taking advantage of quadrupolar interactions between nuclear quadrupolar moments (which are finite for nuclear spins $I>1/2$) and electric field gradients present in a crystal.
This technique is known as nuclear quadrupole resonance (NQR).
NQR is useful in cases when a magnetic field would significantly perturb the intrinsic state of the sample, which is sometimes the case with fragile states of frustrated magnets.
Such measurements have, for instance, revealed the presence of gapless spin dynamics in the spin-liquid ground state of the triangular lattice 1T-TaS$_{2}$~\cite{Klan2017}, and provided direct detection of emergent monopole density in the the spin-ice material Dy$_2$Ti$_2$O$_7$~\cite{PhysRevLett.108.217203}. }

\subsection{Electron spin resonance (ESR)\label{secESR}}
Electron spin resonance (ESR) is another highly informative local-probe technique used to study magnetism in quantum materials, especially when it comes to determining magnetic anisotropy~\cite{zorko2018determination} and detecting any possible symmetry reduction in these systems~\cite{zorko2017symmetry}.
\textcolor{black}{Since the magnetic anisotropy terms in a spin Hamiltonian do not commute with the Heisenberg Hamiltonian they cause a decrease of temporal spin correlations, which is directly reflected in finite ESR linewidth. 
The different possible anisotropy terms (e.g., Dzyaloshinskii-Moriya interaction, symmetric exchange anisotropy, single-ion anisotropy) lead to different angular dependences of the line width~\cite{PhysRevB.4.38, PhysRevLett.107.257203, zorko2013dzyaloshinsky}.}
For finite anisotropy, their contribution to the ESR linewidth can be readily calculated in the high-$T$ exchange-narrowing limit, where the linewidth is $T$-independent and can be evaluated within the context of well-established Kubo--Tomita (KT) theory~\cite{kubo1954general}.
\begin{figure}[b]
	\includegraphics[width=1\linewidth]{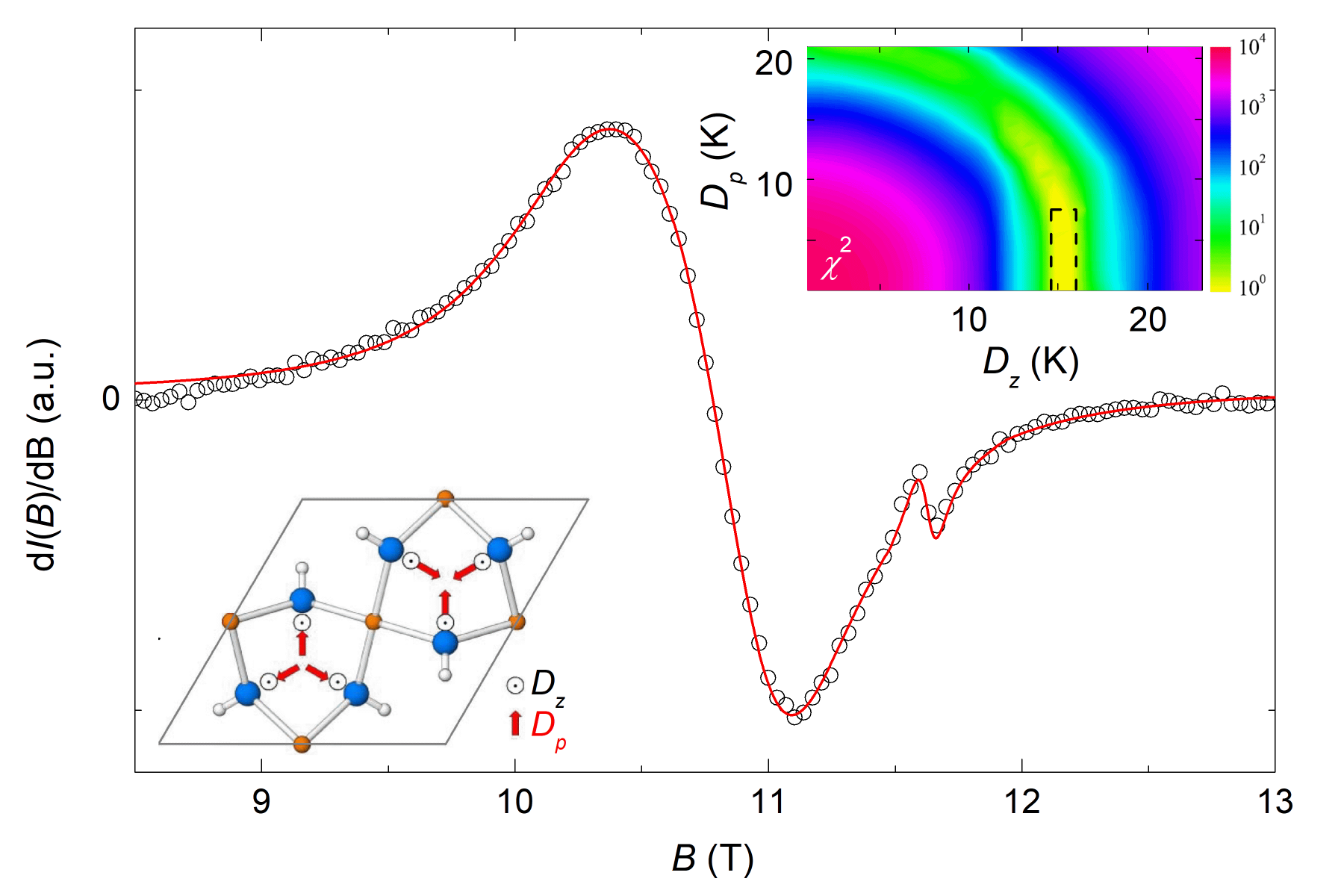} 
	\caption{ESR spectrum of herbertsmithite at 300~K with a fit given by the red solid line.
	The lower inset shows the unit cell with the pattern of out-of-plane Dzyaloshinskii-Moriya (DM) vectors $D_z$ and in-plane DM vectors $D_p$.
	The upper inset shown the quality of the fit as a function of the latter two parameters.
Adapted from~\cite{zorko2008dzyaloshinsky} with permission from APS.}
{\label{ESRherb}}
\end{figure}

\subsubsection{Magnetic anisotropy and symmetry breaking in herbertsmithite\label{DMherb}}
This method was used to determine the magnetic anisotropy of herbertsmithite~\cite{zorko2008dzyaloshinsky}.
For this compound, the analysis of the ESR line shape demonstrated that the dominant anisotropic term in the spin Hamiltonian was the Dzyaloshinskii-Moriya (DM) interaction with a large out-of-plane DM component $D_z/J\simeq0.08(1)$ and a much smaller (if even present) in-plane component $D_p/J\simeq0.01(3)$ (Fig.~\ref{ESRherb}).
The predicted ESR linewidth was also calculated exactly on finite kagome spin clusters~\cite{el2010electron}.
These calculations have revealed that the ESR linewidth should scale with $D_z^2/J$, as predicted by KT theory, but have slightly revised the estimated size of the DM interaction in herbertsmithite to $0.04 \leq D_z/J \leq 0.08$, if the presence of spin diffusion is assumed.

\begin{figure}[t]
	\includegraphics[width=1\linewidth]{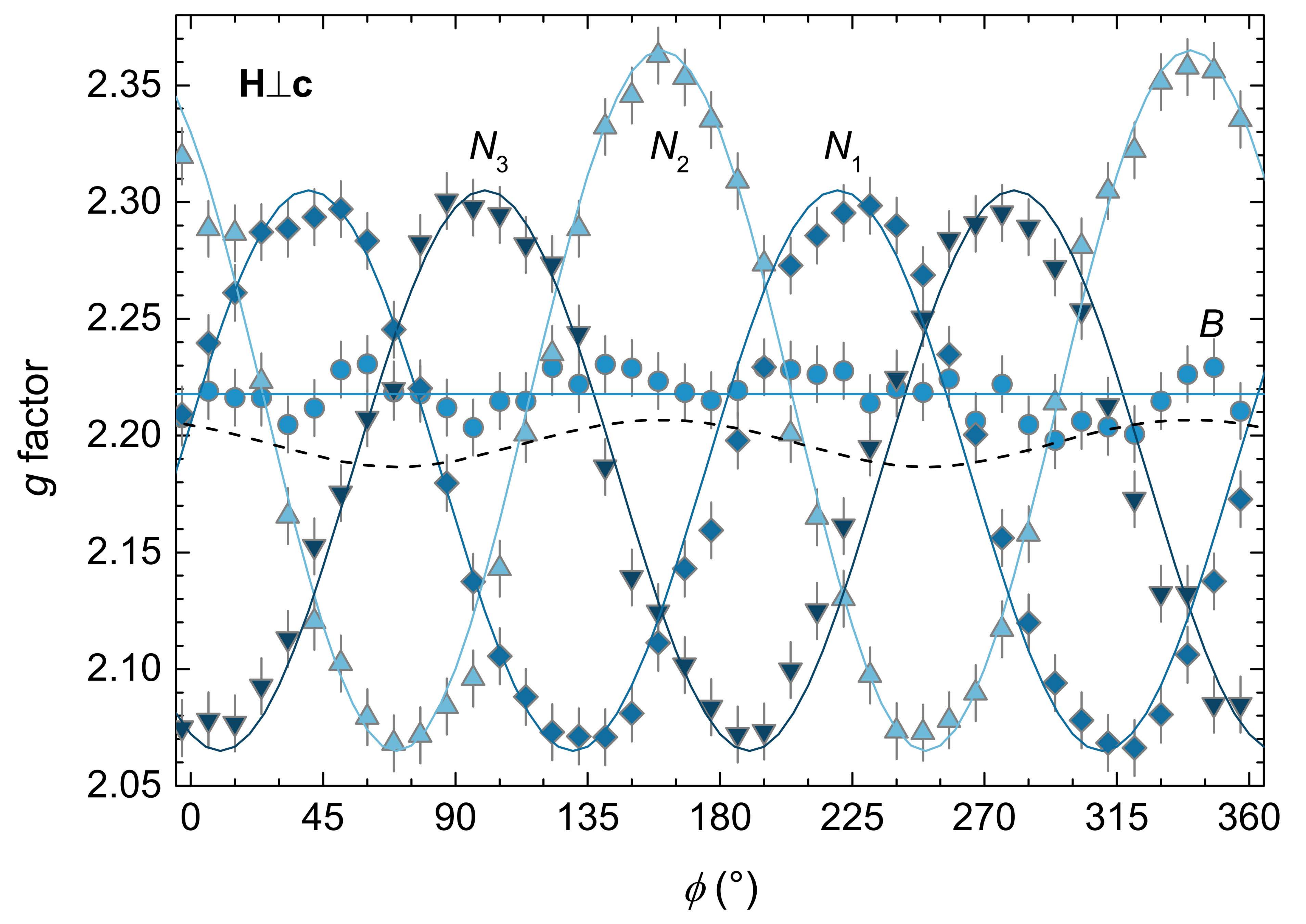} 
	\caption{The Angular dependence of the $g$-factors of three defect ESR lines within the kagome planes in herbertsmithite at 5~K, corresponding to three magnetically inequivalent defect sites under a finite applied magnetic field.
	The dashed line corresponding to the average of these three lines clearly demonstrates the breaking of the three-fold symmetry. 
Adapted from~\cite{zorko2017symmetry} with permission from APS.}
{\label{ESRsym}}
\end{figure}

The ESR technique is also very powerful for determining local symmetry.
An in-depth ESR study accompanied by bulk magnetic torque measurements on herbertsmithite has revealed a global structural distortion~\cite{zorko2017symmetry} that is likely associated with the establishment of the QSL ground state.
ESR lines of intrinsic defects that dominate the ESR response in this material at low temperatures showed evidence of a globally broken three-fold symmetry of the ideal kagome lattice.  
Namely, the average of the three resonance lines from magnetically inequivalent sites of the kagome triangle was found to exhibit a 180$^{\circ}$ angular dependence within the kagome planes (Fig.~\ref{ESRsym}), which is in contradiction with the expected uniaxial symmetry.
The structural distortion in herbertsmithite has recently been confirmed also by complementary optical measurements~\cite{laurita2019evidence} and \textit{ab initio} calculations of magneto-elastic coupling~\cite{PhysRevB.101.161115}, and is also consistent with a change in the quadrupolar frequency $\nu_Q$ at low $T$, as the QSL ground-state is established~\cite{Fu655}. 
Evidence for the presence of two distinct types of magnetic defects in herbertmishite, with different coupling strengths to intrinsic kagome spins, was also found by ESR~\cite{zorko2017symmetry}.

\subsubsection{DM interaction in other kagome representatives}
The KT approach for determining the dominant DM interaction in kagome antiferromagnets was also applied  for a few other compounds based on Cu$^{2+}$ magnetic moments. In vesignieite, BaCu$_3$V$_2$O$_8$(OH)$_2$, it was found that the in-plane DM component is dominant, which likely triggers magnetic ordering in this compound~\cite{zorko2013dzyaloshinsky}, while the magnetic ordering in YCu$_3$(OH)$_6$Cl$_3$ was found to originate from an extremely large out-of-plane DM component $D_z/J=0.25$~\cite{arh2020origin}.

\subsection{Inelastic neutron scattering (INS)\label{INS}}
\begin{figure}[t]
	\centering
	\includegraphics[width=0.8\columnwidth]{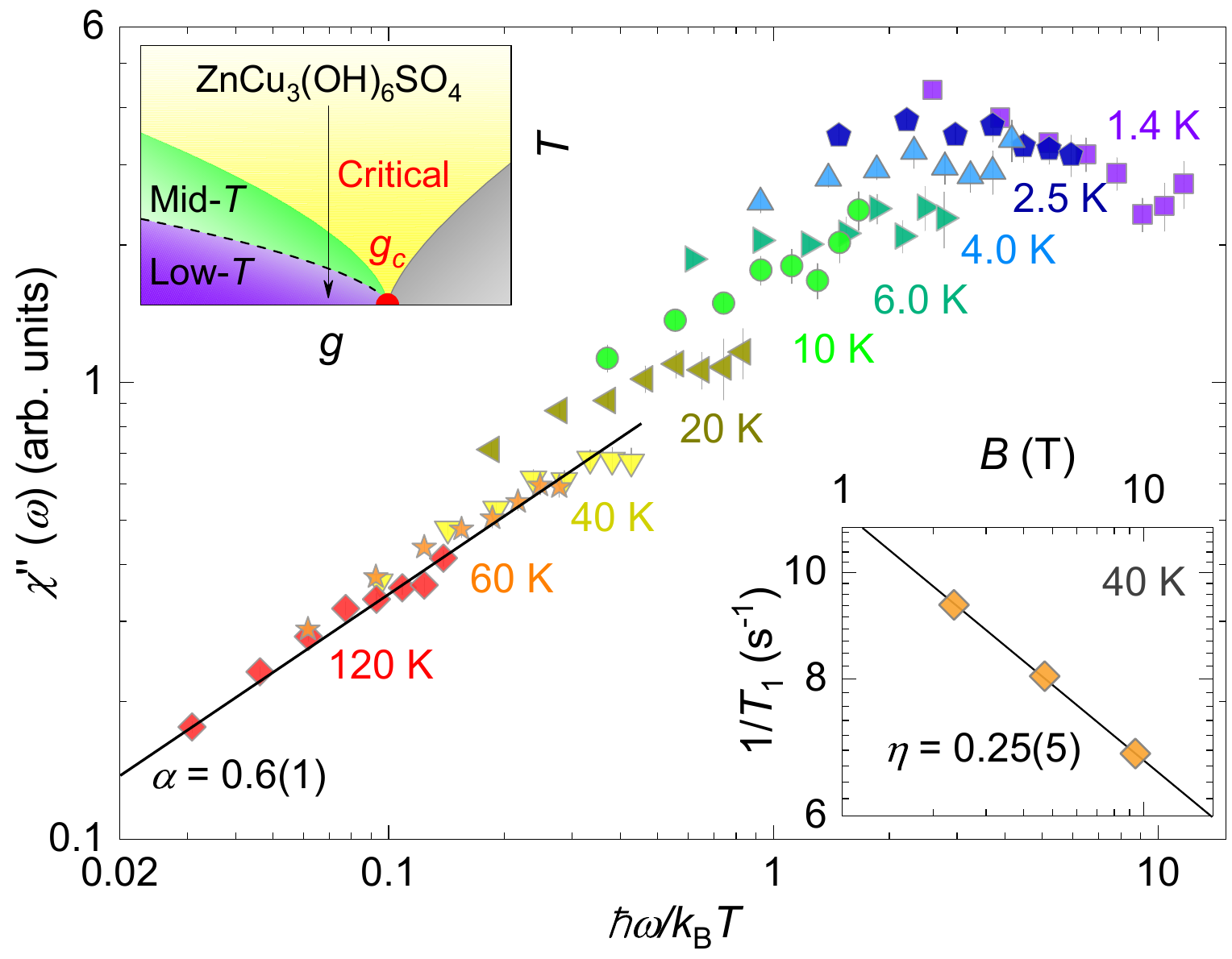}
	\caption{Power-law scaling of the INS dynamical susceptibility $\chi''(\omega) \propto (\hbar\omega / k_{\rm B} T)^\alpha$ of Zn-brochantite for $T$ above $\SI{\sim 40}{\kelvin}$. 
		Lower inset:  field dependence
		of the $^{2}$D NMR $1/T_1 \propto B^{-\eta}$ at $\SI{40}{\kelvin}$. 
		Upper inset: a schematic $T$--$g$ phase diagram with a QCP $g_c$ between the true QSL ground-state (violet) and a neighboring phase (gray) with the associated quantum critical fan (yellow). 
		Two QSL temperature regimes of Zn-brochantite (purple and green) are shown separated by a dashed line.
		Adapted from~\cite{gomilsek2016instabilities,gomilsek2018quantum} with permission from APS.}
	\label{ins_fig1}
\end{figure}

Inelastic neutron scattering (INS) is a very powerful experimental technique that directly probes the dynamical spin--spin correlation function 
\begin{equation}
	S^{\alpha\beta}({\bf q},\omega) = \frac{1}{2\pi} \int_{-\infty}^\infty \dd t~ e^{i \omega t} \langle S^\alpha_{-{\bf q}}(t) S^\beta_{\bf q}(0)\rangle ,
\end{equation}
where $\langle \dots \rangle$ denotes the canonical thermal average, $\alpha$ and $\beta$ are directional indices, and $S^\alpha_{\bf q} = \sum_i e^{i {\bf q} \cdot {\bf R}_i} S^\alpha_i /\sqrt{N}$ is the ${\bf q}$-space spin operator for $N$ spins ${\bf S}_i$ at positions ${\bf R}_i$. 
Specifically, the intensity of coherent magnetic neutron scattering with momentum transfer $\hbar {\bf q}$ and energy transfer $\hbar \omega$  is proportional to $S_\perp^{\alpha\beta}({\bf q},\omega) = \sum_{\alpha \beta} \left( \delta_{\alpha\beta} - q_\alpha q_\beta / q^2 \right) S^{\alpha\beta}({\bf q},\omega)$, up to trivial experimental prefactors~\cite{shirane2006neutron,xu2013absolute}.
In contrast to local-probe techniques such as $\mu$SR, NMR and ESR that sum over all reciprocal space, INS uniquely provides $q$-resolved information on spin correlations and fluctuations.\\ 
In conventional ordered magnets $S^{\alpha\beta}({\bf q},\omega)$ is only large for $\omega$ on a sharp magnon dispersion relation $\omega({\bf q})$, 
while in frustrated and disordered magnets, as well as in systems close to criticality, the intensity of $S^{\alpha\beta}({\bf q},\omega)$ is spread across a continuum of ${\bf q}$ and $\omega$. 
An extreme example of this is quantum criticality where the associated dynamical susceptibility 
\begin{equation}
    \chi''({\bf q},\omega) = \left[ 1 - e^{-\hbar\omega / k_{\rm B} T} \right] S({\bf q},\omega) ,
\end{equation}
is expected to exhibit a scaling relation of the form 
\begin{equation}
    \chi''({\bf q},\omega) \propto |\omega|^\zeta F(\hbar\omega / k_{\rm B} T) ,
    \label{eq_qcp_chi_scaling}
\end{equation}
for some universal power $\zeta$ and universal function $F$~\cite{sachdev1992universal,sachdev2011quantum}.
{This occurs due to the vanishing of any intrinsic preferential energy scale for spin fluctuations at the QCP, which implies that the response of the system can only depend on external parameters like $T$ and $\omega$. 
Due to their lack of an intrinsic energy scale, spin fluctuations at the QCP are sometimes also called scale-free, while the universal function $F$ is often well-approximated by a power-law over broad ranges of its parameter.} 
However, if system Hamiltonian parameters $g$ (e.g., pressure, chemical composition, exchange interaction strengths etc.) are not tuned exactly to the QCP at $g = g_c$, the scaling relation Eq.~(\ref{eq_qcp_chi_scaling}) only holds approximately, at $T$ high enough to lie inside a quantum-critical fan that emanates from the QCP, i.e., for $T$ above the small intrinsic energy scale of spin fluctuations when $g$ is near, but not exactly equal to, $g_c$ (see the upper inset in Fig.~\ref{ins_fig1}).\\
In contrast to scale-free critical fluctuations, an INS continuum in the low-$T$ QSL ground state of frustrated magnets instead arises from the fractionalization of spin excitations. 
This usually takes the form of a spin flip, which would produce a single spin-1 magnon in an ordered magnet, instead producing a pair of spin-1/2 spinon quasiparticles. 
INS thus effectively becomes a three-body scattering process, 
instead of the usual two-body scattering of a neutron on a magnon, 
which means that a continuum of transferred ${\bf q}$ and $\omega$ becomes allowed. 
However, the resulting INS intensity can still exhibit substantial ${\bf q}$ and $\omega$ structure, which can serve as a crucial fingerprint of the material's elusive QSL state.
For example, such INS structure was used to identify the ground state of the triangular-lattice antiferromagnet YbMgGaO$_{4}$ as a gapless spinon Fermi surface QSL~\cite{shen2016evidence,shen2018fractionalized}. 

\subsubsection{Quantum criticality in kagome QSL candidates}
Scale-free spin fluctuations were found
in the kagome antiferromagnet Zn-brochantite at $T$ above $\SI{\sim 40}{\kelvin}$, where a scaling relation $\chi''(\omega) \propto (\hbar\omega / k_{\rm B} T)^\alpha$ with $\alpha = 0.6(1)$ was observed in powder INS measurements performed at $\hbar \omega = 0.3$--$\SI{2}{\milli\electronvolt}$ (Fig.~\ref{ins_fig1})~\cite{gomilsek2016instabilities}. 
The same scaling relation was confirmed to extend to orders of magnitude lower energies $\hbar \omega = \SI{60}{\nano\electronvolt}$ for $T \sim 40$--$200$~K via complementary $^{2}$D NMR measurements. 
Namely, the expected spin--lattice relaxation rate scaling relation $1/T_1 \propto (k_{\rm B} T / \hbar\omega)^\eta \propto (T / B)^\eta$ with $\eta = 1 - \alpha = 0.4(1)$, 
where $\omega = \gamma_n B$ is the nuclear Larmor frequency for an applied field $B$ and $\gamma_n$ is the nuclear gyromagnetic ratio, 
is found both in a power-law $T$ dependence with $\eta = 0.30(3)$ and a power-law $B$ dependence with $\eta = 0.25(5)$ (lower inset in Fig.~\ref{ins_fig1}). 
In contrast, a trivial quasi-free impurity contribution would correspond to a much larger $\eta \approx 1$~\cite{klanjsek2015phonon}. 
{The observed scaling relations would be consistent with the proximity of Zn-brochantite to a QCP of an as-yet undetermined type (see the upper inset in Fig.~\ref{ins_fig1})~\cite{gomilsek2016instabilities}.
To confirm this scenario, and determine the nature of the hypothesized QCP, further experiments under pressure or chemical substitution, which might drive the system across the QCP, would be highly desirable.}

Similar scale-free spin fluctuations were also reported in \herb at high $T$ and $\omega$ via INS~\cite{de2009scale} and might indicate that herbertsmithite is also close to a QCP. 
This was proposed to likely arise from the QCP separating a QSL ground state from a chiral long-range ordered ground state at critical DM strength $D_c/J = 0.10(2)$~\cite{PhysRevB.95.054418,Zhu5437,PhysRevB.78.140405}. 
However, separate, apparently scale-free, spin fluctuations in \herb at low $T$ and $\omega$ in Ref.~\cite{helton2010dynamic} were later suggested to arise from correlated impurities~\cite{han2016correlated}, which dominate the \herb INS response at $T < \SI{10}{\kelvin}$~\cite{nilsen2013low} and $\hbar\omega < \SI{0.8}{\milli\electronvolt}$.
\begin{figure}
	\centering
	\includegraphics[width=0.8\columnwidth]{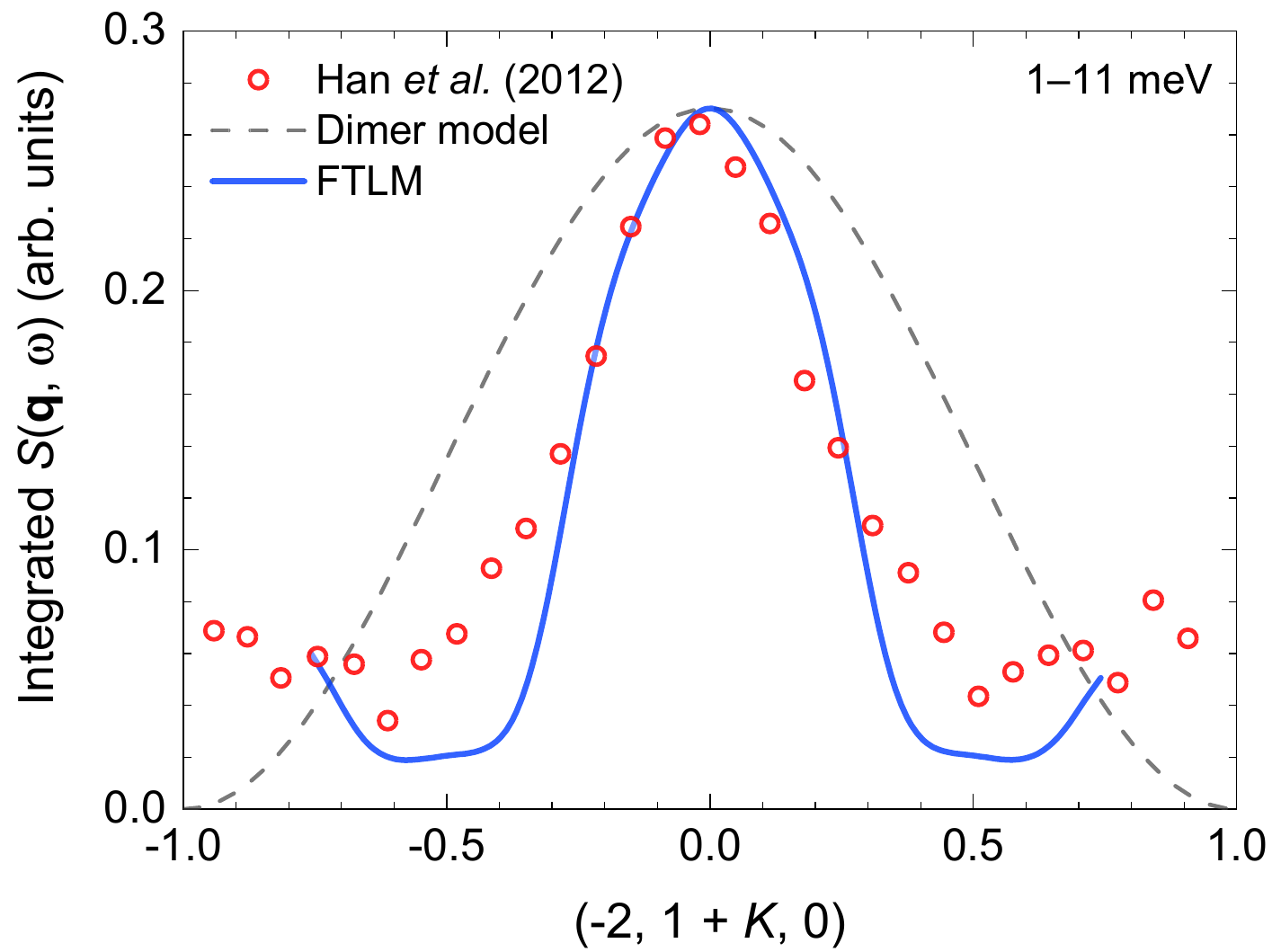}
	\caption{Low-$T$ INS measurements in \herb integrated over $\SI{1}{\milli\electronvolt} < \hbar\omega < \SI{11}{\milli\electronvolt}$ along the $(-2, 1 + K, 0)$ cut in ${\bf q}$-space from Ref.~\cite{han12fractionalized} (symbols) compared to a toy model of independent singlet dimers (dashed line) and full finite temperature Lanczos method (FTLM) calculations from Ref.~\cite{prelovsek2021dynamical} (solid line). 
		Adapted from~\cite{prelovsek2021dynamical} with permission from APS.}
	\label{ins_fig2}
\end{figure}

\begin{figure*}[t]
	\centering
	\includegraphics[width=0.9\linewidth]{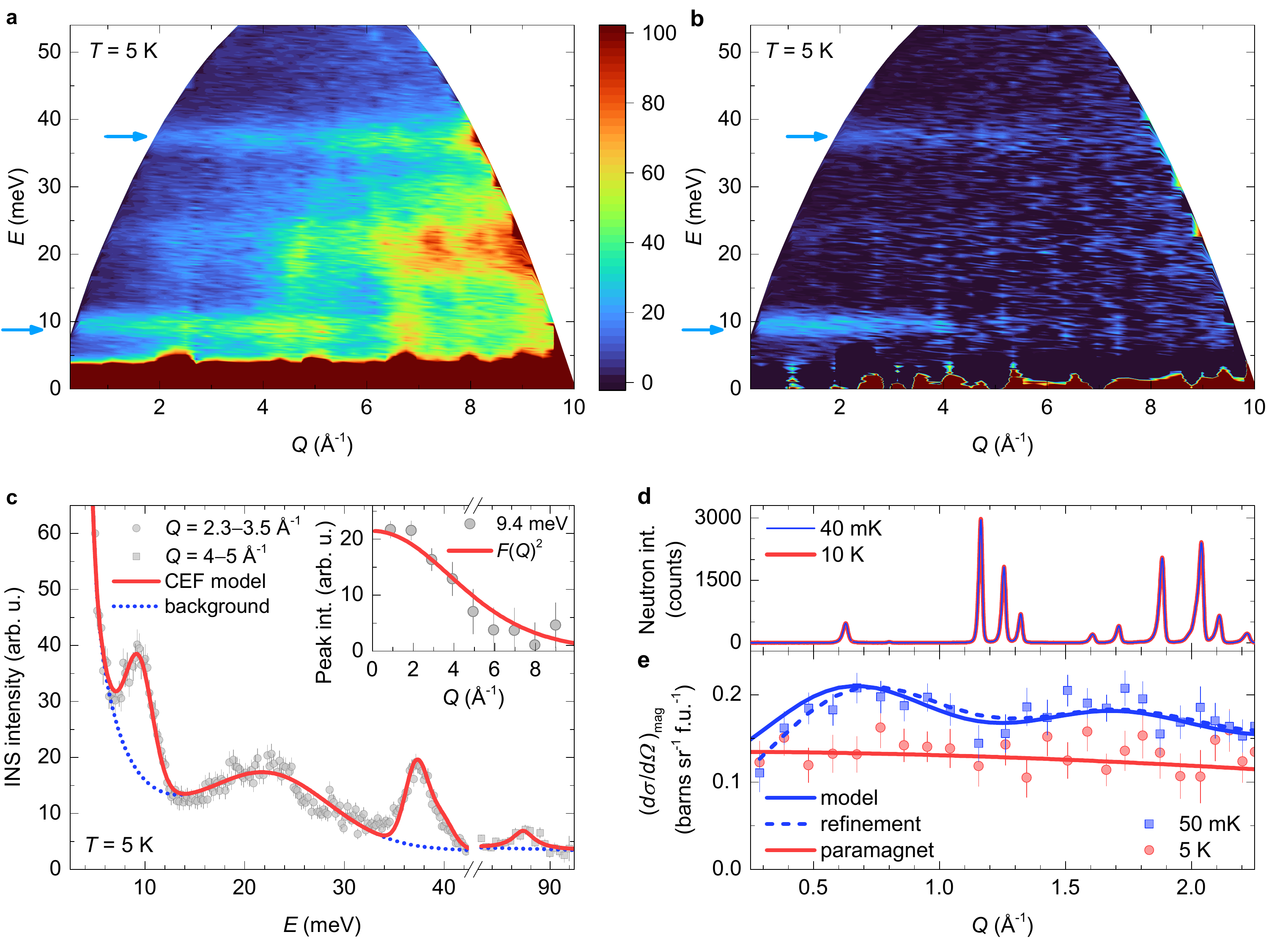}
	\caption{Inelastic neutron scattering intensity of NdTa$_7$O$_{19}$ (a) as measured at 5~K and (b) when the phonon contribution is subtracted by the use of the non-magnetic analogue LaTa$_7$O$_{19}$~\cite{Arh2021}. The arrows indicate the remaining CEF magnetic signal,  which is (c) fit with a CEF model, revealing the
		position and composition of the Kramers doublets. (d) Comparison of neutron diffraction intensity at 10~K and 40~mK reveals absence of any magnetic Bragg peaks. (e) The magnetic neutron scattering cross section exhibits paramagnetic behaviour at 5~K, while it shows diffuse scattering due to nearest-neighbour antiferromagnetic spin correlations which are predominantly of the Ising character at 50~mK.
		Adapted from~\cite{Arh2021} with permission from NPG.}
	\label{FigNTO}
\end{figure*}

\subsubsection{Excitations continuum in herbertsmithite}
In the realm of kagome lattice antiferromagnets, very few materials are available in single-crystal form, which means that their intrinsic INS ${\bf q}$-dependence (and possible directional anisotropy) is often averaged out in experiment. 
Herbertsmithite is a rare exception to this where single-crystal INS at low $T$, and $\omega$ high enough to avoid substantial impurity contributions, was used to demonstrate that $S^{\alpha\beta}({\bf q},\omega)$ features a broad maximum around $\hbar\omega \sim \SI{6}{\milli\electronvolt}$~\cite{han12fractionalized}. 
This feature was reproduced very well by recent kagome antiferromagnet FTLM numerical calculations of Ref.~\cite{prelovsek2021dynamical}, where it was shown to correspond to the lower, $\hbar\omega \approx 0.3 J$, maximum of the two distinct maxima in the $\omega$-dependence of $S^{\alpha\beta}({\bf q},\omega)$. 
It stems from dominant chiral collective spin fluctuations, while the upper frequency-dependence maximum at $\hbar\omega \approx 1.5 J$ corresponds to simple local triplet excitations of isolated spin triangles. 
This conclusion is reinforced by the ${\bf q}$-variation of the INS $S^{\alpha\beta}({\bf q},\omega)$ of \herb when integrated over a broad frequency window $\SI{1}{\milli\electronvolt} < \hbar\omega < \SI{11}{\milli\electronvolt}$, which peaks at the M point of the extended Brillouin zone of the kagome lattice~\cite{han12fractionalized}. 
Both the position and the width of this magnetic peak in ${\bf q}$-space are well reproduced by FTLM calculations (Fig.~\ref{ins_fig2})~\cite{prelovsek2021dynamical}, which demonstrates that the dynamical response of the kagome antiferromagnet is dominated by uniform ($q = 0$) chiral spin fluctuations, i.e., fluctuations of the AFM order parameter for $\ang{120}$ ordered spins on each kagome triangle. 
Furthermore, the substantially narrower width of this peak compared to a simplified toy model of isolated singlet dimers of Ref.~\cite{han12fractionalized} implies that these fluctuations have a nontrivial low-$T$ correlation length $\xi > 1$ that extends beyond the nearest neighbors on the kagome lattice. 

\subsubsection{CEF excitations and Ising correlations in NdTa$_7$O$_{19}$\label{NTOIN}}
The extremely insightful nature of neutron scattering has recently been demonstrated in triangular-lattice antiferromagnet  neodymium heptatantalate,  NdTa$_7$O$_{19}$~\cite{Arh2021}.
INS measurements (Fig.~\ref{FigNTO}(a)) first revealed that at low $T$ the dynamical structure factor, which features arcs extending up to ${\sim}40$~meV, is dominantly due to phonon excitations if energies above the quasi-elastic scattering $E\lesssim 5$~meV are considered.
However, after subtracting the phonon contribution with the help of the non-magnetic analogue compound LaTa$_7$O$_{19}$~\cite{Arh2021}, additional, very flat excitation bands emerge at low $Q$'s (Fig.~\ref{FigNTO}(b)), which can be attributed to magnetic transitions between levels split by the crystal electric field (CEF).
The composition of CEF states and their corresponding magnetic characteristics were determined by a simultaneous fit of INS data (Fig.~\ref{FigNTO}(c)) and the bulk magnetic response with a CEF model~\cite{Arh2021}.
These simulations showed that the ground-state Kramers doublet of the $^4I_{9/2}$ Nd$^{3+}$ multiplet is separated from the lowest excited doublet by 9.4~meV and can be regarded as an effective spin-1/2 state at low $T$.
Furthermore, this state is highly anisotropic, with the magnetic easy-axis perpendicular to the triangular planes, and Ising-like exchange interaction $J_z=0.90(2)$~K~$\gg J_{xy}=0.16(2)$~K, as suggested by ESR measurements~\cite{Arh2021}.

Interestingly, additional neutron diffraction measurements failed to detect any magnetic Bragg peaks even at the lowest temperature of 40~mK (Fig.~\ref{FigNTO}(d)), in agreement with $\mu$SR that also did not detect any signs of magnetic ordering in this compound~\cite{Arh2021}. 
An important additional insight was provided by polarized neutron scattering.
Namely, the three-directional $XYZ$ polarization analysis~\cite{doi:10.1063/1.373364} allowed the authors to separate the magnetic contribution to the neutron scattering from incoherent and nuclear coherent contributions~\cite{Arh2021}.
In accordance with neutron diffraction, these measurements failed to detect any narrow magnetic peaks, but nevertheless revealed broad magnetic diffuse scattering (Fig.~\ref{FigNTO}(e)).
It was found that this diffuse scattering is featureless at 5~K, where the system behaves like an ordinary paramagnet.
However, at 50~mK a clear enhancement of diffuse scattering was observed. 
Additionally, at base $T$ oscillations were found in the diffuse scattering signal, which corresponded well to a nearest-neighbour Nd--Nd distance of $r_1=6.22$~\AA~on the triangular lattice, while their shape revealed antiferromagnetic and predominantly Ising nature of spin correlations~\cite{Arh2021}.
Namely, reverse Monte Carlo refinements using the {\sc spinvert suite}~\cite{Paddison_2013} demonstrated that parallel correlations $Z_1 \langle {\bf S}_{0,\|} \cdot {\bf S}_{1,\|} \rangle/S(S+1)= -0.40(12)$, where $Z_1 = 6$ is the nearest-neighbor coordination number, were much larger than transverse correlations $\frac{1}{2} Z_1 \langle {\bf S}_{0,\perp} \cdot {\bf S}_{1,\perp} \rangle/S(S+1)= -0.09(6)$, in agreement with Ising-like exchanges, ${J}_{xy}/{J}_{z}=0.18$~\cite{Arh2021}.

The discovery that NdTa$_7$O$_{19}$ is a disorder-free Ising-like triangular antiferromagnet featuring no magnetic order at $T$ much lower than the dominant Ising interaction, i.e., a spin liquid, makes this compound unique, as it contrasts all other well-studied triangular system with a QSL ground state~\cite{Arh2021}.
Furthermore persisting spin fluctuations were witnessed in this compound at millikelvin temperatures by $\mu$SR~\cite{Arh2021}, which revealed a potentially crucial role of the smaller non-commuting exchange component ${J}_{xy}$ that lifts the degeneracy of the classical Ising model and introduces quantum dynamics~\cite{PhysRevB.63.224401}.
This resembles the situation encountered in the quantum-spin-ice phase of pyrochlores~\cite{gingras2014quantum}
and thus appears to be universal among various highly anisotropic frustrated spin lattices.
The realm of QSLs borne out of large magnetic anisotropy, similarly as observed in the paradigmatic Kitaev honeycomb model, is an exciting new avenue in the field of frustrated quantum magnets. 

\subsection{Thermal conductivity}
{
Experimental techniques based on heat exchange can reveal important information about excitations and correlations in frustrated magnets, e.g., see Section~\ref{majorana}.
Specific heat is the most commonly used method.
However, given its limitations at low $T$, where it is in many instances dominated by nuclear Schottky contributions, and at high-$T$, where the phonon contribution prevails, it can be quite challenging to extract the magnetic contribution unambiguously. 
Furthermore, in many cases, it is difficult to distinguish between the behavior arising from localized and itinerant excitations. 
Unlike specific heat, the alternative technique of thermal conductivity can be very powerful ~\cite{Yamashita2009,Yamashita2008,PhysRevLett.104.066403,Shimozawa2017,PhysRevX.9.041051,PhysRevLett.123.247204,Czajka2021}, as it is not affected by many of these problems and is only sensitive to itinerant excitations (e.g., spinon in QSLs).
It can directly measure spin transport or phonons scattered by the spinons.
 In the case of a gapless QSL state it takes the form~\cite{Li2020} 
	\begin{equation}
	\kappa/T = \kappa_{0} + \beta T^2,
	\label{Ther}
	\end{equation}
where a constant non-zero value of $\kappa_{0}$ at low $T$ indicates the presence of spinon excitations in case of a gapless QSL state with a spinon Fermi surface~\cite{PhysRevLett.117.267202}, whereas a finite spin gap ($\Delta$) in the QSL state leads to $\kappa_{0} \propto \exp (-\Delta/k_{B}T)$.
The second term in Eq.~(\ref{Ther}) accounts for the phonon contribution to thermal conductivity, where the coefficient $\beta$ is constant~\cite{PhysRevLett.117.267202,PhysRevB.96.081111,Li2020,Yamashita2009}.

	\begin{figure}[t]
		\includegraphics[width=0.85\linewidth]{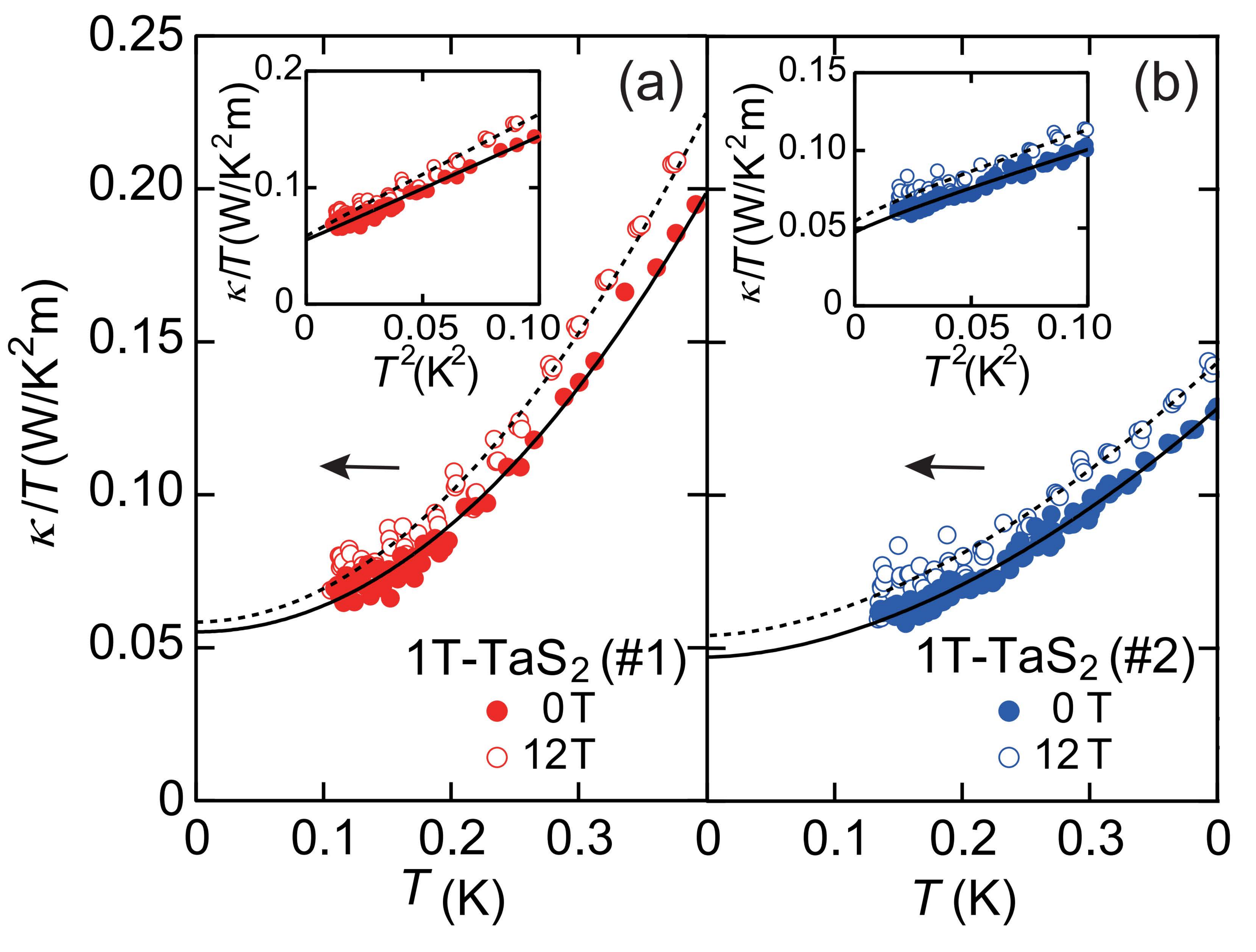} 
		\caption{ \textcolor{black}{The temperature dependence of the thermal conductivity divided by temperature ($\kappa/T$) of the QSL candidate 1T-TaS$_{2}$ under an applied field 0 and 12~T. 
  Panels (a) and (b) refer to two different crystal sizes. 
  The dotted and solid lines correspond to fits using Eq.~(\ref{Ther}), which yield low-$T$ value $\kappa_{0}/T \sim 0.05$~W/K$^{2}$m. The insets display plots of $\kappa/T$ as a function of $T^{2}$.
  Adapted from~\cite{PhysRevResearch.2.013099} with permission from APS.}}
		{\label{TaS2}}
	\end{figure}

Thermal Hall conductivity ($\kappa_{xy}^{2D}$) can also provide important information on charge neutral excitations in frustrated magnets~\cite{Czajka2021}, as different QSL states should show distinct signal~\cite{PhysRevLett.104.066403, Kasahara2018}.
The thermal Hall effect is a thermal analogue of the famous electric Hall effect, where transverse heat flow is induced under an applied thermal gradient and magnetic field. 
In the case of the Majorana fermions in the Kitaev QSL, 
thermal Hall conductivity should be quantized and given by~\cite{Kasahara2018}
\begin{equation}
\frac{\kappa_{xy}^{2D}}{T}=q\frac{\pi k_B^2}{6 \hbar},
\end{equation} 
 with $q = 1/2$.

\begin{figure*}[t]
	\includegraphics[width=0.85\linewidth]{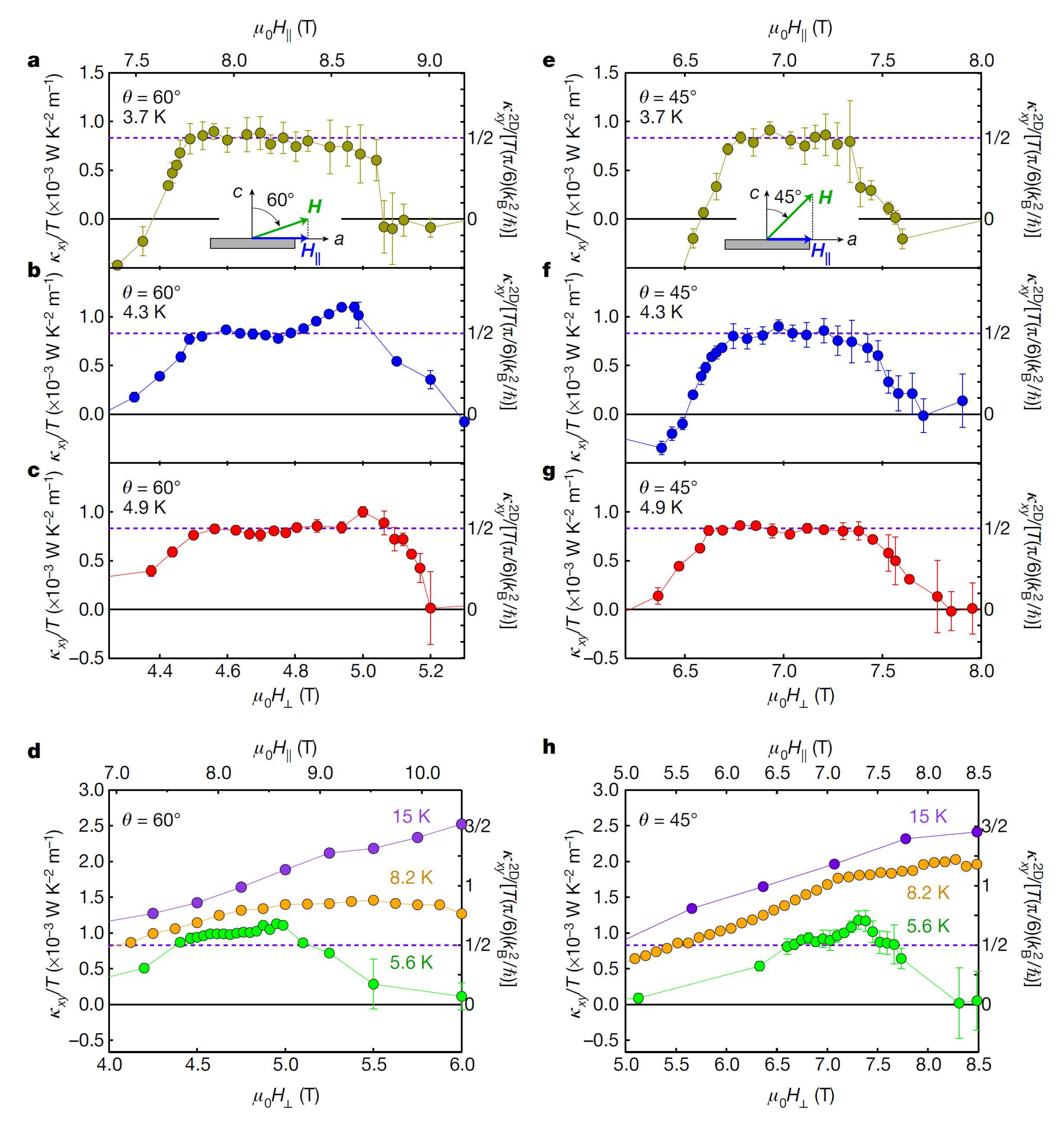} 
	\caption{Thermal Hall conductivity $\kappa_{xy}/T$ of $\alpha$--RuCl$_3$ plotted as a function of $H_{\perp}$ for angles $\theta= 60^{\circ}$ (a--d) and $45^{\circ}$ from the $c$ axis (e--h). 
 Thermal Hall conductivity in units of $\pi k_B^2/6\hbar$ is shown  on the right hand scale and the parallel field component $H_\parallel$ by the top scale. 
 The Half-integer values are represented by  violet dashed lines. 
 Adapted from~\cite{Kasahara2018} with permission from NPG. }{\label{Thermal_conductivity_different_angle}}
\end{figure*}

\subsubsection{Residual linear thermal conductivity in QSL candidates\label{Tcond}}
A finite residual liner term in specific heat was observed in certain QSL candidates, including $\kappa$--(BEDT-TTF)$_{2}$Cu$_{2}$(CN)$_{3}$, herbertsmithite, Zn-brochantite~\cite{li2014gapless,gomilsek2016instabilities}, 1T-TaS$_{2}$, and YbMgGaO$_{4}$~\cite{Valero2021,PhysRevLett.98.107204,Yamashita2008,PhysRevLett.117.267202,PhysRevX.12.011014,PhysRevB.96.195131,Valero2021}, indicating their gapless excitations. 
To determine whether these excitations are itinerant or localized, thermal conductivity measurements have been employed~\cite{PhysRevLett.123.247204,PhysRevB.105.245133,PhysRevX.9.041051,cryst12010102,Yamashita2022}.
The transition metal dichalcogenide 1T-TaS$_{2}$, a QSL candidate on the triangular lattice~\cite{law20171t,PhysRevLett.121.046401}, is an illustrative example.
It exhibits incommensurate charge-density wave with metallic behavior at high $T$.
However, below ${\sim}200$~K, a fully commensurate charge-density wave phase develops in which Ta atoms reside on a triangular lattices of star-of-David clusters, each composed of 13 Ta atoms and having a single unpaired electron (i.e., a total spin-1/2)~\cite{doi:10.1126/sciadv.1500606,Vaskivskyi2016,Valero2021}.  
Despite the strong exchange interaction of several hundreds of	Kelvin between these spins, no evidence of long range magnetic ordering was found in this material by various techniques, including $\mu$SR and NMR~\cite{Klan2017}.  
Furthermore, the NMR spin--lattice relaxation rate  $1/T_{1}$ was found to exhibit power-law behavior, indicating the presence of gapless excitations~\cite{Klan2017}. 
Recent thermal conductivity measurements on a high-quality single crystal of 1T-TaS$_{2}$ have revealed a finite value of $\kappa_{0}$ at low $T$ (see Fig.~\ref{TaS2}), which supports the conjecture of gapless excitations~\cite{PhysRevResearch.2.013099}. 

However, it has to be noted that thermal conductivity can be very sensitive to crystal defects and cooling rates~\cite{Yamashita2022,PhysRevLett.123.247204}, therefore, it is often challenging to detect itinerant excitation unambiguously.
A recent systematic investigation of the pure compound 1T-TaS$_{2}$ and a partially substituted compound 1T-TaS$_{1-x}$Se$_{x}$ demonstrated that the finite value of $\kappa_{0}$ found in pure 1T-TaS$_{2}$, can be tuned to zero by adding a certain amount of defects~\cite{PhysRevResearch.2.013099}. 
This finding potentially has important implications for other QSL candidates where non-zero $T$-linear terms were observed in specific heat, but $\kappa_{0}$ was found to be zero, such as in $\kappa$--(BEDT-TTF)$_{2}$Cu$_{2}$(CN)$_{3}$ and YbMgGaO$_{4}$~\cite{PhysRevLett.117.267202,Li2015,Yamashita2009}.
This may thus well be an effect of disorder.
Likewise, in the kagome lattice QSL herbertsmithite, ZnCu$_{3}$(OH)$_{6}$Cl$_{2}$, where a large amount of defects is intrinsically present, there is also no significant contribution to heat conductivity in the QSL regime, as no residual $\kappa_{0}$ term could be observed in zero applied magnetic field at low $T$~\cite{PhysRevB.106.174406,PhysRevB.107.054434}. 
On the other hand, there is still a discrepancy in thermal conductivity results on the triangular lattice EtMe$_{3}$Sb[Pd(dmit)$_{2}$]$_{2}$, as some measurements showed the residual $\kappa_0/T$ term while others did not~\cite{doi:10.1126/science.1188200,PhysRevX.9.041051,PhysRevLett.123.247204}. 
It has been suggested that defects are to be blamed, as differences in long-range coherence in the electronic state could results from disorder in the crystals that appears during the cooling process~\cite{doi:10.7566/JPSJ.88.083702}.}


\subsubsection{Thermal Hall effect in $\alpha$--RuCl$_3$}
The fractional thermal Hall effect was observed in $\alpha$--RuCl$_3$ in the Kitaev paramagnetic region, i.e., for $T_N < T < T_H$~\cite{PhysRevLett.120.217205,Kasahara2018} (see Subsection~\ref{majorana}). 
Thermal Hall conductivity $\kappa_{xy}$ is negative in the conventional paramagnetic region $T > T_H$ and gains a positive value in the Kitaev paramagnetic region, which indicates the appearance of unusual emergent itinerant excitations in the form of Majorana fermions. 
Due to magnetic anisotropy, a magnetic field applied within the $ab$ plane can suppress magnetic order significantly, whereas a perpendicular field has less of an effect on the low-$T$ ordering temperature $T_N$. 
External magnetic fields of different magnitudes were applied at different angles $\theta$ with respect to the $c$-axis of the single crystal. 
It was observed that at $\theta = 45^{\circ}$ the ordering temperature vanishes for an in-plane component of the applied magnetic field of $\mu_0H_{\parallel}= 6$~T. 
So, in order to observe thermal conductivity in the QSL region in the lower-$T$ regime various fields above  6~T were applied at $\theta = 45^{\circ}$. Fig.~\ref{Thermal_conductivity_different_angle} demonstrates the field dependence of $\kappa_{xy}/T$ with fields applied at different angles ($45^{\circ}$ and $60^{\circ}$). 
In the ordered state, $\kappa_{xy}$/$T$ shows a plateau in the field range 4.5~T $ < \mu_0H_{\perp} < $ 4.8--5.0~T  for $\theta = 60^{\circ}$ and  6.8~T $<\mu_0H_{\perp}< 7.2$--$7.4$~T for $\theta = 45^{\circ}$ (see Fig.~\ref{Thermal_conductivity_different_angle}). 
Note that for the integer quantum Hall effect, this type of plateau also appears, but in $\alpha$--RuCl$_3$ the plateaus occur at approximately half of the value reported for integer quantum Hall systems. 
This  half-integer quantum Hall conductance persists up to 5.5~K, increases rapidly above this temperature, then starts to decrease above 15~K, and, finally, vanishes around 60~K, in agreement with the strength of Kitaev interaction in $\alpha$--RuCl$_3$.

\textcolor{black}{ The initial report of a half-quantized thermal Hall conductivity (Fig.~\ref{Thermal_conductivity_different_angle}) as evidence for Majorana fermions in $\alpha$--RuCl$_{3}$~\cite{Kasahara2018} sparked a significant amount of attention. 
However, several contradictory results have been reported for this material subsequently~\cite{Czajka2023,PhysRevB.106.L060410,PhysRevB.102.220404}.
The difficulty in reproducing thermal conductivity measurements of $\alpha$--RuCl$_{3}$ is potentially due to the different stacking arrangements possible in this van der Waals material when prepared with different crystal growth techniques~\cite{Kee2023}. 
In contrast to the result in Ref.~\cite{Kasahara2018}, Phuan \textit{et al}.~have observed  non-quantized thermal Hall
conductance that could be ascribed to topological bosonic modes in a Chern insulator-like model~\cite{Czajka2023}. 
The search for unambiguous fingerprints of a Kitaev QSL state thus remains a persistent challenge. 
Firmer evidence might come from thermal conductivity measurements of more perfect materials or, perhaps, will await the development of more sophisticated spectroscopic methods that will be sensitive to entanglement in quantum materials~\cite{PhysRevLett.118.227201}.}

\section{\textbf{Topology in frustrated magnets}}\label{top}
As we discussed in the previous sections, topology plays a crucial role in frustrated magnetism. 
In spin ices it manifests as non-local string-order parameters, including Dirac strings discussed in Subsection~\ref{SI_monopoles}~\cite{Castelnovo_2012}. 
In QSLs it arises from the generally anyonic nature of their spin excitations, as seen most clearly in the emergence of non-Abelian anyons in the Kitaev honeycomb model discussed in Subsection~\ref{majorana}~\cite{Jansa2018,Do2017,KITAEV20062}, emergent non-local string order parameters~\cite{RevModPhys.89.025003}, topological ground-state degeneracy in gapped QSLs, etc. 
In fact, most quantum phases of matter, that are often realized in frustrated magnetic systems, can be classified by topological order~\cite{RevModPhys.89.041004,wen2019choreographed,wen2004quantum}. 
This is intimately connected to long-range quantum entanglement of these states, and their classification touches upon deep mathematical concepts like braid groups, algebraic topology, and higher-category theory. 
However, topology is applicable even more broadly, with important topological effects found also in complex, topologically non-trivial spin textures that can arise in frustrated magnets~\cite{lancaster2019skyrmions}.
Due to their topological stability, such spin textures are often highly relevant for practical applications. 
Below, we briefly introduce a general organizing principle in topology, the Berry phase and its associated Berry curvature, as well as the anomalous Hall effect that arises from it in non-collinear spin textures. 
Finally, we introduce one of the most promising examples of topologically-nontrivial spin textures, that of magnetic skyrmions, which are stabilized by frustrated magnetic interactions. 

\begin{figure}[t]
\includegraphics[width=0.5\textwidth]{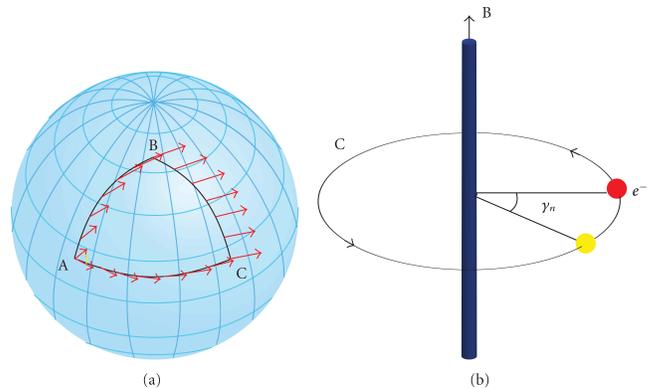}
\caption{(a) Illustration of parallel transport of a vector on the surface of a sphere. Red arrows denotes the vector as it is parallel-transported around
	a loop (A $\rightarrow$ B $\rightarrow$ C $\rightarrow$ A) while keeping the vector always tangential to the surface. After one loop its direction has changed due to the geometrical
	curvature of the enclosed surface around which it was transported. (b) Aharonov--Bohm effect: an Electron encircling a confined magnetic field $\mathbf{B}$ via a loop C acquires a Berry phase $\gamma_n$. The magnetic field on the path of the electron is negligible but interaction of magnetic  vector potential with electron gives rise to the phase shift.  Adapted from~\cite{Boldrin2012} with permission from Hindawi Publishing Corporation.}{\label{Berry}}
\end{figure}
\subsection{Berry Phase, Curvature and anomalous Hall Effect}\label{bbbb}
\begin{figure}[b]
	\includegraphics[width=8cm]{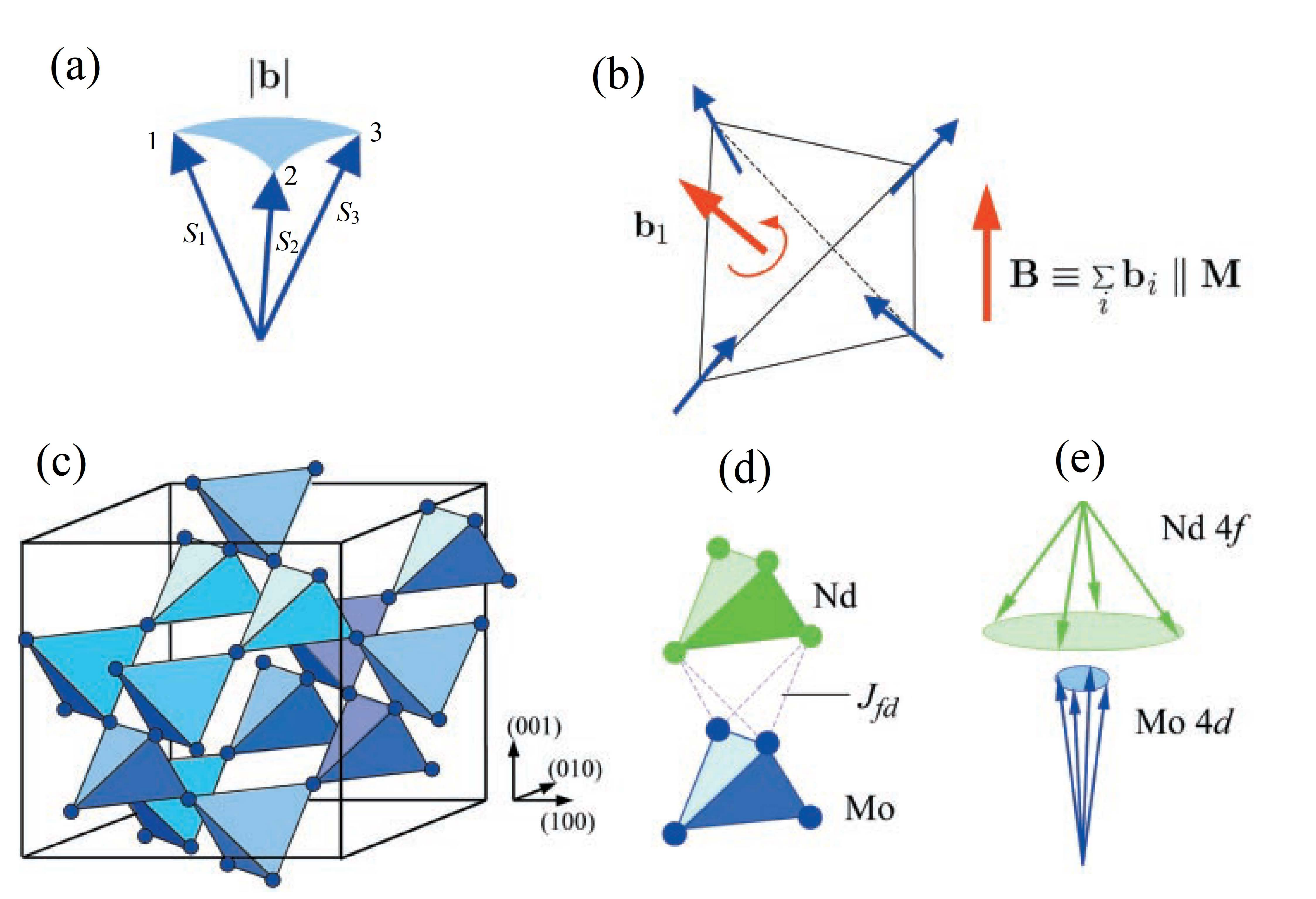} 
	\caption{(a) Fictitious magnetic field of magnitude $|\mathrm{b}|$ is produced from the spin chirality  of three spins ($S_{1}$, $S_{2}$, and $S_{3}$) which are arranged in a non-collinear configuration. 
		The shaded region is the solid angle subtended by these three spins. 
		(b) For the 2-in--2-out structure of the pyrochlore lattice, flux penetrates along $b_1$ and the total fictitious field is a vector sum of all such fluxes trough all plaquette. 
		(c) Schematic presentation of the Mo sublattice of Nd$_2$Mo$_2$O$_7$ and (d)  the relative positions of Nd and Mo sublattices. 
		(e) There are four Nd$^{3+}$ 4\textit{f} moments (\textbf{n$_i$}) and four Mo$^{4+}$ 4\textit{d} moments (\textbf{m$_i$}) in a magnetic unit cell of  Nd$_2$Mo$_2$O$_7$. Adapted from~\cite{Taguchi2573} with permission from AAAS.{\label{Nd2Mo2O7}}}.
\end{figure}

The concept of a Berry or geometrical phase can be explained by analogy with parallel transport on curved surfaces.
When a vector on the surface of a sphere is tangentially (parallelly) transported along a closed loop, it acquires an angular
(or phase) shift upon returning to its initial position, as depicted in Fig.~\ref{Berry}. 
This geometry-dependent phase shift is called the Berry phase~\cite{berry1984quantal}. 
More specifically, when the system evolves adiabatically from an initial eigenstate $\ket{n(\mathbf{R}(0)}$ at $t = 0$ its wavefunction can be written as~\cite{berry1984quantal,xiao2010berry}
\begin{equation}{\label{Berry1}}
	\ket{\Psi_n(t)}= \exp \left(-\frac{i}{\hbar}\int_0^t E_n(\mathbf{R}(t^{'}))\dd t^{'}\right) e^{i\gamma_{n}(t)} \ket{n(\mathbf{R}(t)},
\end{equation} 
where $\mathbf{R} = \mathbf{R}(t)$ represents a vector of all parameters that can be varied to change the Hamiltonian $\hat{H}(\mathbf{R})$, and $\ket{n(\mathbf{R}(t)}$ are its instantaneous eigenstates with eigenenergies $E_n(\mathbf{R})$, i.e.,
\begin{equation}
	\hat{H}(\mathbf{R})\ket{n(\mathbf{R})}=E_n(\mathbf{R})\ket{n(\mathbf{R})}.
\end{equation} 
The first exponential factor in Eq.~(\ref{Berry1}) describes the dynamical phase factor, while the second one, $\gamma_{n}(t)$, is known as the Berry phase. 
{The Berry phase was long thought to be unobservable, as for a generic time-dependent change of parameters $\mathbf{R}(t)$ it depends on the arbitrary (gauge) choice of phase prefactors of the eigenstates $\ket{n(\mathbf{R})}$, which can be chosen independently at each $\mathbf{R}$~\cite{xiao2010berry}.} 
However, Berry's main contribution was to consider what happens when the parameters $\mathbf{R}(t)$ follow a closed loop $C$, i.e., from time $t=0$ to $t=T$ they completes one full circuit such that $\mathbf{R}(T)=\mathbf{R}(0)$. 
Remarkably, it turns out that the Berry phase does not always return to its original value, i.e., after completing a circuit $C$ we can have the situation where $\gamma_{n}(T) \neq \gamma_{n}(0)$. 
{Furthermore, this change is gauge-invariant, and thus observable, and is} given  explicitly by~\cite{xiao2010berry}
\begin{equation}
	\gamma_{n}(C)=i\oint_C \bra{n(\mathbf{R})}\ket{\nabla_{\mathbf{R}}n(\mathbf{R})} \cdot \dd\mathbf{R}.
\end{equation}
{Furthermore, by Stokes' theorem this phase change can be related to Berry curvature, which is defined as~\cite{xiao2010berry}
\begin{equation}
    \Omega_{\mu\nu}^n(\mathbf{R}) = i \left[ \bra{\frac{\partial n(\mathbf{R})}{\partial R^\mu}}\ket{\frac{\partial n(\mathbf{R})}{\partial R^\nu}} - (\mu \leftrightarrow \nu) \right] ,
\end{equation}
via 
\begin{equation}
	\gamma_{n}(C) = i \int_S \dd R^\mu \wedge \dd R^\nu \frac{1}{2} \Omega_{\mu\nu}^n(\mathbf{R}) ,
\end{equation}
where $S$ is an arbitrary surface enclosed by the path $C$, and $\mu$ and $\nu$ are indices of the parameter vector $\mathbf{R}$. 
Unlike the Berry phase $\gamma_{n}$, which becomes gauge-invariant only for closed loops, the Berry curvature $\Omega_{\mu\nu}^n(\mathbf{R})$ is gauge-invariant, and thus observable, at every individual point $\mathbf{R}$ in parameter space. 
In many systems its effect on quasiparticle excitations can be understood as an effective emergent magnetic field~\cite{xiao2010berry}. 
}
A well known phenomenon related to the Berry phase is the Aharonov--Bohm effect~\cite{PhysRev.115.485}. Phenomenologically, it states that  when an electron encircles a path $C$ around a confined  magnetic  field (Fig.~\ref{Berry}) it acquires a  phase shift (i.e.,Berry phase)
\begin{equation}
	\gamma_{n}=2\pi N(C)\frac{q}{h}\Phi,
\end{equation}
where $N(C)$ is the number of windings, $\Phi$ is the magnetic flux enclosed  by the loop, $q$ is the elementary charge, and $h$ is the Planck constant. 
{As $N(C)$ is an integer, this Berry phase is a topological quantity.}
Similarly, in the absence of any external magnetic field, when an itinerant electron hops around a non-collinear spin texture it can acquire a non-zero Berry phase. 
For example, if an electron hops around a loop (\textcolor{black}{1$\rightarrow$ 2 $\rightarrow$ 3 $\rightarrow$ 1}) in a non-collinear spin arrangement it acquires a Berry phase that is the solid angle subtended by constituent spins and is proportional to $\mathbf{S_1}\cdot\mathbf{S_2} \times \mathbf{S_3}$ (Fig.~\ref{Nd2Mo2O7}(a))~\cite{lancaster2019skyrmions}. 
The Berry phase induces a fictitious (emergent) magnetic field which in turn produces an effective Lorentz force on it. 
As a result, the trajectory of the electron picks up a transverse component of momentum that results in transverse voltage drop in a Hall-effect experiment. 
The transverse resistivity  can in general be written as~\cite{RevModPhys.90.015005}
\begin{equation}
	\rho_{yx}= R_0B+R_sM+\rho_{yx}^T,
\end{equation}
where $R_0$ and $B$ are the ordinary Hall coefficient and magnetic field density, respectively, whereas $R_s$ and $M$ are the anomalous Hall coefficient and magnetization, respectively. 
The first two contributions are often observed  in  ferromagnetic metals, such as Fe, Co or Ni~\cite{RevModPhys.25.151,RevModPhys.82.1539,A_Kundt_1893}.  
 However, in the absence of both the external magnetic field and spin–orbit
coupling, a (topological) Hall effect can still be realized if the magnetization of the compound is non-uniform due to it
hosting a non-trivial spin texture, such as a non-collinear antiferromagnetic ordering~\cite{Taguchi2573}, which is denoted by the third term $\rho_{yx}^T$.
\begin{figure*}
	\centering
	\includegraphics[width=0.8\textwidth]{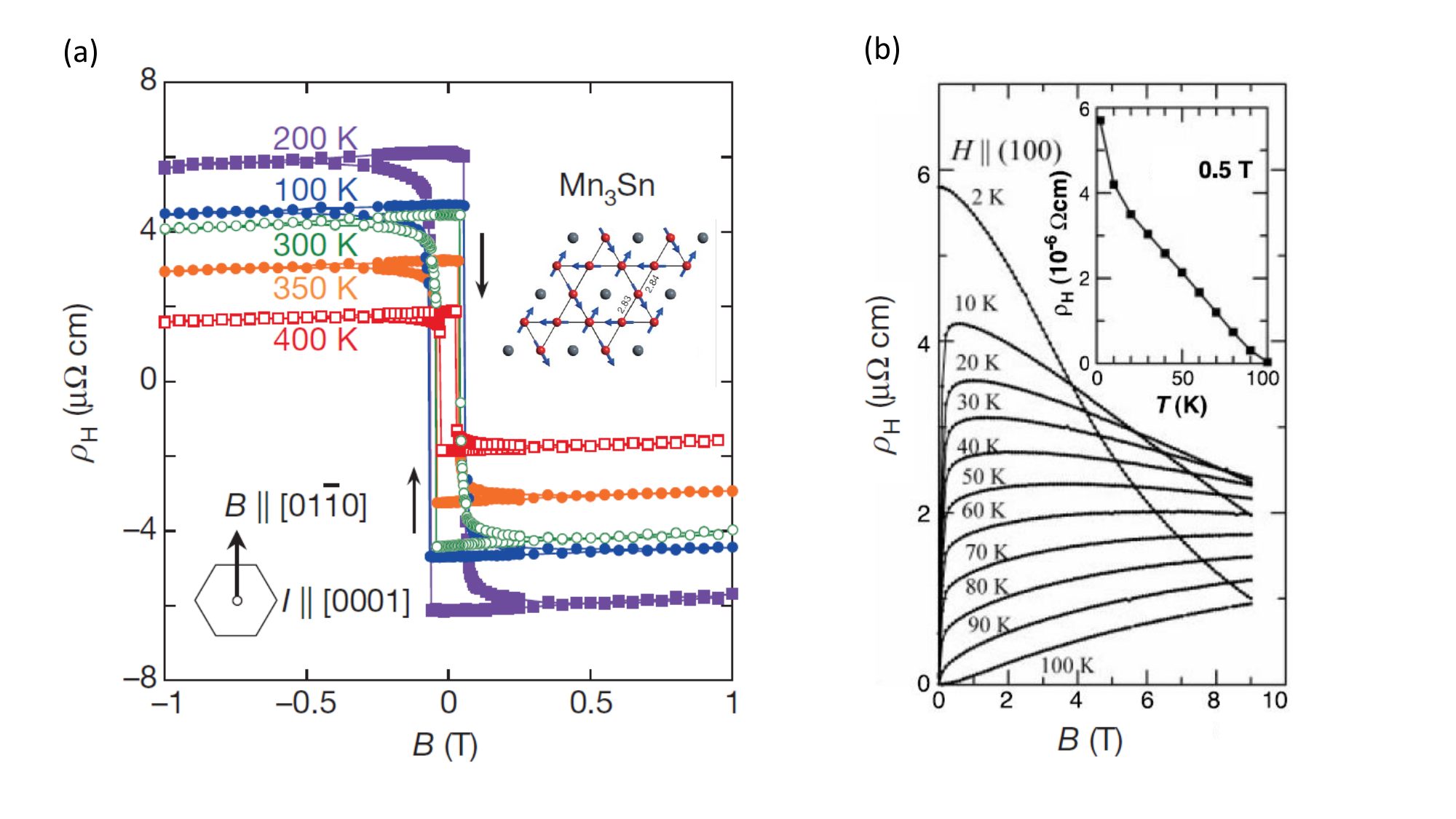} 
	\caption{(a) Field dependent Hall resistivity ($\rho_{xy}$) of the frustrated kagome compound Mn$_3$Sn at various temperatures. 
		The inset shows the kagome lattice of Mn sites. 
		Magnetic field $B$ and electric current $I$ are applied parallel to the directions [01$\bar{1}$0] and [0001], respectively. 
		Adapted from~\cite{nakatsuji2015large} with permission from NPG. 
		(b) Hall resistivity $\rho_{H}$ vs. applied magnetic field at different temperatures for $H\parallel$(100) in Nd$_2$Mo$_2$O$_7$. 
		The sharp decrease of $\rho_H$ at low temperature and under high applied fields is due to anomalous Hall effect. 
		The temperature dependence of $\rho_H$ is shown in the inset for 0.5~T. 
		Adapted from~\cite{Taguchi2573} with permission from AAAS.}
	{\label{AHEF}}
\end{figure*} 
This spin chirality can arise at a finite $T$ from topological spin texture called skyrmions (see Subsection~\ref{skyy}) for example and is responsible for transverse conductivity systems hosting them~\cite{PhysRevLett.83.3737}.

Frustrated magnetic materials are ideal hosts of non-zero Berry curvature and associated exotic topological phenomena.
In manganites, e.g., La$_\frac{2}{3}$(Ca,Pb)$_\frac{2}{3}$MnO$_3$~\cite{PhysRevB.59.11155}, and in conventional ferromagnets the anomalous Hall term vanishes at low $T$~\cite{PhysRevLett.83.3737,SMIT1955877}. 
On the other hand, in the pyrochlore compound Nd$_2$Mo$_2$O$_7$, the anomalous Hall effect increases continuously with decreasing $T$, which can be explained within the picture of the Berry phase~\cite{PhysRevB.62.R6065}. 
Fig.~\ref{Nd2Mo2O7}(e) shows that each magnetic unit cell of Nd$_2$Mo$_2$O$_7$ carries  four inequivalent Nd$^{3+}$ 4\textit{f} moments (\textbf{n$_i$}) and four Mo$^{4+}$ 4\textit{d} moments  (\textbf{m$_i$}) which form a stable umbrella-like structure where (\textbf{n}$_{i}$ $-$ \textbf{n}) $\perp$ \textbf{n} and (\textbf{m}$_{i}$ $-$ \textbf{m}) $\perp$ \textbf{m}. 
Here \textbf{n} and \textbf{m} denote the average moments of four \textbf{n}$_{i}$ 
and \textbf{m}$_{i}$ spins, respectively.
Because of the large tilting angle of Mo spins  in weak magnetic field, this field induces  a large spin chirality,  which in turn gives rise to large anomalous Hall effect. 
The tilting angle of Mo spins reduces in higher magnetic fields leading to a diminishing anomalous Hall response (Fig.~\ref{AHEF}(b)).

Another growing area of interest is the anomalous Hall effect in spin--orbit driven frustrated intermetallic compounds~\cite{PhysRevLett.112.017205,nakatsuji2015large,Nayake1501870,Ghimire2018,PhysRevB.94.075135,doi:10.1063/1.5088173,doi:10.1063/1.4943759}.
It can be observed when a particular symmetry is broken in a system. 
A system with both unbroken
spatial-inversion and unbroken time-reversal symmetry does not possess non-zero Berry curvature. For instance, in the frustrated  magnet Mn$_3$Ir, Mn atoms form 2D kagome planes that are stacked along the (111) direction.  
The mirror-reflection symmetry  ($\mathcal{M}$) with respect to the kagome plane flips all the in-plane spins and the time reversal symmetry ($\mathcal{T}$) rotates the spins in the opposite direction. Thus, the combination (product) of $\mathcal{T}$ and  $\mathcal{M}$ is not broken~\cite{PhysRevLett.112.017205} if we exclude spin--orbit coupling. 
However, a suitably oriented non-zero external magnetic field can break $\mathcal{T}$ symmetry with preserving  $\mathcal{M}$ symmetry.
As a result, the product of the two symmetries can be broken and can result in non-zero
Berry curvature appearing. Furthermore, even in the absence of an external magnetic field, non-zero spin–orbit coupling
can break  $\mathcal{M}$ symmetry without breaking $\mathcal{T}$ symmetry. 
As a result, a non-zero Berry curvature can arise in the momentum space and this leads to large anomalous Hall effect,  as in the case of Mn$_3$Ir. 
In this compound, the theoretical  Hall conductivity was estimated to be $\sigma = 218~\Omega^{-1} \mathrm{cm}^{-1}$~\cite{PhysRevLett.112.017205} with a small net magnetic moment of ${\sim}0.02 \mu_B$ per formula unit along the direction perpendicular to the kagome plane. However, the small magnetic moment does not
have any considerable effect on the observed anomalous Hall effect in this compound. Namely, if the magnetic moments
were forced to lie in the kagome plane, the Hall predicted conductivity would change only by $1~\Omega^{-1} \mathrm{cm}^{-1}$
 i.e., more
than two magnitudes less than the total Hall conductivity. Experimentally, a large anomalous Hall conductivity of about $20~\Omega^{-1} \mathrm{cm}^{-1}$  was also found in the non-collinear antiferromagnet Mn$_3$Sn~\cite{nakatsuji2015large} and $50~\Omega^{-1} \mathrm{cm}^{-1}$ in Mn$_3$Ge at room $T$. Both of these conductivities are consistent with theoretical predictions~\cite{Nayake1501870,K_bler_2014}.
Like in Mn$_{3}$Ir, these compounds there
host a small magnetic moment. However, the presence of a small soft ferromagnetic moment in Mn$_{3}$Ge and Mn$_{3}$Sn is
unique as it can cause a change in the sign of the overall Hall effect even under a weak magnetic field, which makes
these materials highly promissing for potential future application in spintronics. 

\subsection{Magnetic skyrmions}\label{skyy}
Skyrmions are localized topological textures prevalent not just in condensed matter~\cite{lancaster2019skyrmions,fert2017magnetic,posnjak2016points}, but also in other fields, such as cosmology and particle physics. 
In magnetic materials, skyrmions can be stabilized by a competition between frustrated magnetic interactions, with prominent examples being: (i) chiral skyrmions arising from a competition between Heisenberg exchange and anisotropic DM interactions in systems that break space-inversion symmetry~\cite{Nagaosa2013,Fert2013,fert2017magnetic}, and (ii) frustrated skyrmions that can arise from a competition between long-range RKKY and biquardartic exchange interactions in centrosymmetric metals~\cite{doi:10.1063/1.1654968,Yu2014,doi:10.1002/adma.201600889,milde2013unwinding}.
Both kinds of skyrmions strongly respond to electromagnetic stimuli, both magnetic fields and electrical currents, but due to their topological properties are protected from untangling into a topologically-trivial underlying state, and thus behave as stable extended quasiparticles  (solitions)~\cite{lancaster2019skyrmions,Nagaosa2013,Fert2013,fert2017magnetic}. 

\begin{figure}[b]
	\includegraphics[width=0.45\textwidth]{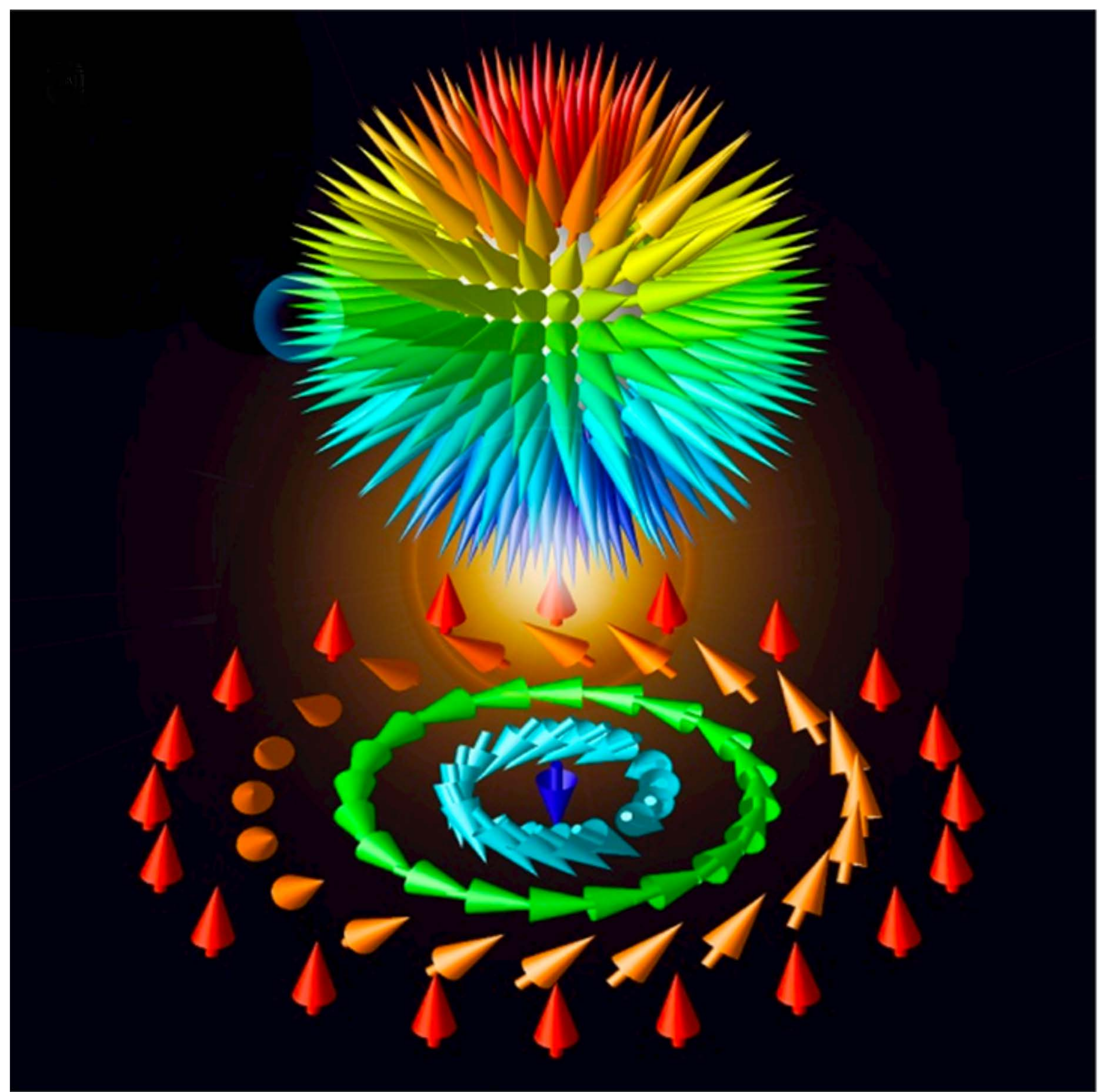}
	\caption{Skyrmion topological charge (top)  representation (bottom). When the moment vectors are assembled with the identical origin, the constituent spin moments of the skyrmion surround the sphere precisely once (top).
 Adapted  from~\cite{doi:10.1021/acs.chemrev.0c00297} with permission from ACS.{\label{scr}}}
\end{figure}

The skyrmion concept was first introduced by Tony Skyrme already in 1960, in an attempt to model the stability of hadrons in nuclear physics via topological arrangement of four-dimensional vector field in space-time~\cite{skyrme1962unified,doi:10.1098/rspa.1961.0018,skyrme1994non}.
Later, the search for a variant of this topological configuration in spin space has attracted a lot of attention in the condensed matter community.
Magnetic skyrmions are  2D whirl-like topological spin textures in a host material~\cite{BOGDANOV1994255,Robler2006,Zhao4918,PhysRevLett.122.107203,Malsch2020,Yu2010real,Yu2010near},
\textcolor{black}{ 
in
which magnetic moment directions wrap around a unit sphere in a topologically non-trivial fashion (see Fig.~\ref{scr}).  
Each
skyrmion-like texture can be characterized by a non-zero topological skyrmion number (also known as its topological
charge) defined  by~\cite{lancaster2019skyrmions} 
\begin{equation}
N_{\rm sk}=\frac{1}{4\pi}\int\int \textbf{n}\cdot \left( \frac{\partial \textbf{n}}{\partial x}\times \frac{\partial \textbf{n}}{\partial y}\right) \dd^{2}\textbf{\rm \textbf{r}} = p W,
\end{equation}
where $\textbf{n}(\textbf{r}) = \textbf{n}(x,y)$  is a unit vector average spin describing the direction at position \textbf{r}, $p$ is defined as the direction of spin at the origin ($p = 1, -1$ for up and down direction, respectively), and $W$ represents the  winding number of the spin
texture $\textbf{n}(\textbf{r})$, which counts how many times its normalized magnetic moments topologically-non-trivially wrap around the
unit sphere~\cite{Nagaosa2013,rajaraman1982solitons,doi:10.1080/00018732.2012.663070}.	
 Skyrmions, which correspond to $N_{\rm sk}$ = $-1$, come a variety of configurations with diverse inner
spin orientations. To differentiate their various spin modulations, two basic varieties of cylindrically-symmetric skyrmions
are defined as (i) Bloch skyrmions, where spins rotate perpendicular to the radial direction as one moves radially out
from the central spin, and (ii) N\'eel skyrmions, where spins rotate in the radial plane as one follows the same radial
path~\cite{PhysRevLett.128.227204,PhysRevLett.102.186602,doi:10.1126/sciadv.abg9551}. 
Closely related are also antiskyrmion $N_{\rm sk} = + 1$, where the Bloch and N\'eel-type moment orientations
interchange. Other, more complicated spin textures with distinct topological charges, such as merons and antimerons
 ($N_{\rm sk}$ = $\mp$ 1/2) ~\cite{Yu2018} and biskyrmions ($N_{\rm sk}$ = $-$ 2) ~\cite{doi:10.1063/1.5048972} can also be found. In 3D there are also additional topological spin
textures, such the hedgehog  ($N_{\rm sk}$ = $+$ 1), antihedgehog ($N_{\rm sk} = -1$), and hopfions, to name just a few ~\cite{PhysRevB.96.220414,doi:10.1021/acs.chemrev.0c00297}.}

$N_{\rm sk}$ is a topological invariant, meaning that its value cannot change under arbitrary continuous deformations of the spin
texture. This topological protection guarantees that skyrmions are long-lived stable quasiparticles and that they usually
have high energy barriers for their creation or destruction~\cite{PhysRevB.99.174421}. 
Experimentally, magnetization, thermodynamic, and
transport measurements can provide signatures of skyrmions, while direct evidence of skyrmions can be obtained by
Lorentz transmission electron microscopy (LTEM) real-space imaging~\cite{Yu2012,Yu2013,Yu2011,Zhao4918}, at least in thin films, as well as by
small angle neutron scattering (SANS)~\cite{PhysRevB.99.174421,PhysRevB.100.014425}, and by small angle and resonant X-ray scattering (SAXS)~\cite{PhysRevResearch.2.013096}, which
provide reciprocal-space information.

\begin{figure*}[t]
	\includegraphics[width=0.75\textwidth]{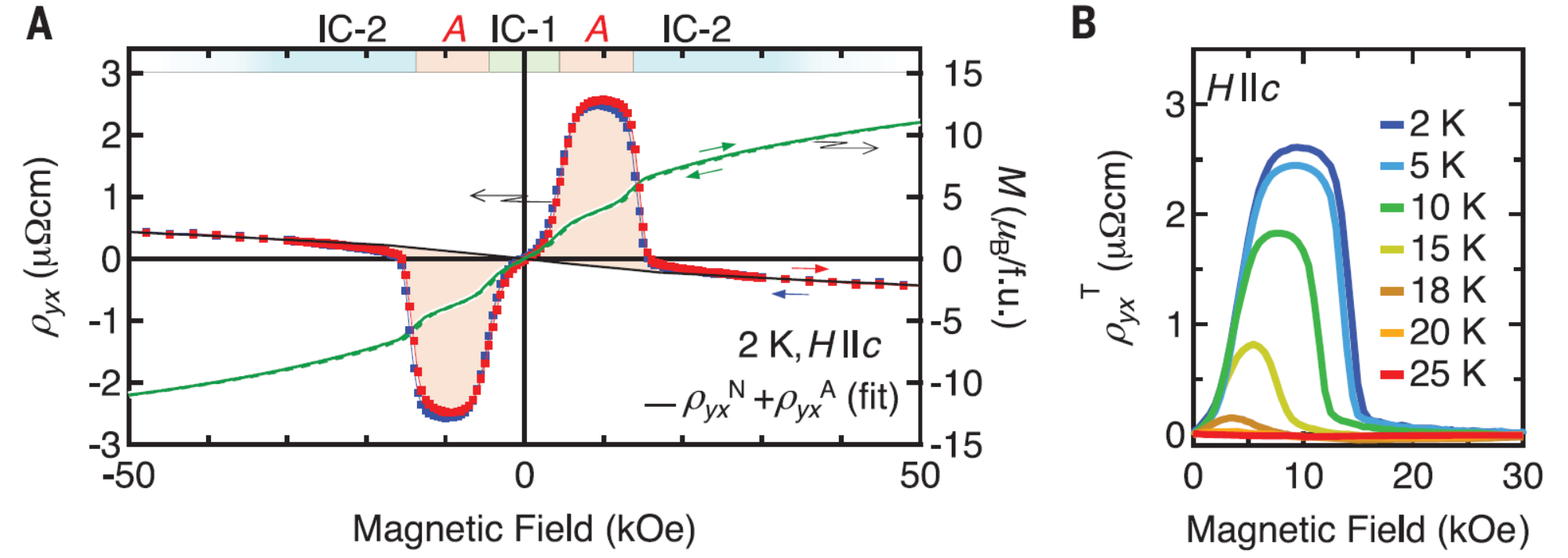}
	\caption{(a) Magnetic field dependence of the topological Hall resistivity in the frustrated triangular lattice Gd$_{2}$PdSi$_{3}$. (b) The value of $\rho_{xy}^{T}$ enhances around 10~K due to the presence of a Bloch skyrmion spin texture.  Adapted from~\cite{Kurumaji914} with permission from AAAS.}{\label{sky}}
\end{figure*}
The existence of such exotic and versatile topological objects was first identified in chiral magnets with non-centrosymmetric crystal lattice, such as MnSi~\cite{hlbauer915,PhysRevB.80.054416}, FeGe~\cite{Yu2010near}, and MnGe~\cite{Tanigaki2015}. 
The crystal structure of these
magnets breaks space-inversion symmetry and thus allows for net-non-zero antisymmetric anisotropic DM interactions,
arising from relativistic spin–orbit coupling, that can stabilize magnetic skyrmions under a narrow range of external
magnetic field at elevated $T$~\cite{DZYALOSHINSKY1958241,PhysRev.120.91,BOGDANOV1994255,Yu2011}.
Phenomenologically, it is observed that the skyrmion phase can exist in ferromagnetic  metals with chiral interactions and describes all possible phases within a continuum model defined by free energy $F$ containing a gradient term, an anisotropic term $D\mathcal{L}(m)$ and a Landau term $F_{0}(m)$, 
\begin{equation}
	F=Am^{2}\sum_{ij}(\partial_{i}n_{j})^{2}+\eta A (\nabla m)^{2}+D\mathcal{L}(m)+F_{0}(m),
\end{equation}
where $A$ is the isotropic exchange parameter and $m$ is the magnetization along the unit vector $\boldsymbol{\mathbf{n}}$, while $\eta$ is the ratio between the  longitudinal and the transverse magnetic stiffness, which can be obtained experimentally~\cite{Robler2006}.
 skyrmion phases have lately also been experimentally detected in some oxide-based frustrated
magnets, such as Cu$_2$OSeO$_3$ and VOSe$_{2}$O$_5$~\cite{Seki2012,PhysRevLett.108.237204,skyrme1994non,PhysRevMaterials.2.111402,PhysRevB.103.024428,PhysRevB.100.014425,PhysRevB.99.174421,PhysRevB.100.224426,PhysRevB.94.094409}, ferroelectric compounds~\cite{Nahas2015}, semiconductors~\cite{Kezsmarki2015}, and magnetic multilayers~\cite{Jiang283,Woo2016,Romming636}.

Recently, a N\'eel skyrmion phase has also been observed in a spinel compound GaV$_4$S$_8$ with additional cycloidal and ferromagnetic phases~\cite{Ruffe1500916,PhysRevResearch.2.032001}. GaV$_{4}$S$_{8}$ belongs to the lacunar spinel family AM$_4$X$_8$ ($A$ = Ga and Ge; $M$ = V, Mo; $X$ = S and Se).
Due to strong correlation effects, these spinels  harbor interesting phenomena, including large negative magnetoresistance~\cite{Dorolti2010}, orbital-driven ferroelectricity~\cite{PhysRevLett.113.137602}, and pressure-induced superconductivity~\cite{Pocha2000}. 
Furthermore, it is predicted that enhanced electrical polarization occurs close to the skyrmions core of  GaV$_{4}$S$_{8}$, which is due to an exchange striction mechanism, opening up the possibility of manipulating electric polarization by magnetic skyrmions and vice versa~\cite{Ruffe1500916,doi}. 
In addition to the pristine material, Se doped  GaV$_4$S$_{8-y}$Se$_y$ has also been studied. 
For $y$ close to 0 or 8 the spin dynamics closely resemble that of skyrmions. 
This behaviour, however, extends to much lower $T$ than in pristine samples, indicating the presence of (at least) low-$T$ skyrmion precursor states under low levels of quenched substitutional disorder~\cite{PhysRevResearch.2.032001}. 
This dynamical skyrmion-like signature is accompanied by the formation of localized regions of increased static spin density. 
Under higher levels of substitution $y = 2$ or 4  skyrmion and long-range-ordered phases give way to a frozen disordered spin-glass state at low $T$~\cite{PhysRevB.98.054428}. 
Furthermore, doping of skyrmion materials often leads to
increased lifetimes of metastable skyrmion phases~\cite{PhysRevB.100.014425,PhysRevB.99.174421,PhysRevResearch.2.013096}, which are obtained by rapid cooling of the system from
the (usually) rather high-$T$ skyrmion phase (close to the magnetic ordering temperature) to much lower temperatures.
Such a cooling protocol freezes-in topologically-protected metastable skyrmions, which gain extremely long life-times for
$T$ below the energy barrier for their unwinding and can drastically increases their region of (meta)stability at low $T$ in
the $T$–$B$ phase diagram.

In contrast to DM interactions in non-centrosymmetric chiral magnets, uniaxial magnetic anisotropy with frustration provides another route to stabilize magnetic skyrmions  in materials with centrosymmetric spin lattices~\cite{doi:10.1063/1.1654968,Yu2014,doi:10.1002/adma.201600889,milde2013unwinding}. 
\textcolor{black}{Magnetic frustration provides an exciting alternative way for a centrosymmetric material to host nanometric
	skyrmions ~\cite{PhysRevLett.108.017206,Hirschberger2019,PhysRevB.106.094434,PhysRevX.9.041063,PhysRevB.105.184426}.}
In some frustrated magnets, due to complex non-collinear order arising from competing exchange
interaction, skyrmion spin-textures can be stabilized with the same energy irrespective of their helicity (e.g., left- or
right-winding of spins in a Bloch skyrmion spin texture; see Fig.\ref{scr})~\cite{Leonov2015,PhysRevB.93.184413}.  This stands in stark contrast to the
usual preference for a single skyrmion helicity in intrinsically chiral non-centrosymmetric lattices, which is prefered by
the overall DM anisotropy. Recently, both experimental and theoretical efforts have shown that a disordered skyrmion
phase can be stabilized by magnetic frustration in chiral magnets~\cite{Karubeeaar7043,PhysRevB.100.060407}.
In addition, it has been shown theoretically
that in frustrated magnets uniaxial magnetic anisotropy can lead to various spin structures including isolated magnetic
skyrmions with two additional degrees of freedom: spin helicity and vorticity~\cite{Leonov2015}. 
\textcolor{black}{Besides in frustrated insulating
	magnets~\cite{Yao_2020,PhysRevX.9.041063,PhysRevB.106.224406,PhysRevB.105.184426,PhysRevB.100.245106}, magnetic-frustration-driven a skyrmion phase is also proposed for 4$f$ -based magnets, where
	competing interactions emerge as a result of a coupling between localized spins and conduction electrons on the Fermi
	surface~\cite{Kurumaji914,Hirschberger2019,PhysRev.96.99,PhysRev.106.893}.} 
 The current challenge is to search for new materials that can host nanoscale skyrmions at room
$T$, which would unlock the potential of skyrmions for applications in next-generation spintronic devices.

An interesting phenomenon closely related to skyrmions is the topological anomalous Hall effect. 
Here, a system can induce a finite Hall voltage even in the absence of magnetic field owing to the topological nature of the spin arrangement. 
Recently, Kurumaji \textit{et al}. have found a Bloch-type skyrmion phase near the ordering temperature in a centrosymmetric frustrated magnets Gd$_{2}$PdSi$_{3}$ where triangular planes of Gd$^{3+}$ ions are separated by honeycomb layers of Pd/Si ions~\cite{PhysRevB.84.104105,Kurumaji914,PhysRevLett.108.017206}. 
This discovery confirms that chiral DM interactions are not the only possible stabilization mechanism
behind skyrmion phases and that on a triangular lattice these unconventional topological spin textures can also arise from
the interplay between frustrated RKKY and higher-order biquadratic interactions~\cite{PhysRevB.84.104105,PhysRevB.98.019903,PhysRevB.97.054408,PhysRevLett.108.017206,Leonov2015}. 
Remarkably, the observation
of a topological Hall effect in a frustrated centrosymmetric magnet due to the topological nature of skyrmions (Fig.~\ref{sky})
reveals an intimate connection between spin topology and magnetic frustration. Similar frustration-induced skyrmion
phases have subsequently also been found in the in the breathing kagome lattice magnet Gd$_3$Ru$_4$Al$_{12}$~\cite{Hirschberger2019} and triangular lattice magnets GdRu$_2$Si$_2$~\cite{Khanh2020} and $R_2$RhSi$_3$ ($R$ = Gd, Tb, Dy)~\cite{PhysRevB.101.144440}. 
Spectroscopic techniques, such as $\mu$SR, ESR, and NMR, can provide invaluable microscopic insights into skyrmion dynamics in frustrated quantum materials, possibly akin to Abrikosov vortex dynamics of high-$T_{c}$, type-II superconductors~\cite{PhysRevB.91.224408,Chen2007,Hnsel2001,PhysRevB.98.054428,PhysRevResearch.2.032001,PhysRevMaterials.2.111402,PhysRevB.103.024428}.
Of particular interest are
static and dynamic properties of both frustrated skyrmions \cite{Zhang2017,PhysRevLett.130.106701}, as well as even more exotic skyrmioniums \cite{10.1063/5.0012706},
bimerons \cite{PhysRevB.101.144435}, and bimeroniums \cite{10.1063/5.0034396}. All of these topological spin textures in frustrated magnets usually also have
multiple degrees of freedom that can be controlled by external stimuli. For example, the helicity of a frustrated skyrmion
can be tuned by a current pulse \cite{Zhang2017,PhysRevLett.130.106701}. It has even been proposed that these frustrated spin textures could potentially
be employed as quantum computing bits (qubits) \cite{PhysRevLett.130.106701}.
\section{\textbf{Outlook}}\label{out}
\textcolor{black}{Emergent quantum and topological states that arise in frustrated magnetic materials, including spin ices with magnetic
	monopole excitations, quantum spin liquids with fractionalized spinon excitations, and non-trivial topological spin
	textures, such as skyrmions, remain to be fully understood. Several illustrative examples of the non-intuitive complex
	phenomena that have been discovered in this field in recent years have been presented in this review, with special
	emphasis put on their experimental signatures and detection. The plethora of novel magnetic states, quasiparticles, and
	excitations in frustrated magnets is truly exceptional, but where can we advance from here?
}

\textcolor{black}{Despite the fact that the field of frustrated magnetism has seen enormous progress over the last two decades or so,
	there remain many challenges to be conquered and questions to be addressed. The first challenge is related to materials
	themselves, mainly their quality and purity. In many candidate QSL materials, structural defects such as stacking faults,
	anti-site disorder, site vacancies, or sample off-stoichiometry are, unfortunately, unavoidable and perhaps lead to bond
	randomness, which can crucially influence their ground states. The most obvious way forward is via synthesis and crystal
	growth of purer samples~\cite{chamorro2021chemistry}. 
A current trend in this direction is to focus on compounds where constituent ions are too different, either in the physical and/or chemical sense, to allow for their intermixing during sample synthesis.
Rare-earth based frustrated magnets are especially promising in this respect~\cite{Liu_2018, Arh2021}.
On the other hand, the presence of
defects does not necessarily always have to be detrimental.
One possible future direction is to employ impurities as \textit{in situ} probes of the host QSL state~\cite{Gomilsek2019}, 
in an approach similar to the well-established use of impurities to probe superconductors~\cite{alloul2009defects, RevModPhys.78.373}.
Specifically, a characteristic local response of impurities can potentially be used to efficiently differentiate between different candidate QSL ground states~\cite{PhysRevB.74.165114,florens2006kondo,kim2008kondo,ribeiro2011magnetic,vojta2016kondo}.
Furthermore, details of impurity magnetization can even reveal very subtle information about emergent  spinon--spinon interactions and the corresponding emergent gauge fields~\cite{Gomilsek2019}, which are not directly accessible via any other currently existing experimental approach. 
}\\
\textcolor{black}{Complementary to the growth of bulk samples, advanced deposition techniques offer a platform for designing oxide heterostructures with suitable control of defects~\cite{doi:10.1146/annurev-conmatphys-062910-140445,Hwang2012,RevModPhys.86.1189}. 
 Design and growth of frustrated spin lattices with a chosen
number of atomic planes and well-defined crystallographic orientations, may, in the future, provide an excellent route for
realizing novel magnetic states with non-trivial spin dynamics~\cite{Ruan2021}.
Importantly, this approach allows for systematic control of structural and chemical aspects of the samples. 
Furthermore, artificially designed nanoscale frustrated magnet are an interesting alternative to natural materials.
Lithographically fabricated single-domain ferromagnetic islands arranged on a square lattice, for example, were shown to lead to a 2D analogue of spin ice~\cite{Wang2006} and gave rise to characteristic
associated topological defects~\cite{drisko2017topological}.
Finally, highly tunable optical lattices of interacting Rydberg atoms arranged into frustrated patters could provide a unique opportunity to study the quantum effects under frustration in a more systematic and controlled manner and gain crucial insights~\cite{Li:22,doi:10.1126/science.abi8794}.
}

\textcolor{black}{The next challenge in the field of frustrated magnets relates to experimental detection, characterization, and manipulation of the exotic and usually highly elusive spin states and emergent excitations in these systems.
For instance, QSL states are notoriously difficult to detect and thoroughly characterize~\cite{Wen2019} because they do not spontaneously break any lattice symmetries and develop no local order parameter.
Although many powerful experimental approaches already exist, some of which have been presented in this review, a thorough investigation usually requires a combination of complementary techniques, where each can illuminate a specific aspect of the spin state.
However, by providing only indirect proofs, such as the lack of magnetic ordering or the presence of peculiar spin correlations, even these techniques usually struggle when trying to unambiguously determine the true nature of the QSL state.
Development of novel experimental approaches, and the extension of existing techniques beyond their usual limitations, may finally allow for unambiguous characterization of these states and will therefore be crucial for making further headway in the field. 
}

\textcolor{black}{Spectroscopic tools capable of direct detection of quantum entanglement or of fractionalized excitations, the two
	defining properties of QSLs, would be of particular interest, as they do not currently exist.
Successful approaches will require both theoretical calculations and new experiments. 
Some progress has lately been achieved in determining the entanglement entropy by numerical methods~\cite{Zhao2022, PhysRevA.106.042417}.
Another possibility lies in more sophisticated data analysis to extract lower entanglement bounds from dynamical susceptibility accessible via neutron spectroscopy~\cite{hauke2016measuring, PhysRevB.103.224434}.
The latter technique could possibly also provide the first direct measurement of spin entanglement in QSLs by employing two neutrons prepared in an entangled state that scatter off of different areas of a sample~\cite{shen2020unveiling}.
Fractionalized spinons, on the other hand, could potentially be probed by higher order susceptibilities (beyond linear response), either via nonlinear optical techniques such as THz spectroscopy~\cite{PhysRevLett.122.257401, PhysRevLett.124.117205} or measured directly via spin currents when QSL materials are incorporated into manufactured spintronic devices~\cite{barkeshli2014coherent, PhysRevX.10.031014}.
}

\textcolor{black}{The physics of frustrated magnetism is very rich and diverse, which demands an interdisciplinary approach encompassing broad knowledge from sample synthesis to advanced theory. 
An effort should be devoted towards establishing a greater synergy between theory and experiment that could widen our understanding of the various facets of frustrated magnetic materials.
Establishing more realistic Hamiltonians and understanding the influence of external perturbations will become possible through more refined studies.
Theoretical and numerical studies can be used to guide experiments and to fully understand them, for example on issues related to symmetry or topology.
However, limitations of such studies need to be properly taken into account.
From the experimental perspective, tuning of ground states by external stimuli, such as applied pressure, chemical substitution, or magnetic fields, is a key and very promising
approach that could provide further invaluable insights.
}\\
Extending the theoretical concepts of emergent quasiparticles and gauge fields found in frustrated systems, which we
described in this review, by considering frustrated systems that possess higher-order symmetries, leads us naturally to
frontier theoretical concepts such as fractons~\cite{pretko2020fracton,nandkishore2019fractons,mcgreevy2022generalized}.
These are novel emergent quasiparticles that are immobile in isolation, but can often move by forming bound states, and are predicted to be found in a variety of physical settings, including in QSLs.
What sets them apart is that, unlike spinons and gauge fields of ordinary QSLs which can be described by an emergent topological quantum field theory~\cite{wen2004quantum,RevModPhys.89.041004,wen2019choreographed}, fractons require a novel theoretical description in terms of higher- and categorical symmetries, which are an area of active research  both in theoretical physics and
in pure mathematics. 
Fascinatingly, such models are found to naturally exhibit quantum phases of matter that host emergent gravity (described by emergent general relativity) and exhibit holographic properties related to the AdS/CFT correspondence~\cite{pretko2020fracton,yan2019hyperbolic}.
As similar phenomena are found in fundamental particle physics and quantum gravity research, 
investigations of frustrated magnets, and especially fractons, could thus inform our search for an eventual theory that would merge quantum field theory with general relativity into one all-encompassing description of the natural world.

\textcolor{black}{Apart from their fundamental appeal, the emergent states of frustrated magnets also possess enormous potential for a wide
	range of practical applications.
Therefore, the search and identification of various states and their exotic quasiparticles is also motivated by their potential future use.
The configurational degeneracy of spin ice can be used to store and manipulate data~\cite{doi:10.1126/science.aad8037}, control of magnetic monopoles with external fields is a possible route for creating new spintronics devices~\cite{Khomskii2012}, while artificial spin ice can implement reservoir computing~\cite{heyderman2022spin}, an emerging paradigm for next generation, beyond-neural-network artificial intelligence (AI) \cite{LUKOSEVICIUS2009127}.
Furthermore, non-Abelian anyons arising from the Kitaev honeycomb model and other QSLs provide a potential platform for fault-tolerant quantum computation, as their braiding --- the order in which their position is swapped in space and time --- and fusion are topologically protected due to the non-local encoding in the QSL quantum state~\cite{KITAEV20032,Nayak_2008}.
This appears to be a promising route to building scalable quantum computers and is currently being pursued by major players in the field~\cite{gibney2016inside,google2023non,iqbal2023creation}.
However, development of suitable materials still appears to be a crucial challenge.
Furthermore, the search for topological skyrmion phases has recently drawn great attention in the context of future spintronics devices, such as racetrack~\cite{Fert2013,M_ller_2017} and random-access memories~\cite{Nagaosa2013}, skyrmion-based logic gates and transistors~\cite{Kang2016SkyrmionElectronicsAO}, microwave to radio frequency detectors and oscillators~\cite{Finocchio_2016}, and as a natural platform for reservoir computing and AI~\cite{PhysRevApplied.14.054020,D0MH01603A}.
Even if only a fraction of this potential
transforms into actual technology, frustrated magnets will have become our everyday companions.}

\textcolor{black}{The thriving field of frustrated magnets is extremely rich and diverse.
Their incredible potential for hosting emergent phenomena beyond our wildest imagination has been proven over and over again.
It would certainly be unreasonable to expect anything less in the future.}

\acknowledgments
We acknowledge insightful discussions with Philippe Mendels. P.K. acknowledges the funding by the Science and
Engineering Research Board, and Department of Science and Technology, India through Research Grants. A.Z. and M.G.
acknowledge the financial support of the Slovenian Research and Innovation Agency through Program No. P1-0125 and
Projects Nos. N1-0148, J1-2461, Z1-1852, J1-50008, and J1-50012.
\bibliographystyle{apsrev4-2}
\bibliography{RP_BIB}
\end{document}